\documentclass[11pt]{article}
\usepackage{amssymb,amsmath,amsthm}
\usepackage{amsfonts,bbm}
\usepackage{blindtext}
\usepackage{booktabs}
\usepackage{authblk}
\usepackage{bm}
\usepackage{caption}
\usepackage{comment}
\usepackage[shortlabels]{enumitem}
\usepackage{float}
\usepackage{graphicx}
\usepackage{bbm}
\graphicspath{ {../figures/} }
\usepackage{indentfirst}
\usepackage[left=.8in,right=.8in]{geometry}
\usepackage{mathtools}
\usepackage{mathrsfs}
\usepackage{multirow}
\usepackage{upgreek}
\usepackage{siunitx} 
\usepackage[normalem]{ulem}
\usepackage{cancel}
\usepackage[backend=biber,style=apa, natbib=true,doi=false, url=true, backend=biber, uniquelist=false, uniquename=false, sorting=nyt]{biblatex}
\renewbibmacro{in:}{}
\DeclareNameAlias{default}{last-first/first-last}
\DeclareMathOperator*{\argmin}{arg\,min}

\DeclareMathOperator*{\plim}{plim}
\newcommand{\indicator}[1]{\mathbbm{1}\{#1\}}

\usepackage{array}
\newcolumntype{L}[1]{>{\raggedright\arraybackslash}m{#1}}
\usepackage{rotating}
\usepackage{setspace}
\usepackage{subcaption}
\usepackage{tikz}
\usepackage{ctable}
\usepackage[colorlinks,citecolor=blue,urlcolor=blue,linkcolor=blue]{hyperref}
\usepackage{cleveref}
\crefname{enumi}{}{} 
\usepackage{dcolumn}
\usepackage{float}
\usepackage{pdflscape}

\usepackage[utf8]{inputenc}
\usepackage{longtable}
\usepackage{threeparttable}

\makeatletter

\newcommand{\Rmnum}[1]{\expandafter\@slowromancap\romannumeral #1@}
\newcommand{\E}{\mathbb{E}}

\makeatother

\newtheorem{theorem}{Theorem}[section]

\newtheorem{lemma}{Lemma}[section]

\newtheorem{assumption}{Assumption}[section]

\newtheorem{remark}{Remark}[section]

\crefname{figure}{Figure}{Figures}
\crefname{assumption}{Assumption}{Assumptions}
\crefname{proposition}{Proposition}{Propositions}
\crefname{condition}{Condition}{Conditions}
\crefname{lemma}{Lemma}{Lemmata}
\crefname{example}{Example}{Example}
\crefname{remark}{Remark}{Remark}
\numberwithin{lemma}{section}
\numberwithin{result}{section}
\numberwithin{definition}{section}
\numberwithin{remark}{section}
\numberwithin{proposition}{section}
\numberwithin{condition}{section}

\newfloat{sfigure}{tbp}{sfig}
\allowdisplaybreaks[4]

\begin{document}
	\title{A Consistent ICM-based $\chi^2$ Specification Test}
        \date{\today}
	\author[1]{Feiyu Jiang\footnote{Email: jiangfy@fudan.edu.cn}}
 \author[2]{Emmanuel Selorm Tsyawo\footnote{Corresponding Author. Email: estsyawo@gmail.com}}
 \affil[1]{Department of Statistics and Data Science, Fudan University, Shanghai, China}
 \affil[2]{Department of Economics, Finance and Legal Studies, Culverhouse College of Business, University of Alabama}
	\maketitle
        
        \begin{center}
            First Draft: \texttt{August 30, 2022} 
        \end{center}
	
 \begin{refsection}
		\begin{abstract}
			In spite of the omnibus property of Integrated Conditional Moment (ICM) specification tests, they are not commonly used in empirical practice owing to, e.g., the non-pivotality of the test and the high computational cost of available bootstrap schemes, especially in large samples. This paper proposes specification and mean independence tests based on ICM metrics. The proposed test exhibits consistency, asymptotic $\chi^2$-distribution under the null hypothesis, and computational efficiency. Moreover, it demonstrates robustness to heteroskedasticity of unknown form and can be adapted to enhance power towards specific alternatives. A power comparison with classical bootstrap-based ICM tests using Bahadur slopes is also provided. Monte Carlo simulations are conducted to showcase the excellent size control and competitive power of the proposed test.
			
			\vspace{0.5cm}
			
			\noindent \textit{Keywords:} specification test, mean independence, omnibus, pivotal
			
			\vspace{0.5cm}
			
			\noindent \textit{JEL classification: C12, C21, C52}
		\end{abstract}
		
		\newpage
		
		\section{Introduction}
		 
Model misspecification is a major source of misleading inference in empirical work.  This issue is further compounded when various competing models are available. It is thus imperative that model-based statistical inference be accompanied by proper model checks such as specification tests \citep{stute1997nonparametric}.

Existing tests in the specification testing literature can be categorized into three classes, namely, conditional moment (CM) tests, non-parametric tests, and integrated conditional moment (ICM) tests. The class of CM tests, such as those proposed by \textcite{newey1985maximum,tauchen1985diagnostic}, is not consistent as it relies on only a finite number of moment conditions implied by the null hypothesis \citep{bierens1990consistent}. 
The class of non-parametric tests is therefore proposed as a remedy, see e.g. \citet{wooldridge1992test,yatchew1992nonparametric,hardle1993comparing,hong1995consistent,zheng1996consistent,li1998simple,fan2000consistent,su2007consistent,li2022learning}. The key idea of these test statistics is to nonparametrically estimate the conditional moments -- such as through local smoothing techniques -- and then compare them to their parametric counterparts under the null hypothesis. 

The class of non-parametric tests may encounter challenges such as non-parametric smoothing and suboptimal performance stemming from over-fitting the non-parametric alternative.  In contrast, the class of ICM tests, such as those introduced by \citet{bierens1982consistent,bierens1990consistent,delgado1993testing,bierens1997asymptotic,stute1997nonparametric,delgado2006consistent,escanciano2006consistent,dominguez2015simple,su2017martingale,antoine2022identification}, has gained popularity due to its ability to avoid these issues and detect local alternatives at faster rates. ICM metrics, on which ICM tests are based, also appear in other contexts:  martingale difference hypothesis tests \citep{escanciano2009lack}, joint coefficient and specification tests \citep{antoine2022identification}, model-free feature screening \citep{zhu2011model, shao2014martingale,li2023generalized}, model estimation \citep{escanciano2018simple,tsyawo2023feasible}, specification tests of the propensity score \citep{sant2019specification}, and tests of the instrumental variable (IV) relevance condition in ICM estimators \citep{escanciano2018simple,tsyawo2023feasible}.
 
Despite their advantages, ICM tests are not widely used in empirical research \citep{escanciano2009simple,dominguez2015simple}. First, ICM test statistics are not pivotal under the null hypothesis, thus critical values cannot be tabulated analytically \citep{bierens1997asymptotic,dominguez2015simple}. Second, ICM tests—when implemented via the wild bootstrap—tend to be computationally costly, as they require estimations on resampled data to compute $p$-values. {    Moreover, the wild bootstrap-based ICM test is arguably unsuitable for limited dependent outcome variable models such as the logit, because bootstrap replicates of the outcome may fail to respect the outcome variable's limited support.} While the more recent multiplier bootstrap approach to ICM specification testing \citep[e.g.,][]{escanciano2009simple,li-song-consistent-2022,escanciano-2024-gaussian} offers a substantial computational advantage relative to the wild bootstrap by avoiding model re-estimation, it still entails non-negligible computational complexity due to re-sampling and the removal of the effect of estimating nuisance parameters. Third, although ICM tests are omnibus \citep{bierens1982consistent,stute1997nonparametric,dominguez2015simple}, they only have substantial local power against alternatives in a finite-dimensional space \citep{escanciano2009lack}. Moreover, it is not obvious how to leverage prior knowledge of potential directions under the alternative to enhance the power of existing ICM tests.

This paper proposes a consistent $\chi^2$-test for the unified ICM framework of mean independence and specification testing. The key idea is to augment the ICM metric with a user-specified  non-degenerate transformation of the conditioning covariates as in CM tests, which removes the first-order degeneracy inherent in classical ICM tests and yields a pivotal test statistic. 
The approach accommodates endogenous regressors, instrumental variables (IV), and heteroskedasticity of unknown form in both linear and non-linear models. As an ICM-based test, it is omnibus and capable of detecting a wide range of model misspecifications, including violations of IV exogeneity. Similar to the multiplier bootstrap method \citep{escanciano2009simple}, our approach is suited for models with possibly non-additively separable errors or non-continuous outcomes. 

Compared to existing ICM tests, our test has three advantages. First, it can be implemented as a $\chi^2$- or two-sided $t$-test, which can be interpreted more easily when compared to bootstrap-based tests. Second, the proposed test does not require bootstrap calibration of critical values; hence, it is computationally fast and remains feasible even in very large samples. Third, although our proposed test is not optimal, its power can be enhanced with the knowledge of directions under the alternative or, more generally, with directions the researcher may have in mind, whereas ICM tests lack this property. Therefore, we consider our test as a bridge between CM tests and ICM tests.  We also acknowledge that the implementation of our test involves a tuning parameter, which is used in computing the generalized inverse that forms the Wald-type test statistic.
	
This paper is not the first attempt at circumventing the non-pivotality of ICM test statistics. \textcite{bierens1982consistent} approximates the critical values of the ICM specification test using Chebyshev's inequality for first moments under the null hypothesis, which is subsequently improved upon by \textcite{bierens1997asymptotic}.  \textcite{bierens1982consistent} also proposes a $\chi^2$-test based on two estimates of Fourier coefficients and a carefully chosen tuning parameter. Simulation evidence therein shows a high level of sensitivity to the tuning parameter. Besides, estimating Fourier coefficients no longer makes the test ICM as the test statistic is no longer ``integrated". Another attempt in the literature is the conditional Monte Carlo approach of \textcite{hansen1996inference,de1996bierens}. These are, however, computationally costly as \textcite{bierens1997asymptotic} notes. {Recent studies have also revisited the goal of constructing pivotal and computationally feasible specification tests within the conditional moment framework. \textcite{raiola2024testing} draws inspiration from the classical Pearson’s $\chi^2$ test by partitioning the sample space into cells. Although this test enjoys pivotality and excellent computational scalability, its power is inherently tied to the chosen partition and is therefore not fully omnibus: smooth model deviations that do not necessarily manifest as changes in constructed cells may go undetected. \textcite{li2025powerful} leverages sample splitting and learns the optimal projection direction in the reproducing kernel Hilbert space via support vector machines (SVM). This means that it can achieve high local power when the learned projection aligns with the alternative; however, sample splitting and hyper-parameter tuning (in both kernels and SVM) may introduce finite-sample variability.}

The rest of the paper is organized as follows. \Cref{Sect:Preliminaries} provides a brief presentation of ICM  metrics, a new metric, and its omnibus property. \Cref{Sect:Test} proposes the $\chi^2$-test statistic and derives its limiting distribution under the null, local, and alternative hypotheses within the unified framework of mean independence and specification tests. Monte Carlo simulations in \Cref{Sect:MC_Sim_Spec} compare the empirical size and power of the $ \chi^2 $ specification test to the multiplier and wild bootstrap-based ICM specification tests, and  \Cref{Sect:Conclusion} concludes. All technical proofs and additional simulation results are relegated to the supplementary material.

\paragraph{Notation:} For $a\in\mathbb{R}^p$, we denote its transpose by $a^{\top}$, and its Euclidean norm as $\|a\|$.  We denote $\mathrm{i}$ as the imaginary unit which satisfies $\mathrm{i}^2=-1$. ``$\overset{p}{\rightarrow}$" and ``$\overset{d}{\rightarrow}$" denote convergence in probability and distribution, respectively. Throughout the paper, for a random vector $W$, we denote $W^\dagger$ as its independent and identically distributed ($iid$)  copy, and write $\E_n W=n^{-1}\sum_{i=1}^n W_i$ as the empirical mean for $iid$ copies $\{W_i\}_{i=1}^n$ of $W.$  To cut down on notational clutter, $\widetilde{W}$ is sometimes used to denote the centered version of a random variable $W$, i.e., $\widetilde{W}:=W-\E W.$

\section{The Framework}\label{Sect:Preliminaries}

In this section, we briefly discuss ICM metrics, their omnibus property, and a new omnibus metric that characterizes mean independence.

\subsection{ICM Metrics}

For a random variable $U\in\mathbb{R}$ and  a random vector $Z\in \mathbb{R}^{p_z},$ we say $U$ is mean-independent of $Z$, if
		\begin{equation}\label{eq:CM}
		    		\E[U \mid Z]=\E[U] \ almost \ surely\ (a.s.),
		\end{equation}
 otherwise, $U$ is mean-dependent on $Z$. To characterize the relationship \eqref{eq:CM}, the existing literature puts much effort into studying ICM mean dependence metrics of the form:
 \begin{flalign*}\label{ICMmetric}
 T(U \mid Z;\nu)=\int_{\Pi} \Big|\E[(U-\E U)w(s,Z)]\Big|^2\,\nu(ds)
\end{flalign*}
where $w(s,Z)$ is a weight function indexed by $s \in \Pi$, and $\nu$ is a measure on $\Pi$. A notable feature of $T(U \mid Z;\nu)$ is its \emph{omnibus property}, namely, $T(U \mid Z;\nu)=0$ if and only if \eqref{eq:CM} holds,  see e.g., \textcite[Theorem 1.2]{shao2014martingale}. The omnibus property guarantees the consistency of ICM tests. Therefore, a larger value of $T(U \mid Z;\nu)$ indicates a stronger mean dependence of $U$ on $Z$. 

Under suitable conditions, the ICM metric has the general form
\begin{align*}
    T(U \mid Z;\nu) 
    &= \int_{\Pi} \Big(\E[(U-\E U)w(s,Z)]\Big)\overline{\Big(\E[(U-\E U)w(s,Z)]\Big)}\nu(ds)\\
    &= \E\Big[(U-\E U)(U^\dagger-\E U)\int_{\Pi} \big(w(s,Z)\overline{w(s,Z^\dagger)}\big)\nu(ds) \Big]\\
    &:= \E\Big[(U-\E U)(U^\dagger-\E U)K(Z,Z^\dagger)\Big]\\
    &:= \mathrm{ICM}(U \mid Z),
\end{align*}where $\overline{w(\cdot,\cdot)}$ denotes the complex conjugate of $w(\cdot,\cdot)$ and $K(\cdot,\cdot)$ denotes the kernel. \Cref{tab:kernels} provides a few examples of commonly used ICM kernels, see \citet{li2023generalized} for a recent review.
\begin{table}[H]
\centering
\caption{Examples of ICM Kernels}
\label{tab:kernels}
\begin{tabular}{@{}llll@{}}
\toprule
 $w(s,Z)$ & $\nu(ds)$ & $K(Z,Z^\dagger)$ & Reference \\
\midrule
$\exp( \mathrm{i} Z^\top s)$ & $ \prod_{l=1}^{p_Z}\phi(s_l)ds_l$ & $\exp\left(-0.5 \| Z - Z^\dagger\|^2\right)$ & \citet{bierens1982consistent} \\
$1(Z \le s)$ & $dF_Z(s)$ & $1 - F_Z(Z \lor Z^\dagger)$ & \citet{stute1997nonparametric} \\
$\exp( \mathrm{i} Z^\top s)$ & $\big(c_{p_Z}||s||^{1+p_Z}\big)^{-1} ds$ & $ -\| Z-Z^\dagger\| $ & \citet{shao2014martingale} \\
\bottomrule
\end{tabular}
\footnotesize

        \textit{Notes: $\phi(x) = \frac{1}{\sqrt{2\pi}} e^{-\frac{1}{2}x^2} $ is the standard normal probability density function, and $ c_p = \frac{\pi^{(1+p)/2}}{\Gamma((1+p)/2)}, \ p \geq 1 $ where $ \Gamma(\cdot) $ is the complete gamma function.}
\end{table}

Examples of weight functions from the literature include the step function $w(s,Z) = \mathrm{I}(Z\leq s) $, e.g.,  \textcite{stute1997nonparametric,dominguez2004consistent,delgado2006consistent,escanciano2006goodness,zhu2011model}; a one-dimensional projection in the step function $w(s,Z) =\mathrm{I}(Z^{\top}s_{-1}\leq s_1)$, e.g.,  \textcite{escanciano2006consistent,kim2020robust}; the real exponential $w(s,Z) = \exp(Z^{\top}s)$, e.g., \textcite{bierens1990consistent}; and the complex exponential $w(s,Z)=\exp(\mathrm{i}Z^{\top}s)$, e.g., \textcite{bierens1982consistent,shao2014martingale,antoine2022identification}. The space $\Pi$ in \textcite{escanciano2006consistent,kim2020robust} is given by $\Pi = \mathbb{R}\times \mathbb{S}_{p_z}$ where $\mathbb{S}_{p_z}$ denotes the space of $p_z\times 1$ vectors with unit Euclidean norm while $\Pi=\mathbb{R}^{p_z}$ for the other works mentioned above. See \textcite[Lemma 1]{escanciano2006goodness} for a general characterization of ICM weight functions.

The omnibus property of ICM metrics thus translates as 
\begin{equation}\label{eqn:Prop_Omnibus}
   \mathrm{ICM}(U \mid Z)=0 \quad \text{if and only if }~~ \E[U \mid Z]=\E[U] \ almost \  surely \ (a.s).
 \end{equation}
\noindent When \( U \) and \( Z \) are observed, a natural empirical estimator of \( \mathrm{ICM}(U \mid Z) \) is given by the (modified) U-statistic
\begin{equation*}
\widehat{\mathrm{ICM}}_n(U \mid Z) = \frac{1}{n(n-1)} \sum_{i \neq j} (U_i - \E_n[U])(U_j - \E_n[U]) K(Z_i, Z_j).
\end{equation*}
As with most existing ICM-based tests, the asymptotic null distribution of the corresponding test statistic (under \eqref{eq:CM}) is non-pivotal \citep[Theorem 3]{bierens1997asymptotic}:
\begin{equation}\label{eqn:ICM_asymp}
    n\widehat{\mathrm{ICM}}_n(U \mid Z) \overset{d}{\rightarrow}  \sum_{k=1}^{\infty}\lambda_k G_k^2,
\end{equation} granted $\E [U^2U^{\dagger 2}K^2(Z,Z^{\dagger})]<\infty$. Here, $\{G_k\}_{k=1}^{\infty}$ is a sequence of $iid$ standard Gaussian random variables, and $\{\lambda_k\}_{k=1}^{\infty}, \ \lambda_k\geq 0$, is a sequence of non-increasing coefficients that depend on the distribution of $[U,Z^\top]^\top$ and the kernel $K(\cdot,\cdot)$. Thus, the limiting distribution of \( n\widehat{\mathrm{ICM}}_n(U \mid Z) \) under \eqref{eq:CM} is non-pivotal, and re-sampling/bootstrap techniques are usually required to obtain valid \( p \)-values for inference.

\subsection{A new characterization of mean independence}\label{Sub_Sect:Test_Mean_Indep}

Consider the following test hypotheses of mean independence:
	\begin{flalign}\label{eqn:hypothesis}
		\begin{split}
		\mathbb{H}_o&: \E[U \mid Z] - \E[U] = 0 \ a.s.;\\
		\mathbb{H}_a&: \mathbb{P}\big(\E[U \mid Z]=\E[U]\big) < 1.
	    \end{split}
	\end{flalign}
The above hypotheses of interest, in view of the omnibus property \eqref{eqn:Prop_Omnibus}, can be restated as testing $ \mathrm{ICM}(U \mid Z)=0.$ Indeed, by the Law of Iterated Expectations (LIE), 
\begin{equation}
\begin{split}
    \mathrm{ICM}(U \mid Z)=& \E\left\{K(Z,Z^\dagger) (U^\dagger-\E U) \E[(U-\E U) \mid Z,Z^\dagger,U^\dagger]\right\} \\
    =& \E\left\{K(Z,Z^\dagger)(\underbrace{\E[U \mid Z]-\E U}_{=0 \, a.s. \text{ under } \mathbb{H}_o})(U^\dagger-\E U)\right\}.
\end{split}\label{eqn:LIE2}
\end{equation}

\noindent Under $\mathbb{H}_o$,  $\E[U \mid Z]$ is degenerate, i.e., $\E[U \mid Z]-\E U=0\ a.s.$ This is the first-order degeneracy problem in U/V-statistic-based estimator for  \eqref{eqn:LIE2} that accounts for the non-pivotal limiting distribution in \eqref{eqn:ICM_asymp}. In this regard, we propose to replace the role of  $U$ in the ICM metric with a random vector $V\in \mathbb{R}^{p_v}$ constructed such that $\E[V \mid Z]$ is non-degenerate under both the null and alternative hypotheses while preserving the omnibus property \eqref{eqn:Prop_Omnibus}. In other words, we propose to measure the mean independence of $U$ on $Z$ with the assistance of a random vector $V$ by 
$$
\delta(V):= \E\left[K(Z,Z^\dagger) (U^\dagger-\E U) (V - \E V) \right].
$$
The resulting test relies on the empirical estimator of $\delta(V)$.

 The two key properties---\textit{first-order non-degeneracy} and \textit{omnibusness}---ensure a pivotal limiting distribution under the null and the consistency of the test, respectively. Specifically, we require  (1) $V$ to be mean dependent on $Z$ to rule out first-order degeneracy, and (2) $V$ to be chosen such that $\delta(V)\neq 0$ under $\mathbb{H}_a$.  The first condition is easy to satisfy by selecting $V$ to be a measurable function of $Z$. In contrast, the second condition is non-trivial: randomly selecting $V$ may lead to the same limitation as the classical CM test if it happens to be orthogonal to the direction of departure under $\mathbb{H}_a$. This trade-off mirrors the strengths and weaknesses of existing methods: while the CM test is first-order non-degenerate, it may lack consistency under $\mathbb{H}_a$; conversely,  the ICM test is consistent under $\mathbb{H}_a$, but suffers from first-order degeneracy, resulting in a non-pivotal limiting distribution under $\mathbb{H}_o$. 

To this end, the proposed testing procedure integrates the respective strengths of two existing approaches in the literature: the pivotality of the CM test and the omnibus property of the ICM test. 
In particular, we focus on the quantity \begin{equation}\label{eqn:V_h}
    V_h = \big[ h(Z), \ U - h(Z) \big]^\top,
\end{equation}such that  
\begin{equation}\label{eqn:delta}
    \delta_h:=\delta(V_h)= \begin{bmatrix}
        \E[K(Z,Z^\dagger) (h(Z)-\E[h(Z)]) ({U}^\dagger-\E U)]\\
        \mathrm{ICM}(U \mid Z) - \E[K(Z,Z^\dagger) (h(Z)-\E[h(Z)]) ({U}^\dagger-\E U)]
    \end{bmatrix}.
\end{equation}

The two components in $V_h$, namely $h(Z)$ and $U - h(Z) $, serve different purposes. First, one can view $h(Z)$ as the assistant function where prior information about the alternative can be incorporated, as done in the classical CM test. Second,  the inclusion of $ U $ in the second element draws inspiration from ICM tests, which gives the omnibus property of our test, that is, it has non-trivial power in all possible directions of the alternative. This structure acts as a safeguard in scenarios where the chosen assistant $h(Z)$ is (nearly) orthogonal to the alternative, a situation in which traditional CM tests suffer from trivial power. Moreover, the $-h(Z)$ in the second element ensures that $\delta_h$ is first-order non-degenerate and that the proposed test statistic is pivotal ($\chi^2$ distributed)  under the null hypothesis. We remark that the choice of 
$V$ can be readily extended beyond two dimensions, although in this paper we focus on the specific two-dimensional case for clarity and analytical tractability.

The following fundamental result establishes the omnibus and first-order non-degeneracy properties of  $\delta_h $.

\begin{lemma}\label{lem_key}
Let \( h(Z) \) denote an arbitrary measurable and non-degenerate function of $Z$, in the sense that \( \mathbb{P}\big( h(Z) = \mathbb{E}[h(Z)] \big) < 1 \).  Then:     
(a) \( \delta_h \) defined in \eqref{eqn:delta} satisfies the \emph{omnibus property}, that is, \( \|\delta_h\| = 0 \) if and only if \eqref{eqn:Prop_Omnibus} holds; (b) \( \delta_h \) exhibits \emph{first-order non-degeneracy}.
\end{lemma}
\begin{proof}
Let $\delta_h^{(1)}=\E[K(Z,Z^{\dagger})(h(Z) - \E[h(Z)])({U}^\dagger-\E U)]$ and $\delta_h^{(2)}=\mathrm{ICM}(U \mid Z)-\delta_h^{(1)}$. Then from \eqref{eqn:delta} we have that $\delta_h=[\delta_h^{(1)},\delta_h^{()}]^{\top}$, and 
$\delta_h^{(1)} + \delta_h^{(2)} = \mathrm{ICM}(U \mid Z) > 0$ under $\mathbb{H}_a$. This implies that under $\mathbb{H}_a$, at least one element in $\delta_h$ is strictly positive (hence strictly different from zero); $\delta_h$ satisfies the omnibus property. Note that $h(Z)$ is non-degenerate, and that under $\mathbb{H}_o$, $\E \big[V_h - \E[V_h] \mid Z\big] = \big[(h(Z) - \E[h(Z)]),\ \E [U \mid Z] - \E U - (h(Z) - \E[h(Z)]) \big]^{\top} = (h(Z) - \E[h(Z)])[1,-1]^{\top}\neq 0 $ with probability one. It follows that $\delta_h$ avoids first-order degeneracy.
\end{proof}

\Cref{lem_key} implies that, with our construction of $V_h$ in \eqref{eqn:V_h}, $\delta_h$ is non-zero  under the alternative. Therefore,   $\|\delta_h\|$ can be interpreted as a metric for conditional mean independence,  i.e. $\|\delta_h\|=0$ if and only if $\mathbb{E}[U\mid Z]=\mathbb{E}[U] \ a.s. $ While the performance of $\delta_h$ depends on the choice of $h(\cdot)$, we note that similar concerns arise in  ICM tests, particularly regarding the selection of kernels. A practitioner who is agnostic about the alternative can use  $h(Z) = \exp\big(Z^{\top}\mathbbm{1}_{p_z}/\sqrt{p_z}\big) $ where $\mathbbm{1}_{p}$ denotes a $p\times 1$ vector of ones.\footnote{The specification is appropriate when \( p_z \) is small. For larger values of \( p_z \), it is advisable to de-mean each element of \( Z \) to ensure the variance of \( h(Z) \) is robust to \( p_z \).} The Maclaurin's series expansion, namely $\exp(z) = \sum_{l=0}^{\infty} z^l/{l!}$ shows the potential of $h(Z) = \exp\big(Z^{\top}\mathbbm{1}_{p_z}/\sqrt{p_z}\big) $ to capture different directions under the alternative.\footnote{ { The covariance matrix of $Z$ ought to be positive definite, i.e. $\mathbb{P}(Z^{\top}a=c)<1$  $\forall\, a\in\mathbb{R}^{p_z}$ and $ \forall \, c\in\mathbb{R}$, to avoid  possible degeneracy in the linear combination $Z^\top\mathbbm{1}_{p_z}$.}}

A key message of \Cref{lem_key} is that symmetrizing any non-degenerate measurable function $h(\cdot)$ of $Z$ about $U$ guarantees the omnibus and first-order non-degeneracy properties of $\delta_h$.  
The requirement on $h(Z)$ in \Cref{lem_key} can hardly be termed a ``condition," as it imposes only measurability and non-degeneracy. The proof of \Cref{lem_key} provides an insight into the inclusion of $U$ linearly in the construction of $V$. This brings in the term $\mathrm{ICM}(U \mid Z)$ in $\delta_h$ that is strictly positive under $\mathbb{H}_a$ and contributes power under the alternative.  The proposed test thus draws its consistency from the ICM metric.

Since $h(Z)$ can be chosen arbitrarily, the practitioner does not bear the burden of ``carefully" selecting functions such as polynomials that provide power by approximating $\E[U \mid Z]$ under $\mathbb{H}_a$,  as required in non-parametric specification tests, e.g., \citet{wooldridge1992test,yatchew1992nonparametric,zheng1996consistent}. In addition, the user-specified $h(Z)$ provides extra flexibility as the practitioner can use it to augment the power of the test in given directions, unlike existing ICM-based bootstrap tests.

\begin{remark}
We assume $p_z$ is fixed in this study. In high-dimensional settings where $p_z \rightarrow \infty $,  \citet[Remark 2.2]{zhang2018conditional} demonstrates that $\mathrm{ICM}$ criteria only capture linear dependence. Our $\chi^2$-test derives its consistency property and part of its power from the $\mathrm{ICM}$ metric so it cannot be expected to perform better in high dimensions. Recent studies have explored the challenge of testing independence between high-dimensional random vectors using distance covariance-based statistics \citep{szekely2007measuring}, revealing that the performance of such tests depends critically on the relative growth rates of the dimensions of the random variables and the sample size $n$ \citep{zhu2020distance,chakraborty2021new,gao2021asymptotic,zhang2024statistical}. Consequently, extending the current testing procedure to high dimensions requires a separate study that is deferred to
future research.
 \end{remark}

\subsection{Extension to  Testing the Nullity of $\E[U \mid Z]$}\label{subsec:nullity1}
In some applications, such as specification testing, one may be interested in the almost sure nullity of $\E[U \mid Z]$ directly, i.e.
\begin{flalign}\label{eqn:hypothesis2}
		\begin{split}
		\mathbb{H}_o^*&: \E[U \mid Z]=0 \ a.s.;\quad
		\mathbb{H}_a^* : \mathbb{P}(\E[U \mid Z]=0) <1.
	    \end{split}
	\end{flalign}
This is an augmented version of \eqref{eqn:hypothesis} which further imposes $\E[U]=0$, i.e., a joint hypothesis of conditional mean independence and nullity of the unconditional mean. To this end, we follow \citet{su2017martingale} by augmenting $\delta_h$  with an additional quantity that accounts for $\E[U] = 0$ under $\mathbb{H}_o^*$. In particular, one may consider the metric  
$$
\delta_h^*=\delta_h + \E|K(Z,Z^{\dagger})|\E U \E\big[h(Z), \ U - h(Z)) \big]^\top.
$$

The following result shows that ${\delta}_h^*$ has the omnibus property with respect to \eqref{eqn:hypothesis2}.
\begin{lemma}\label{lem:key_nullity}
    ${\delta}_h^*=0$ if and only if $\E[U \mid Z]=0 \ a.s. $
\end{lemma}
\noindent To conserve space, the discussion below focuses on $\delta_h$. Theoretical properties of the $\chi^2$-test based on $\delta_h^*$ are presented in Section~\ref{sec:deltastar} of the supplementary material.

\section{Test Statistic and Theoretical Properties}\label{Sect:Test}
In this section, we construct the test statistic based on $\delta_h$. In Section~\ref{SubSect:Uni_frame}, we present a framework that unifies the tests of mean independence and model specification. We analyze the asymptotic behavior of $\widehat{\delta}_h$, the empirical estimator for $\delta_h$, and the $\chi^2$-distributed test statistic under the null hypothesis, as well as under local and fixed alternatives in Sections \ref{SubSect:Test_MeanIndep} and \ref{Subsec:test}, respectively. Section~\ref{SubSect:Bahadur} uses Bahadur slopes to compare the power of the $\chi^2$-test and bootstrap-based $\mathrm{ICM}$ tests.

\subsection{A unified framework} \label{SubSect:Uni_frame}

The mean independence testing problem and the specification testing problem are two closely related problems in econometric analyses. Here, we present a unified framework for both tasks. Let $X\in\mathbb{R}^{p_x}$ and $Z\in\mathbb{R}^{p_z}$ be two random vectors. We consider an econometric model defined by the following conditional moment restriction \begin{equation}\label{moment_res}
    \mathbb{E}[U(X;\theta_o)|Z]=0\quad a.s.,
\end{equation}
for a unique model parameter $\theta_o\in\Theta  \subset  \mathbb{R}^k$. Here, $U:\mathbb{R}^{p_x}\times \Theta\mapsto \mathbb{R}^{p_u}$ is the econometric model that is assumed to be known, and $\Theta$ is the parameter space. Given $iid$ observations $\{X_i,Z_i\}_{i=1}^n$, and an empirical estimator $\widehat{\theta}_n$ for $\theta_o$, we are interested in  assessing the model specification in \eqref{moment_res}.

The above framework encompasses a wide range of applications, including treatment effect analyses (e.g., \citealp{callaway2022treatment,sant2019specification}), Euler and Bellman equations \citep{hansen1982generalized, escanciano2018simple}, the hybrid New Keynesian Phillips curve \citep{choi2021generalized}, forecast rationality \citep{hansen1980forward}, and tests of conditional equal and predictive ability \citep{giacomini2006tests}; see \citet{li2022learning} for a comprehensive discussion.

In the mean independence testing problem outlined in Section \ref{Sub_Sect:Test_Mean_Indep}, the expressions for \( \mathrm{ICM}(U \mid Z) \) and \( \delta_h \) involve the nuisance parameter \( \mathbb{E} U \). In this case, we can simply set $X=U$, $\theta_o=\mathbb{E}U$, and $U(X;\theta)=X-\theta$,  then testing for  \eqref{moment_res} reduces to the hypothesis \eqref{eqn:hypothesis}. In the class of nonlinear models with additively separable errors $X^{(1)} = g(X^{(-1)};\beta_o) + U $, where $X^{(1)}$ denotes the first element of $X$, $X^{(-1)}$ is the sub-vector of $X$ that excludes $X^{(1)}$, and $U$ is the model error (up to a location shift), we let 
 \begin{align*}
     U(X;\theta) &= X^{(1)} - g(X^{(-1)};\beta) - \theta_c, 
 \end{align*} 
where $\theta = [ \theta_c,\ \beta^\top ]^\top $, and   $\theta_c$ is the location parameter such that $\mathbb{E}[X_1-g(X_{-1};\beta)-U]=\theta_c$.  Including the location parameter $\theta_c$ in $\theta$  becomes necessary as the ICM metric and $\delta_h$ can only identify mean independence up to an unknown nuisance mean shift. Nullity testing is discussed in Section \ref{subsec:nullity1}.

In practice, a natural estimator for $\delta_h$ in \eqref{eqn:delta} is given by
\begin{equation}\label{eqn:deltaV}
    \widehat{\delta}_h = \frac{1}{n(n-1)}\sum_{i=1}^n\sum_{j\neq i}K(Z_i,Z_j)U(X_i;\widehat{\theta}_n)\big[(h(Z_j) - \E_n[h(Z)]) , \ U(X_j;\widehat{\theta}_n) - (h(Z_j) - \E_n[h(Z)]) \big]^\top.
\end{equation}
We remark that, in the mean independence testing \eqref{eqn:hypothesis}, we have $U(X_i;\widehat{\theta}_n)=U_i-\widehat{\theta}_n$ with $\widehat{\theta}_n=\E_nU.$
\subsection{Asymptotic Distribution} \label{SubSect:Test_MeanIndep}

In this section, we analyze the asymptotic behavior of \( \widehat{\delta}_h \) under the null hypothesis,  local alternatives, and fixed alternatives. Let \( D := [U, X^\top, Z^\top]^\top \), and define \( \widetilde{U} := U(X; \theta_o) \) such that $\E\widetilde{U}=0.$ We consider the following sequences of local alternatives: Pitman local alternatives \(
    \mathbb{H}_{an}: \mathbb{E}[\widetilde{U} \mid Z] = n^{-1/2} a(Z)\) and milder (non-Pitman) local alternatives \(
    \mathbb{H}_{an}': \mathbb{E}[\widetilde{U} \mid Z] = n^{-1/4} a(Z)
    \), where \( a(Z) \) is a non-degenerate measurable function of \( Z \) satisfying \( \mathbb{E}[a(Z)] = 0 \). For completeness, we impose the following regularity conditions. 

\begin{assumption}\label{ass:Sampling_iid}
$\{D_i\}_{i=1}^n$ are independently and identically distributed ($iid$).
\end{assumption}

\begin{assumption}\label{ass:bound_psiD}
$\E\big[K(Z,Z^{\dagger})^4\big] + \E\big[U^4\big] + \E\big[h(Z)^4\big] < \infty$.
\end{assumption}

\begin{assumption}\label{ass:diff_Utheta}
    
\( U(X; \theta) \) is differentiable in \( \theta \) and admits the following first-order expansion:
\[
U(X; \theta) = \widetilde{U} + \frac{\partial U(X; \bar{\theta})}{\partial \theta'} (\theta - \theta_o) := \widetilde{U} - G(X; \bar{\theta})^\top (\theta - \theta_o),
\]
where \( G(x; \theta) \) is continuous in \( \theta \) for all \( x \) in the support of \( X \), \( \E\big[\|G(X; \theta)\|^2\big] <\infty \) for all $\theta$, and \( \bar{\theta} \) lies on the line segment between \( \theta \) and \( \theta_o \), i.e., \( \| \bar{\theta} - \theta_o \| \leq \| \theta - \theta_o \| \).
\end{assumption}

\begin{assumption}\label{ass:theta_asymp_lin}
    $\widehat{\theta}_n$ is a $\sqrt{n}$-consistent estimator of $\theta_o$ with the asymptotically linear representation
\[
    \sqrt{n}(\widehat{\theta}_n - \theta_o) = \frac{1}{\sqrt{n}}\sum_{i=1}^n \xi_{\theta,i} + o_p(1)
\]where $ \xi_{\theta,i}\in\mathbb{R}^k$ satisfies that $\E[\xi_{\theta,i}]=\bf{0}$, $ \E\big[\|\xi_{\theta,i}\|^2\big] < \infty $.
\end{assumption}

\noindent The assumption of $iid$ observations in \Cref{ass:Sampling_iid} simplifies the theoretical analyses.\footnote{Our results are extensible to weak temporal dependence, panel data, and clustered data settings, but this lies beyond the scope of the current paper.} \Cref{ass:bound_psiD} is standard, e.g., \citet[Sect. 5.5.1 Theorem A]{serfling1980}; it is needed to establish the asymptotic normality of $\sqrt{n}(\widehat{\delta}_h-\delta_h)$. Although the moment restriction on $U$ cannot be relaxed, that of $K(\cdot)$ can be relaxed through the choice of bounded kernels or suitable transformations of $Z$. \Cref{ass:diff_Utheta} regulates the smoothness and moment conditions of the possibly nonlinear parameterized error $U(X;\theta)$. \Cref{ass:theta_asymp_lin} is akin to \citet[Assumption A3]{escanciano2009lack}; it assumes the estimator $\widehat{\theta}_n$ is consistent with respect to its probability limit $\theta_o$. The Bahadur linear representation holds for a broad class of estimators, including commonly used estimators such as (nonlinear) least squares, maximum likelihood, and the (generalized) method of moments, under standard regularity conditions.

The result on the limiting behavior of the proposed test statistic relies on the following asymptotically linear representation of 
$\widehat{\delta}_h$ in the following lemma. Define $\widetilde{h}(Z):=h(Z)-\mathbb{E}h(Z)$.
\begin{lemma}\label{lem:del_asymp_lin}
    Under \Cref{ass:bound_psiD,ass:Sampling_iid,ass:theta_asymp_lin,ass:diff_Utheta}, $\widehat{\delta}_h$ has the following asymptotically linear representation:
    \begin{align*}
        \sqrt{n}\big(\widehat{\delta}_h - \delta_h\big)
    = \frac{1}{\sqrt{n}}\sum_{i=1}^{n}\xi_h(D_i) + o_p(1),
\end{align*}
where $ \xi_h(D_i):= 2\big[\xi_{h,1}(D_i), \ \xi_{h,2}(D_i) - \xi_{h,1}(D_i) \big]^\top $, $
    \delta_h = \big[\delta_h^{(1)},\delta_h^{(2)}\big]^{\top}
$, 
\begin{align*}
    \xi_{h,1}(D_i): &= \psi_h^{(1)}(D_i) - \delta_h^{(1)} - H_1^\top[\xi_{\theta,i}^\top, \widetilde{h}(Z_i)]^\top , \\
    \xi_{h,2}(D_i): &= \psi_U^{(1)}(D_i)-\mathrm{ICM}(U\mid Z) - H_2^\top\xi_{\theta,i},\\
    \psi_h^{(1)}(D_i) : &=\E[\psi_h(D_i,D_j)|D_i],\\
    \psi_U^{(1)}(D_i): &=\E[\psi_U(D_i,D_j)|D_i],
\end{align*}
$\delta_h^{(1)}=\E[K(Z,Z^{\dagger})\widetilde{h}(Z)\widetilde{U}^\dagger], $  $\delta_h^{(2)}=\mathrm{ICM}(U \mid Z)-\delta_h^{(1)}$, $ \psi_h(D_i,D_j):= K(Z_i,Z_j)\big\{ \widetilde{h}(Z_j)\widetilde{U}_i + \widetilde{h}(Z_i)\widetilde{U}_j \big\}/2 $, $\psi_U(D_i,D_j) := K(Z_i,Z_j)\widetilde{U}_i\widetilde U_j$, 
$ H_1= (1/2)\Big[ \E\big[K(Z, Z^\dagger)\widetilde{h}(Z^\dagger)G(X;\theta_o)\big]^\top
        ,\ \E\big[K(Z,Z^\dagger)\widetilde{U}\big]
    \Big]^\top $, and 
$H_2= \E\big[K(Z,Z^\dagger){\widetilde{U}^\dagger}G(X;\theta_o)\big] $.
\end{lemma}

The proof of \Cref{lem:del_asymp_lin} exploits the classical Hoeffding decomposition  \citep{serfling1980} for U-statistics, and provides intuition for the two different rates of local alternatives. If the user-defined assistant $h(Z)$ is non-orthogonal to the alternative direction $\E\big[a(Z^{\dagger})K(Z,Z^{\dagger}) \mid Z\big]$ under the Pitman local alternatives $ \mathbb{H}_{an}: \ \mathbb{E}[\widetilde{U}|Z]=n^{-1/2}a(Z)$, then $\sqrt{n}\delta_h^{(1)}$ is non-diminishing, i.e.  
$$
\sqrt{n}\delta_h^{(1)} = \E\big[K(Z,Z^{\dagger})\widetilde{h}(Z)\sqrt{n}\mathbb{E}(\widetilde{U}^\dagger\mid Z^{\dagger})\big] = \E[K(Z,Z^{\dagger})\widetilde{h}(Z)a(Z^{\dagger})] \not \to 0,
$$ 
which drives the power in a manner similar to the classical CM test. If instead $h(Z)$ is (nearly) orthogonal to the alternative with $\sqrt{n}\delta_h^{(1)}\to 0$, then   for the milder local alternatives $ \mathbb{H}_{an}': \ \mathbb{E}[\widetilde{U}|Z] = n^{-1/4}a(Z)$, $$\sqrt{n}\delta_h^{(2)}\to \sqrt{n}\mathrm{ICM}(U\mid Z)= \sqrt{n}\mathbb{E}[K(Z,Z^{\dagger})\mathbb{E}(\widetilde{U}|Z)\mathbb{E}(\widetilde{U}^{\dagger}|Z^{\dagger})]=\mathbb{E}[K(Z,Z^{\dagger})a(Z)a(Z^{\dagger})]>0.$$
In this case, the classical ICM test drives the power. As such, it provides robustness to the user’s choice of direction, though this comes at the cost of a slower detection rate under local alternatives.

To study the limiting behavior of $\sqrt{n}(\widehat{\delta}_h-\delta_h)$, we 
define  $$\Omega_h:= \operatorname{Var}[\xi_h(D_i)].$$ We distinguish $\Omega_{h,o}$ and $\Omega_{h,a}$, corresponding to specific expressions of $\Omega_h$ under $\mathbb{H}_o$ and  $\mathbb{H}_a$, respectively. Both are generically referred to as $\Omega_h$ whenever the distinction is not necessary. In particular, under $\mathbb{H}_o$, $\psi_U^{(1)}(D_i)=\mathrm{ICM}(U\mid  Z)=H_2^\top\xi_{\theta,i}=0 \ a.s.$ by the LIE, and   $\sqrt{n}\big(\widehat{\delta}_h - \delta_h\big)$ reduces to $ \sqrt{n}\widehat{\delta}_h = [1,  -1]^\top \frac{2}{\sqrt{n}}\sum_{i=1}^{n} \xi_{h,1}(D_i) + o_p(1) $, with 

\begin{equation}\label{eqn:Omega_o}
    \Omega_{h,o}: = 4\E\big[\xi_{h,1}(D)^2\big]\times \begin{bmatrix}
    1 & -1\\ -1 & 1
\end{bmatrix}.
\end{equation}

 The following result builds upon \Cref{lem:del_asymp_lin}.

\begin{theorem}\label{thm_all}
Suppose \Cref{ass:Sampling_iid,ass:bound_psiD,ass:diff_Utheta,ass:theta_asymp_lin} hold, then 
\begin{enumerate}[(i)]
    \item under $\mathbb{H}_o$,
 $
 \sqrt{n}\widehat{\delta}_h\overset{d}{\rightarrow}\mathcal{N}(0,\Omega_{h,o}) $;
\item under $\mathbb{H}_{an}$,  
 $
 \sqrt{n}\widehat{\delta}_h\overset{d}{\rightarrow}\mathcal{N}(a_o,\Omega_{h,o})
 $;
 \item under $\mathbb{H}_{an}'$,  
 $
 \sqrt{n}\widehat{\delta}_h\overset{d}{\rightarrow}\mathcal{N}(a_o',\Omega_{h,o})
 $ if $\E[\widetilde{h}(Z)a(Z^{\dagger})K(Z,Z^{\dagger})]=0$; and 
      \item under $\mathbb{H}_a$, 
     $
 \sqrt{n}(\widehat{\delta}_h-\delta_h) \overset{d}{\rightarrow}\mathcal{N}(0,\Omega_{h,a}); $
\end{enumerate}
where  $a_o := [1,-1]^{\top}\E[\widetilde{h}(Z)a(Z^{\dagger})K(Z,Z^{\dagger})]$ and $a_o':= \big[ 0, \, \mathrm{ICM}\big(a(Z)\mid Z\big) \big]^\top $.
\end{theorem}

\Cref{thm_all} offers a comprehensive asymptotic analysis of $\widehat{\delta}_h$ under null, local alternatives, and fixed alternatives. In particular,   part (i) shows that $\sqrt{n}\widehat{\delta}_h$ is non-degenerate and converges in distribution, which forms the basis of our $\chi^2$-distributed test. Under the Pitman local alternatives $\mathbb{H}_{an}$, the test is only non-trivial in directions where the user-defined assistant $h(Z)$ is non-orthogonal to $\E\big[a(Z^{\dagger})K(Z,Z^{\dagger}) \mid Z\big]$, i.e., $\E[\widetilde{h}(Z)a(Z^{\dagger})K(Z,Z^{\dagger})] \neq 0$. This is expected, as the use of the assistant $h(Z)$ draws inspiration from classical conditional moment (CM) tests. Part (iii) of \Cref{thm_all} partially resolves this limitation: the local power becomes non-trivial in all directions due to the presence of $\mathrm{ICM}\big(a(Z)\mid Z\big)$ in the term $a_o' := [0, \ \mathrm{ICM}\big(a(Z)\mid Z\big)]^\top$ under a sequence of alternatives converging to the null at the $n^{-1/4}$ rate. In other words, the test becomes omnibus for alternative directions that converge to the null at a rate slower than $n^{-1/4}$, thereby enhancing its ability to detect a broader class of local alternatives.  The two types of local alternatives stem from our novel construction of the test, which combines components that operate at two different convergence rates. Part (iv) suggests the resulting test is consistent under the fixed alternative. 

\begin{remark}
    The power performance of the proposed $\chi^2$-test under the Pitman local alternative $\mathbb{H}_{an}$ is sensitive to the choice of $h(Z)$ and the ICM kernel $K(Z,Z^\dagger)$. A potential way to enhance the local power of the proposed test is to extend the estimand $\delta_h$ to a vector 
$\big\{ \delta_{h,K} \big\}_{\{h,K\} \in \mathcal{H} \times \mathcal{K}}$, where the subscript in $\delta_{h,K}$ emphasizes the dependence of the parameter on both the function $h$ and the ICM kernel $K$. Here, $\mathcal{H}$ and $\mathcal{K}$ are finite sets of measurable, non-degenerate functions of $Z$ and valid ICM kernels, respectively.\footnote{Another possibility is to extend $V_h$ with a vector comprising a dictionary of transformations, e.g., polynomials of $Z$ -- see \Cref{subsec:pv_more} for some simulation evidence. A treatment of these extensions is omitted for considerations of space and scope.}
\end{remark}

\subsection{Test Statistic}\label{Subsec:test}
Theorem \ref{thm_all}  justifies the asymptotic normality  of $\widehat{\delta}_h$, even under $\mathbb{H}_o$, which  naturally motivates the Wald test statistic:  
\begin{equation}\label{eqn:Chitest_stat1}
\widetilde{T}_{V,n}=n\widehat{\delta}_h^{\top}\widetilde{\Omega}_{h,n}^{-1}\widehat{\delta}_h,
\end{equation}
 
 \noindent where $\widetilde{\Omega}_{h,n}$ is a consistent estimator of $\Omega_h$ under both $\mathbb{H}_o$ and $\mathbb{H}_a$. This testing procedure is valid when $\Omega_h$ is positive definite. However, from \eqref{eqn:Omega_o}, $\Omega_h$ is singular under $\mathbb{H}_o$ with $\mathrm{rank}(\Omega_{h,o})=1$.

Although it is tempting to replace the inverse matrix $\widetilde{\Omega}_{h,n}^{-1}$ in \eqref{eqn:Chitest_stat1} with a generalized inverse matrix $\widetilde{\Omega}_{h,n}^{-}$, e.g., the Moore-Penrose inverse, the resulting Wald statistic may still not have an asymptotic $\chi^2$ distribution unless the rank condition, $\mathbb{P}\big(\mathrm{rank}(\widetilde{\Omega}_{h,n})=\mathrm{rank}(\Omega_h)\big) \to 1 \quad \text{as }n\to\infty$ is satisfied \citep{andrews1987asymptotic}. To ensure the rank condition is satisfied and thereby address this problem, we adopt the thresholding technique described in \citet{lutkepohl1997modified} (see also \citet{duchesne2015multivariate,dufour2016rank}). Let \begin{equation*}\label{Consis}
    \widetilde{\Omega}_{h,n}:=\frac{1}{n-1}\sum_{i=1}^n \widehat{\xi}_h(D_i)\widehat{\xi}_h(D_i)^{\top},
\end{equation*}
be a consistent estimator of $\Omega_h$ \citep{sen1960some}, where $\widehat{\xi}_h(D_i)$ denotes the sample analog of $\xi_h(D_i)$ with parameters replaced by estimates.

By the singular value decomposition,
\begin{equation}\label{SVD}   \widetilde{\Omega}_{h,n}=\widetilde{\Gamma}_n\widetilde{\Lambda}_n\widetilde{\Gamma}_n^{\top},
\end{equation}
where $\widetilde{\Lambda}_n = \mathrm{diag}(\widetilde{\lambda}_1,\widetilde{\lambda}_2)$ is the diagonal matrix comprising the eigenvalues $\widetilde{\lambda}_1\geq \widetilde{\lambda}_2 \geq 0$ of $\widetilde{\Omega}_{h,n}$, and the columns of $\widetilde{\Gamma}_n$ are the corresponding  eigenvectors. Let \( c_n = C n^{-1/2 + \iota} \), where \( \iota \in (0, 1/2) \) is a small positive constant and \( C \in (0, \infty) \) is a fixed constant, as in \citet{dufour2016rank}. Since \( \mathrm{rank}(\Omega_h) \geq 1 \) regardless of whether the null hypothesis \( \mathbb{H}_o \) holds, we define the regularized estimator of \( \Omega_h \) as
\begin{equation} \label{hatOmega}
    \widehat{\Omega}_{h,n} := \widetilde{\Gamma}_n \widehat{\Lambda}_{n,c_n} \widetilde{\Gamma}_n^{\top}, 
    \quad \text{where} \quad 
    \widehat{\Lambda}_{n,c_n} = \mathrm{diag}\big( \widetilde{\lambda}_1, \widetilde{\lambda}_2 \mathbf{1}(\widetilde{\lambda}_2 > c_n) \big).
\end{equation}
The corresponding Moore–Penrose inverse is defined as
\[
    \widehat{\Omega}_{h,n}^{-} := \widetilde{\Gamma}_n \widehat{\Lambda}_{n,c_n}^{-} \widetilde{\Gamma}_n^{\top}, 
    \quad \text{where} \quad 
    \widehat{\Lambda}_{n,c_n}^{-} = \mathrm{diag}\big( \widetilde{\lambda}_1^{-1}, \widetilde{\lambda}_2^{-1} \mathbf{1}(\widetilde{\lambda}_2 > c_n) \big).
\]
Finally,  we define the regularized Wald test statistic: 
\begin{equation}\label{eqn:chi}
T_{h,n} := n\widehat{\delta}_h^{\top}\widehat{\Omega}_{h,n}^{-}\widehat{\delta}_h.
\end{equation}
\noindent In practice, $\mathbb{H}_o$ is rejected when $T_{h,n} > \chi^2_{1,1-\alpha}$, where $\chi^2_{1,1-\alpha}$ denotes the $1-\alpha$ quantile of the $\chi^2_1$ distribution at a pre-specified significance level $\alpha\in(0,1)$. We remark that the ranks of $\Omega_{h,o}$ and $\Omega_{h,a}$ can differ: $\mathrm{rank}(\Omega_{h,o}) = 1$, while $\mathrm{rank}(\Omega_{h,a}) = 2$ if $\mathbb{P}(a(Z)= c \cdot \widetilde{h}(Z))<1$ for any $c \in \mathbb{R}$, and $\mathrm{rank}(\Omega_{h,a}) = 1$ otherwise. We use $c_n=\widetilde{\lambda}_1 n^{-1/3}$ following \citet{lutkepohl1997modified} and \citet{dufour2016rank}.\footnote{For the robustness of the $\chi^2$-test to variations of \( c_n = \widetilde{\lambda}_1 n^{-\iota} \), \( \iota \in (0, 1/2) \), and other common truncation criteria, see Section \ref{subsec:robustness_cn} in the supplement.}

\begin{remark}\label{rem_rank}
$\mathrm{rank}(\Omega_{h,o}) = 1$ and $\mathrm{ICM}_n(U \mid Z) = O_p(n^{-1}) $ under $\mathbb{H}_o$. From the proof of part (i) of \Cref{thm_all}, $ \sqrt{n}\widehat{\delta}_h = \sqrt{n} \widehat{\delta}_h^{(1)} [1,\ -1]^{\top} + O_p(n^{-1/2}) $, and 
    \[ T_{h,n} = n\widehat{\delta}_h^{\top}\widehat{\Omega}_{h,n}^{-}\widehat{\delta}_h = \Bigg(\frac{\widehat{\delta}_h^{(1)}}{\sqrt{\operatorname{Var}(\widehat{\delta}_h^{(1)})}}\Bigg)^2 + o_p(1) \xrightarrow{d} \chi_1^2 \quad \text{as}\quad  n\rightarrow \infty
    \]under $\mathbb{H}_o$. This implies $\sqrt{T_{h,n}}$ converges in distribution to the half standard normal under $\mathbb{H}_o$, and shares the interpretability (without a formal hypothesis test) of a two-sided $t$-test.
\end{remark}

The next theorem justifies using \eqref{eqn:chi} under the null, local alternative, and fixed alternative hypotheses.
\begin{theorem}\label{thm_wald}
 If $\,\E\left|\xi_h(D)\right|^{4+\varepsilon}<\infty$ for some $\varepsilon>0$, and let $c_n= C n^{-1/2+\iota} $ for some constants $\iota\in(0,1/2)$ and $C\in(0,\infty)$ independent of $n$, then 
\begin{enumerate}[(i)]
\item under $\mathbb{H}_o$, $
\widehat{\Omega}_{h,n}^-\overset{p}{\rightarrow} \Omega_{h,o}^-,
$ and 
$$
T_{h,n}\overset{d}{\rightarrow} \chi^2_1;
$$
    \item under $\mathbb{H}_{an}$, $
\widehat{\Omega}_{h,n}^-\overset{p}{\rightarrow} \Omega_{h,o}^-
$, and the asymptotic local power is given by 
$$ 
\lim_{n\to\infty}\mathbb{P}(T_{h,n}>\chi^2_{1,1-\alpha})=\mathbb{P}\left(\chi^2_{1}(b_o)>\chi^2_{1,1-\alpha}\right),
$$
where $ \displaystyle b_o:=a_o^{\top}\Omega_{h,o}^-a_o= \frac{(\E[\widetilde{h}(Z)a(Z^{\dagger})K(Z,Z^{\dagger})])^2}{4\E[\xi_{h,1}(D)^2]} $, $a_o$ is defined in \Cref{thm_all}, and $\chi^2_1(b_o)$ is a non-central $\chi_1^2$ random variable;

    \item under $\mathbb{H}_{an}'$, $
\widehat{\Omega}_{h,n}^-\overset{p}{\rightarrow} \Omega_{h,o}^-
$, and the asymptotic local power is given by 
$$ 
\lim_{n\to\infty}\mathbb{P}(T_{h,n}>\chi^2_{1,1-\alpha})=\mathbb{P}\left(\chi^2_{1}(b_o')>\chi^2_{1,1-\alpha}\right),
$$ 
if $\E[\widetilde{h}(Z)a(Z^{\dagger})K(Z,Z^{\dagger})]=0$ where $ \displaystyle b_o':=a_o^{'\top}\Omega_{h,o}^-a_o' = \frac{\mathrm{ICM}^2\big( a(Z) \mid Z \big)}{16\E[\xi_{h,1}(D)^2]} >0$, and $a_o'$ is defined in \Cref{thm_all}; and

    \item under $\mathbb{H}_a$, $
\widehat{\Omega}_{h,n}^-\overset{p}{\rightarrow} \Omega_{h,a}^-,
$
 and   if $\delta_h\not\in\mathcal{M}_0$, where $\mathcal{M}_0$ is the eigenspace associated with the null eigenvalue of $\Omega_{h,a}$,   $$
    \lim_{n\to\infty}\mathbb{P}(T_{h,n}>\chi^2_{1,1-\alpha})=1.
    $$
\end{enumerate}
\end{theorem}
\Cref{thm_wald} establishes the consistency of the thresholded estimator of the Moore–Penrose inverse of the covariance matrix, and characterizes the asymptotic behavior of the resulting test statistic, in parallel with the analysis in \Cref{thm_all}. In particular, under the local alternatives $\mathbb{H}_{an}$ of part (ii), when $a_0 = \E[\widetilde{h}(Z)a(Z^{\dagger})K(Z,Z^{\dagger})] = 0$, the test exhibits trivial power because $b_o = 0$. However, this limitation is partially remedied in part (iii), where the modified limiting direction $b_o'$ remains positive.

\begin{remark}\label{rem_m0}
Under $\mathbb{H}_a$, $\delta_h\not\in\mathcal{M}_0$ always holds for the mean independence test. The proof is provided in \Cref{lem_del_null_space} of the supplementary material.
\end{remark}

\subsection{Power Comparison with the bootstrap-based ICM test}\label{SubSect:Bahadur}

In this section, we examine the efficiency of the proposed test statistic compared with the bootstrap-based ICM test by adopting the approach of \citet{bahadur1960stochastic}. Under a fixed alternative, both tests demonstrate consistency as $n$ tends to infinity. The Bahadur slope in \citet{bahadur1960stochastic} allows us to further compare the rate of convergence of \( p \)-values to zero as $n$ increases. 

Let \( \displaystyle S_G(t):=\mathbb{P}\Big( \sum_{k=1}^{\infty}\lambda_k G_k^2>t \Big) \) and \( \displaystyle S_T(t):=\mathbb{P}\big( \chi^2_{1} > t \big) \) be the survival functions of the asymptotic null distributions of the ICM test statistic  $n\mathrm{ICM}_n(U \mid Z)$, and the pivotal test statistic, $T_{h,n}$, respectively.
Their Bahadur slopes are  respectively given by
$$
c_G= \lim_{n\to\infty}-\frac{2}{n} \log S_G(n\mathrm{ICM}_n(U \mid Z)) \ \text{ and } \ c_T= \lim_{n\to\infty}-\frac{2}{n} \log S_G(T_{h,n}).
$$

\begin{theorem}\label{thm_bahadur}
Suppose $
\widehat{\Omega}_{h,n}^-\overset{p}{\rightarrow} \Omega_{h,a}^-,
$ and the conditions of \Cref{thm_wald} hold, then under $\mathbb{H}_a$, the (approximate) Bahadur slopes of the ICM test statistic $n\mathrm{ICM}_n(U \mid Z)$ and the pivotal test statistic $T_{h,n}$ are, respectively, given by 
    $$
    c_G= \frac{\mathrm{ICM}(U \mid Z)}{\lambda_1} \quad \text{ and } c_T= \delta_a^{\top}\Omega_{h,a}^{-}\delta_a
    $$
where $\lambda_1$ is the leading eigenvalue associated with the limiting null distribution in \eqref{eqn:ICM_asymp}, and 
$$
\delta_a = \Big[\E\big[\widetilde{h}(Z)a(Z^{\dagger})K(Z,Z^{\dagger})\big], \ \mathrm{ICM}(U \mid Z) - \E\big[\widetilde{h}(Z)a(Z^{\dagger})K(Z,Z^{\dagger})\big] \Big]^\top. 
$$
\end{theorem}

\noindent The approximate Bahadur slopes presented in Theorem \ref{thm_bahadur} are primarily of theoretical interest. Conducting a comprehensive comparison of these slopes is challenging as they depend on data-dependent quantities such as $\lambda_1$, $a(Z)$, and user-specified variables such as $h(Z)$ and the kernel $K(\cdot,\cdot)$.

\begin{remark}
    The main implication of \Cref{thm_bahadur} is that, even with a fixed ICM kernel, neither test is uniformly more powerful across all data-generating processes satisfying \Cref{ass:bound_psiD}. This limitation arises because the true alternative is unknown and the function \( h(\cdot) \) is user-specified. \Cref{Tab:Bahadur} presents a numerical illustration of this result.
\end{remark}

\section{Monte Carlo Experiments - Specification Test}\label{Sect:MC_Sim_Spec}
This section examines the size and power of the proposed $\chi^2$ specification test compared to bootstrap-based ICM procedures via simulations. The specification $\widehat{V}_h = [h(Z),\ \widehat{U}-h(Z)]^\top$ with $h(Z) = \exp\big(Z^\top\mathbbm{1}_{p_z}/\sqrt{p_z} \big) $. For the proposed $\chi^2$-test, the regularized inverse in \eqref{eqn:Chitest_stat1} is computed using $c_n=\widetilde{\lambda}_1 n^{-1/3}$ where $\widetilde{\lambda}_1$ is the leading eigenvalue of $\widetilde{\Omega}_{h,n}$. We consider two commonly used ICM kernels: the Gaussian $K(Z,Z^\dagger) = \exp\left(-0.5 \|Z - Z^\dagger\|^2\right) $ , e.g., \citet{bierens1982consistent} and the negative Euclidean $K(Z,Z^\dagger) = -\|Z - Z^\dagger\|$ \citep{shao2014martingale}. For a given ICM kernel, the proposed $\chi^2$-test is compared to ICM tests based on the multiplier bootstrap (MB) of \citet{escanciano-2024-gaussian} and the standard wild bootstrap, e.g., \citet{escanciano2006consistent,su2017martingale}. The empirical size and power curves are based on $1000$ Monte Carlo replicates. $999$ bootstrap samples are used for the bootstrap procedures.\footnote{Both bootstrap procedures are conducted using the \citet{mammen1993bootstrap} two-point distributed auxiliary variables.} Other simulation results on specification and mean independence tests are available in  Section \ref{Sect_Appendix:MC} of the supplementary material.

\subsection{Specifications}
We consider the linear model 
$$
Y= \theta_c + \sum_{l=1}^5X_l\theta_l + U,$$
with and without excluded instruments.\footnote{{\Cref{Sect_Suppl:MC_NLM} provides simulation results on the specification of propensity scores using the logit model.}} The data-generating processes (DGPs) are varied through $U$: 
\begin{enumerate}[label=(\roman*)]
\item[LS1:] $\displaystyle U = \frac{\mathcal{E}}{\sqrt{1+Z_1^2}} $;
			
\item[LS2:] $ \displaystyle U = \frac{ \gamma }{5\sqrt{2}} \sum_{l=1}^5 Z_l^2 + \frac{\mathcal{E}}{\sqrt{1+Z_1^2}} $; 
			
\item[LS3:] $ \displaystyle U = \gamma \sum_{l=1}^5 \frac{\cos(2Z_l)}{\sqrt{2(1-\exp(-8))}} + \frac{\mathcal{E}}{\sqrt{1+Z_1^2}} $; and

\item[LS4:] $ \displaystyle U = \gamma\sum_{l=1}^5 \big(\exp(-Z_l^2/3) - \sqrt{3/5}\big)  + \frac{\mathcal{E}}{\sqrt{1+Z_1^2}} $
\end{enumerate}
where $ \mathcal{E} \sim \mathcal{N}(0,1)$ independently of $Z$ which follows the multivariate normal distribution with mean zero and covariance matrix $\Sigma_{ll'} = 0.25^{|l-l'|}$, $ [\theta_c,\theta_1,\ldots,\theta_5]^\top = \mathbbm{1}_6 $, and $\gamma\in[0,1]$ tunes the deviation away from $\mathbb{H}_o$. $X=Z$ in DGP LS1 and $X_1 = (1.5Z_1 + \widetilde{\mathcal{E}}) / \sqrt{3.25} $ with  $\widetilde{\mathcal{E}}\sim \mathcal{N}(0,1)$ and $\mathrm{Cov}(\widetilde{\mathcal{E}},{\mathcal{E}})=0.25$ in DGPs LS2 through LS4.\footnote{The non-parametric quantity $\E[X_1 \mid Z_1]$ is needed for the multiplier bootstrap under endogeneity -- see \citet[Section 5]{escanciano-2024-gaussian}. It is estimated using a third-order polynomial.} Thus, $X$ is exogenous under DGP LS1 and endogenous under DGPs LS2 through LS4. LS1  is estimated via Ordinary Least Squares (OLS) while the remaining DGPs are estimated using the Instrumental Variable (IV) estimator with $Z$ as the instrument. Heteroskedasticity of arbitrary form is imposed in all DGPs.
 
Different sample sizes $n \in \{200, 400, 600, 800\}$ serve to examine the empirical size and local power of the test using DGPs LS1 and LS2 (with $\gamma=0$).  For the analyses of power in DGPs LS2 through LS4, the sample size is kept at $n=400$ while $\gamma$ is varied in order to study the power of the proposed $\chi^2$-test in comparison to the bootstrap-based ICM procedures.

\subsection{Empirical Size and Power}

\Cref{Tab:Sim_Spec_LM12} presents the empirical sizes corresponding to DGPs LS1 (under strictly exogenous $X$) and LS2 (under endogenous $X$ instrumented with $Z$) in addition to the local power under LS2 at three nominal levels: 10\%, 5\%, and 1\%.  One observes comparably good size control of the proposed $\chi^2$-test and the bootstrap-based procedures across all the sample sizes considered. This provides evidence in support of the proposed test's validity under both exogenous $X$ and endogenous $X$ with valid instruments $Z$.

\begin{table}[!htbp]
\centering
\caption{Empirical Size \& Local Power}
\begin{tabular}{lcccccccccc}
\toprule
& & \multicolumn{3}{c}{$ 10\% $} & \multicolumn{3}{c}{$ 5\% $} & \multicolumn{3}{c}{$ 1\% $}  \\
\cmidrule(lr){3-5} \cmidrule(lr){6-8}\cmidrule(lr){9-11}
$n$ & Kernel & $\chi^2$ & MB & WB  & $\chi^2$ & MB & WB & $\chi^2$ & MB & WB \\ \midrule

\multicolumn{1}{c}{\textcolor{blue}{LS1}} & & \multicolumn{9}{c}{Empirical Size} \\ 
\cmidrule(lr){1-1}  \cmidrule(lr){3-11}

200   &Gauss  &0.096 &0.103 &0.103 &0.053 &0.049 &0.052 &0.007 &0.012 &0.012 \\ 
      &Euclid &0.101 &0.072 &0.057 &0.048 &0.031 &0.021 &0.008 &0.003 &0.003 \\ \cmidrule[0.25pt](lr){2-11}

400   &Gauss  &0.103 &0.098 &0.096 &0.046 &0.048 &0.046 &0.007 &0.009 &0.006 \\ 
      &Euclid &0.104 &0.074 &0.056 &0.053 &0.032 &0.029 &0.005 &0.003 &0.004 \\ \cmidrule[0.25pt](lr){2-11}

600   &Gauss  &0.089 &0.102 &0.106 &0.053 &0.050 &0.047 &0.009 &0.011 &0.008 \\ 
      &Euclid &0.100 &0.084 &0.085 &0.047 &0.036 &0.034 &0.006 &0.006 &0.005 \\ \cmidrule[0.25pt](lr){2-11}

800   &Gauss  &0.090 &0.104 &0.097 &0.039 &0.053 &0.055 &0.008 &0.007 &0.006 \\ 
      &Euclid &0.098 &0.088 &0.078 &0.042 &0.040 &0.041 &0.006 &0.004 &0.006 \\ 
\midrule

\multicolumn{1}{c}{\textcolor{blue}{LS2}} & & \multicolumn{9}{c}{Empirical Size} \\ 
\cmidrule(lr){1-1}  \cmidrule(lr){3-11}

200   &Gauss  &0.095 &0.093 &0.100 &0.055 &0.052 &0.053 &0.007 &0.011 &0.014 \\ 
      &Euclid &0.097 &0.070 &0.055 &0.045 &0.032 &0.021 &0.008 &0.003 &0.002 \\ \cmidrule[0.25pt](lr){2-11}

400   &Gauss  &0.102 &0.103 &0.099 &0.046 &0.051 &0.048 &0.007 &0.007 &0.006 \\ 
      &Euclid &0.103 &0.079 &0.056 &0.056 &0.031 &0.028 &0.005 &0.005 &0.003 \\ \cmidrule[0.25pt](lr){2-11}

600   &Gauss  &0.089 &0.094 &0.107 &0.052 &0.044 &0.048 &0.008 &0.011 &0.008 \\ 
      &Euclid &0.100 &0.090 &0.083 &0.046 &0.040 &0.032 &0.006 &0.010 &0.006 \\ \cmidrule[0.25pt](lr){2-11}

800   &Gauss  &0.090 &0.099 &0.095 &0.039 &0.051 &0.055 &0.009 &0.008 &0.006 \\ 
      &Euclid &0.099 &0.089 &0.079 &0.042 &0.041 &0.037 &0.006 &0.004 &0.006 \\ 
\midrule

\multicolumn{1}{c}{\textcolor{blue}{LS2}} & & \multicolumn{9}{c}{Local Power: $\gamma = 5/\sqrt{n}$} \\ 
\cmidrule(lr){1-1}  \cmidrule(lr){3-11}

200   &Gauss  &0.549 &0.486 &0.480 &0.418 &0.391 &0.371 &0.164 &0.223 &0.201 \\ 
      &Euclid &0.883 &0.656 &0.625 &0.803 &0.543 &0.507 &0.535 &0.295 &0.262 \\ \cmidrule[0.25pt](lr){2-11}

400   &Gauss  &0.605 &0.545 &0.522 &0.459 &0.413 &0.390 &0.202 &0.226 &0.182 \\ 
      &Euclid &0.887 &0.707 &0.689 &0.813 &0.601 &0.575 &0.588 &0.376 &0.341 \\ \cmidrule[0.25pt](lr){2-11}

600   &Gauss  &0.592 &0.521 &0.509 &0.461 &0.423 &0.408 &0.221 &0.237 &0.215 \\ 
      &Euclid &0.896 &0.723 &0.714 &0.830 &0.614 &0.621 &0.606 &0.410 &0.398 \\ \cmidrule[0.25pt](lr){2-11}

800   &Gauss  &0.606 &0.547 &0.539 &0.473 &0.441 &0.428 &0.243 &0.246 &0.225 \\ 
      &Euclid &0.912 &0.743 &0.742 &0.840 &0.651 &0.644 &0.632 &0.420 &0.422 \\
\bottomrule
\end{tabular}    
\label{Tab:Sim_Spec_LM12}
\end{table}

\begin{figure}[H]
\centering 
\caption{DGP LS2 -- Gaussian Kernel -- $n=400$.}
\begin{subfigure}{0.32\textwidth}
\centering
\includegraphics[width=1\textwidth]{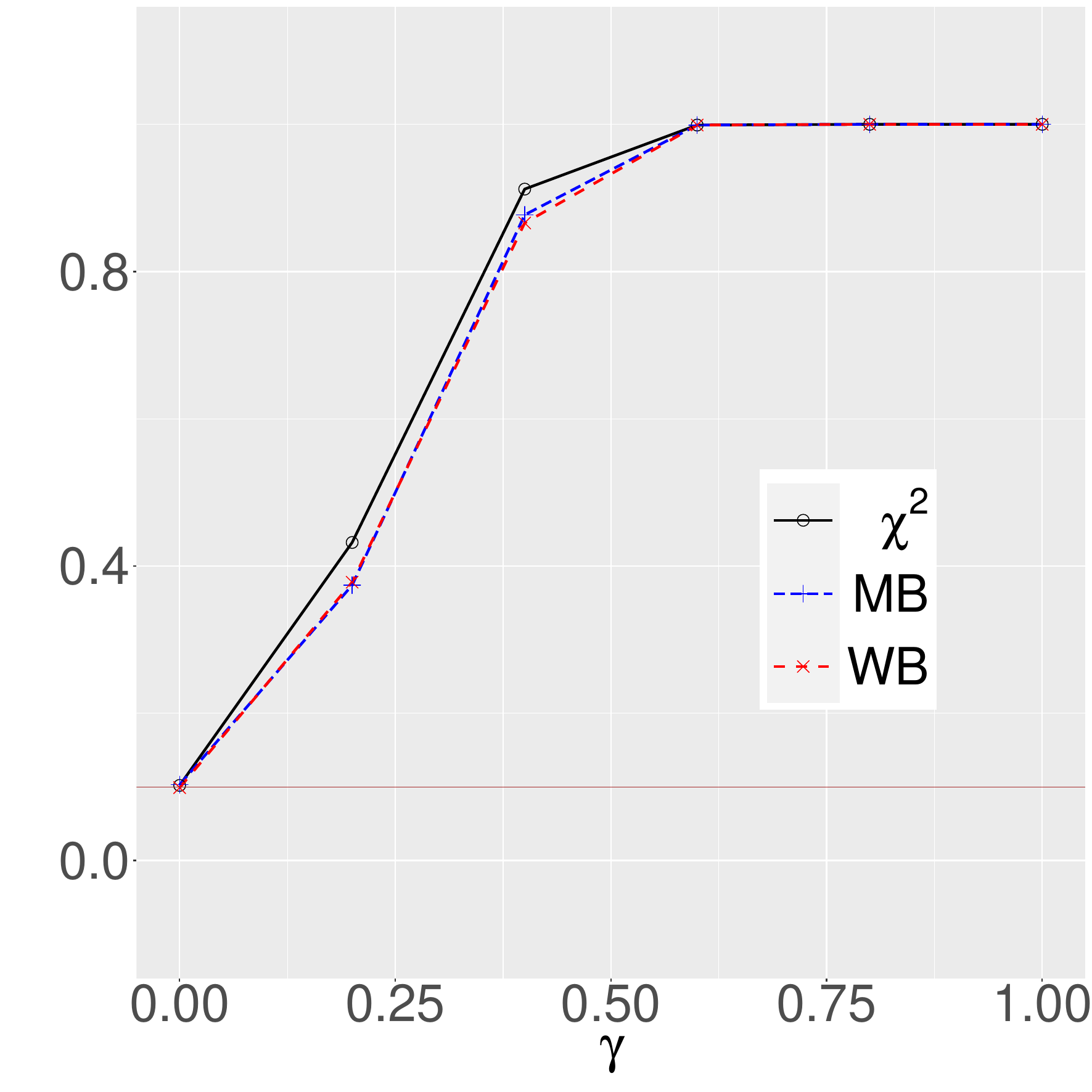}
\caption{10\%}
\end{subfigure}
\begin{subfigure}{0.32\textwidth}
\centering
\includegraphics[width=1\textwidth]{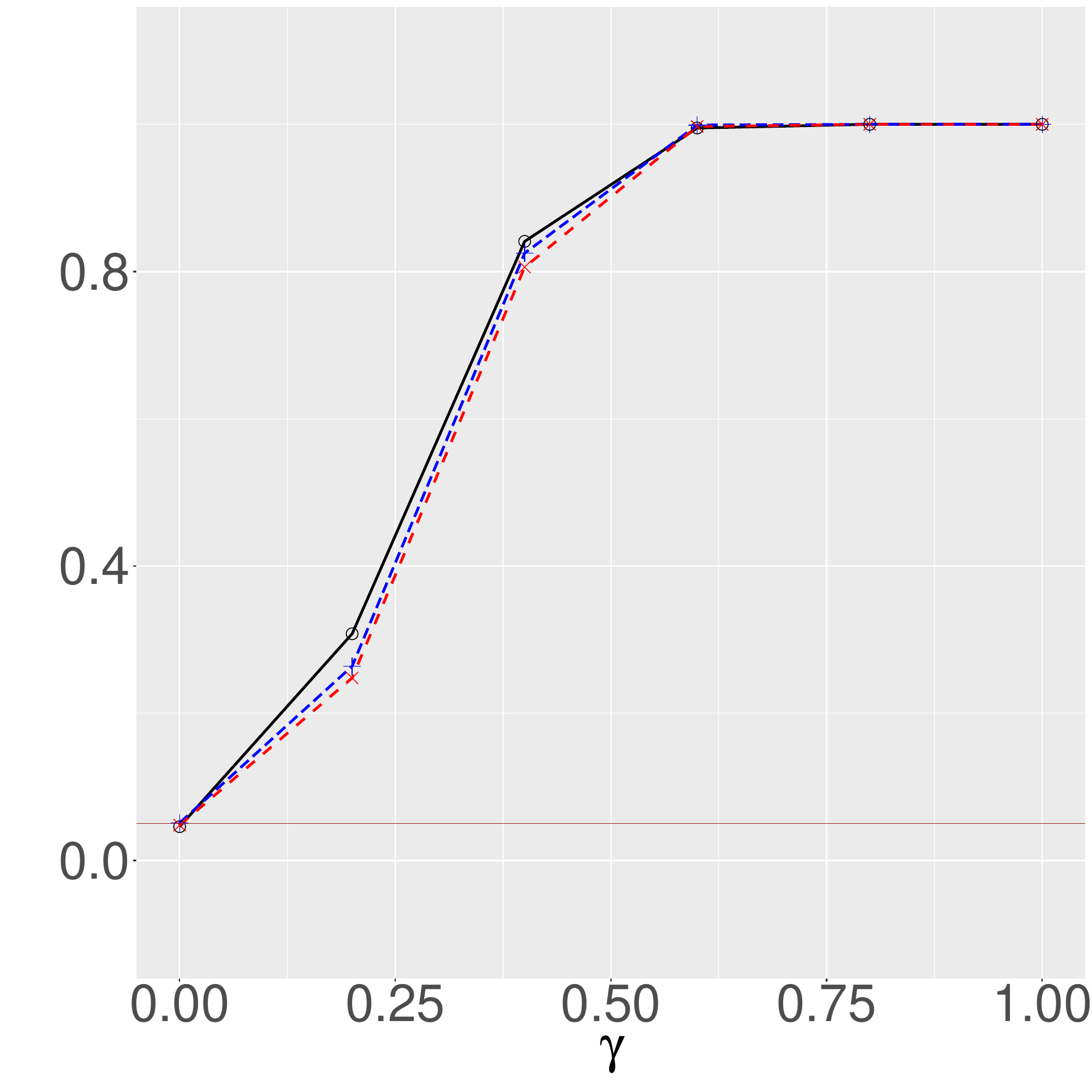}
\caption{5\%}
\end{subfigure}
\begin{subfigure}{0.32\textwidth}
\centering
\includegraphics[width=1\textwidth]{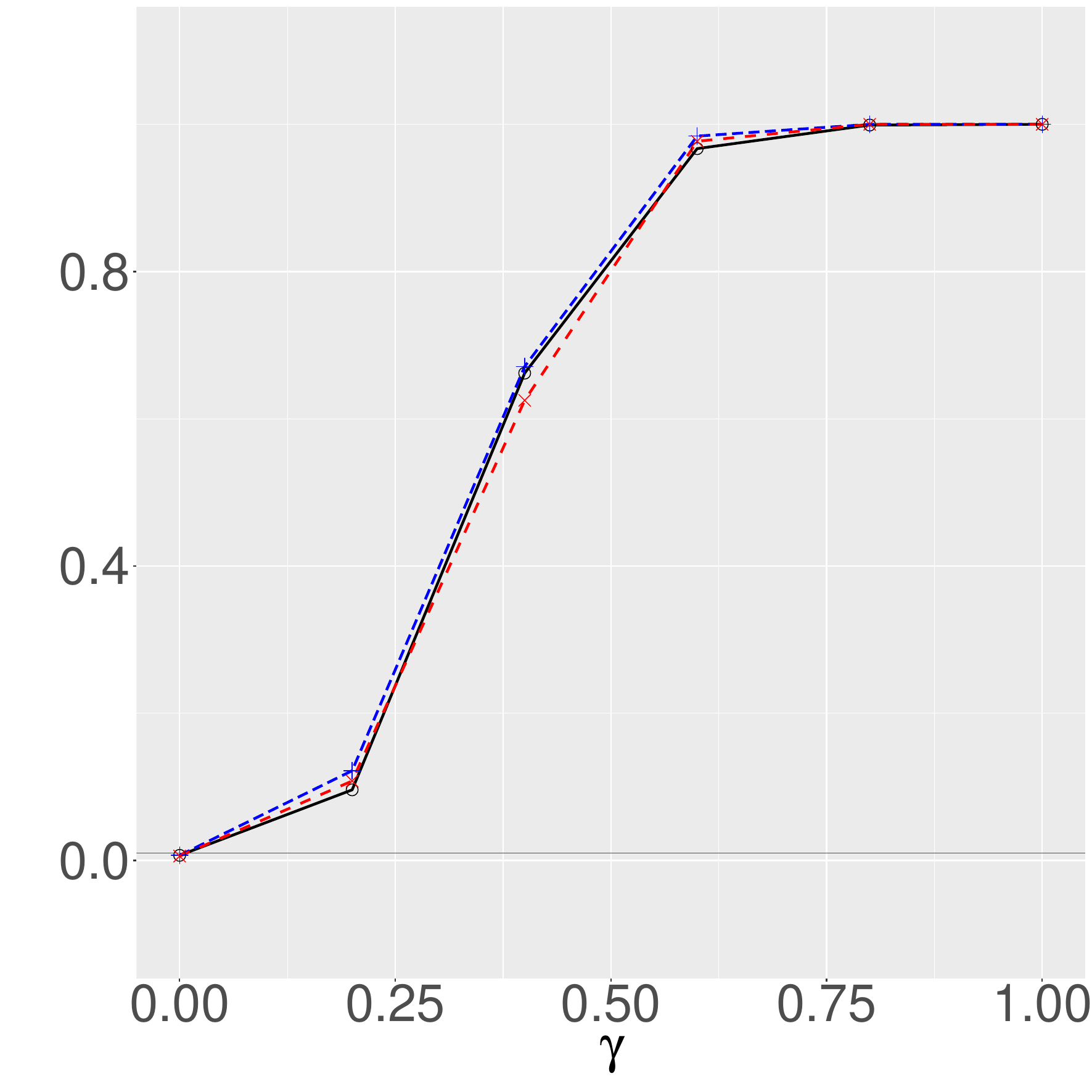}
\caption{1\%}
\end{subfigure}
\label{Fig:LS2_Gauss}
\end{figure}

\begin{figure}[H]
\centering 
\caption{DGP LS2 -- Negative Euclidean -- $n=400$.}
\begin{subfigure}{0.32\textwidth}
\centering
\includegraphics[width=1\textwidth]{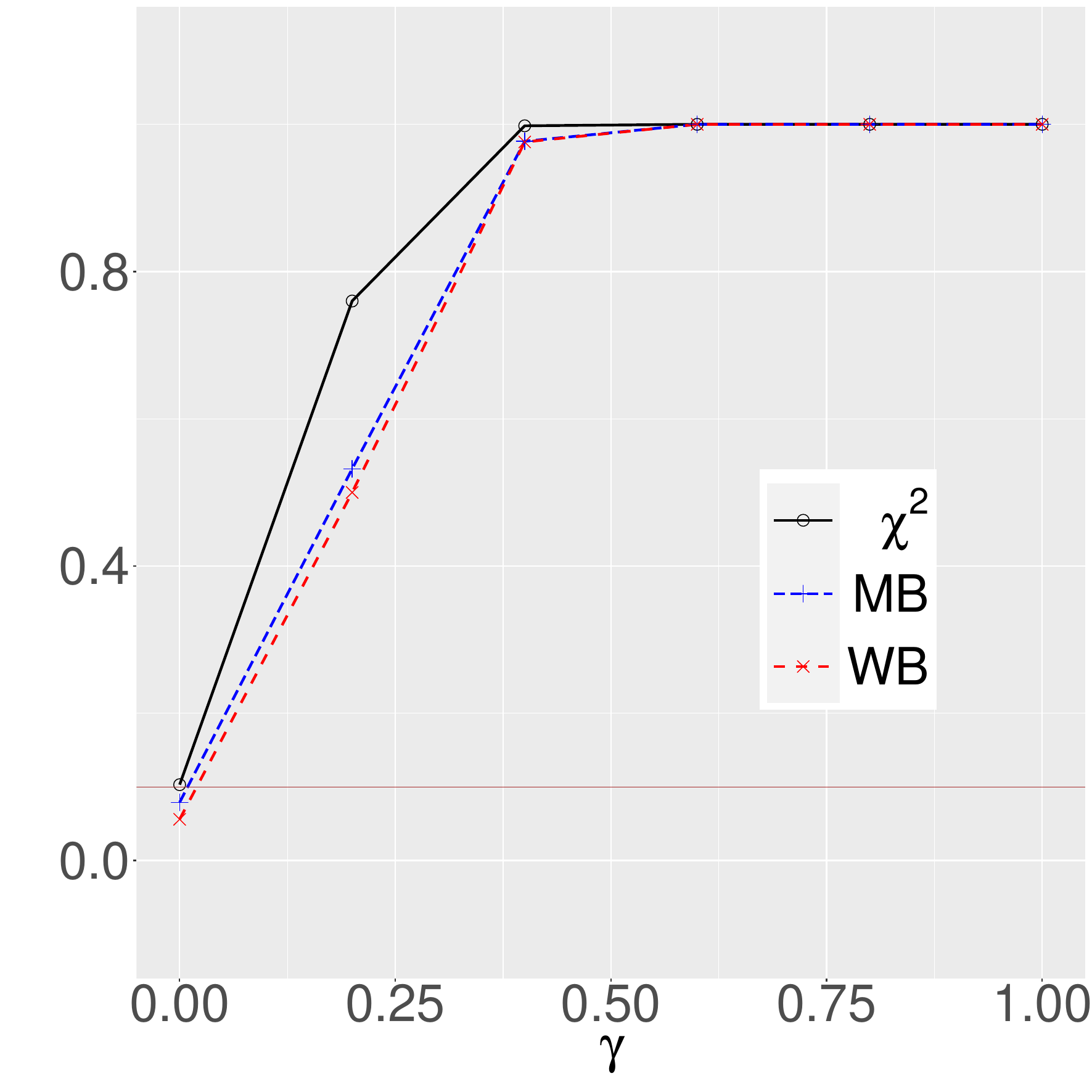}
\caption{10\%}
\end{subfigure}
\begin{subfigure}{0.32\textwidth}
\centering
\includegraphics[width=1\textwidth]{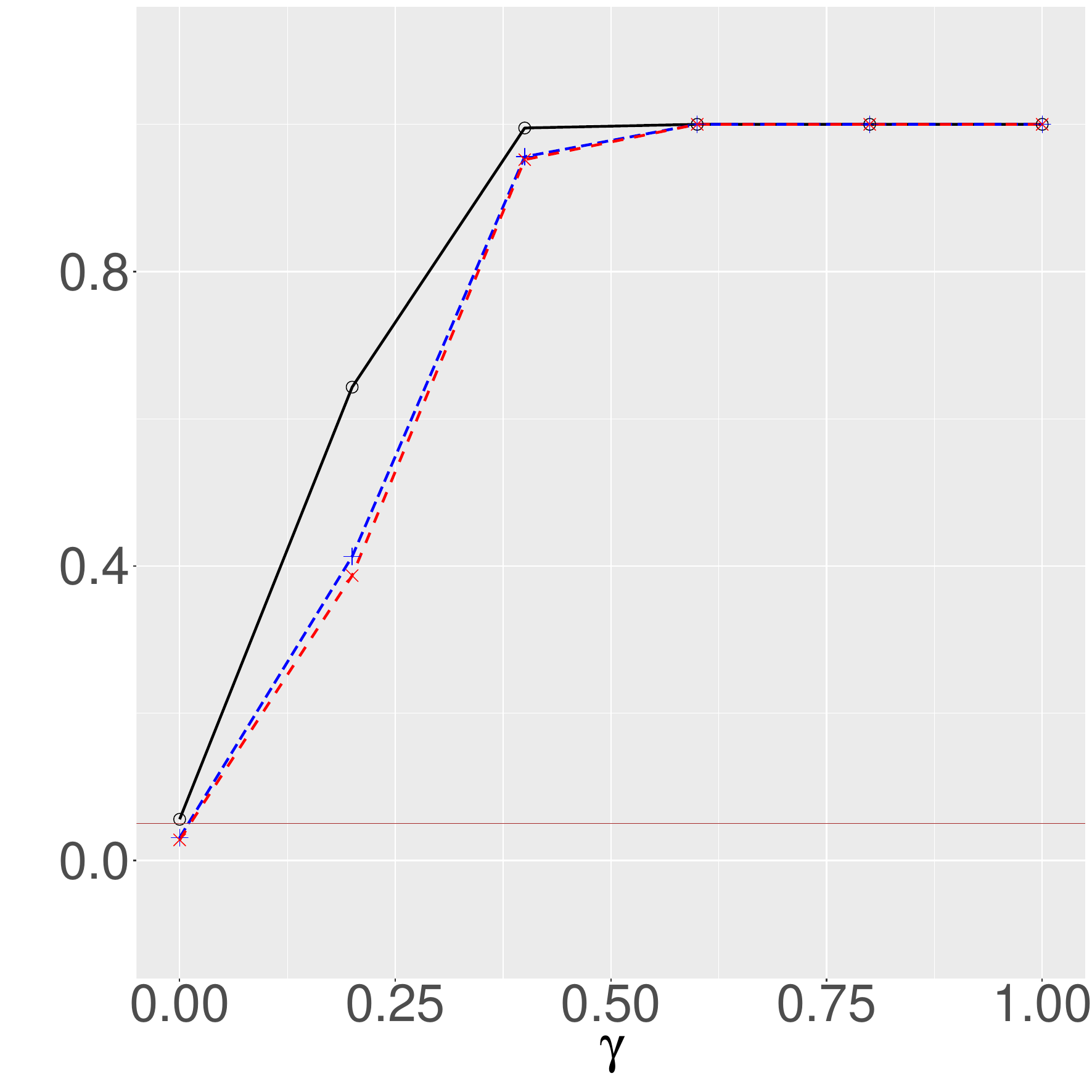}
\caption{5\%}
\end{subfigure}
\begin{subfigure}{0.32\textwidth}
\centering
\includegraphics[width=1\textwidth]{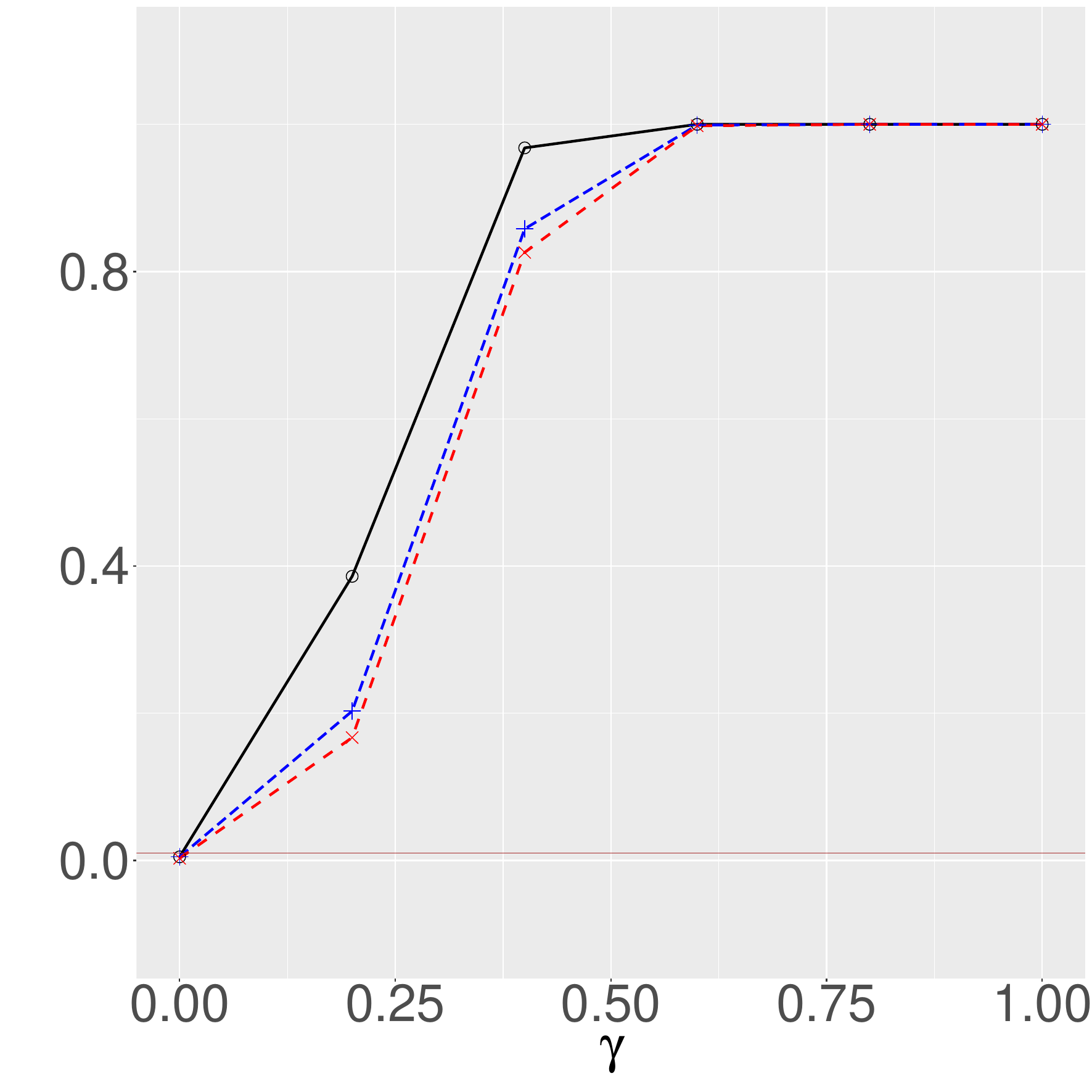}
\caption{1\%}
\end{subfigure}
\label{Fig:LS2_Euclid}
\end{figure}

\begin{figure}[H]
\centering 
\caption{DGP LS3 -- Gaussian Kernel -- $n=400$.}
\begin{subfigure}{0.32\textwidth}
\centering
\includegraphics[width=1\textwidth]{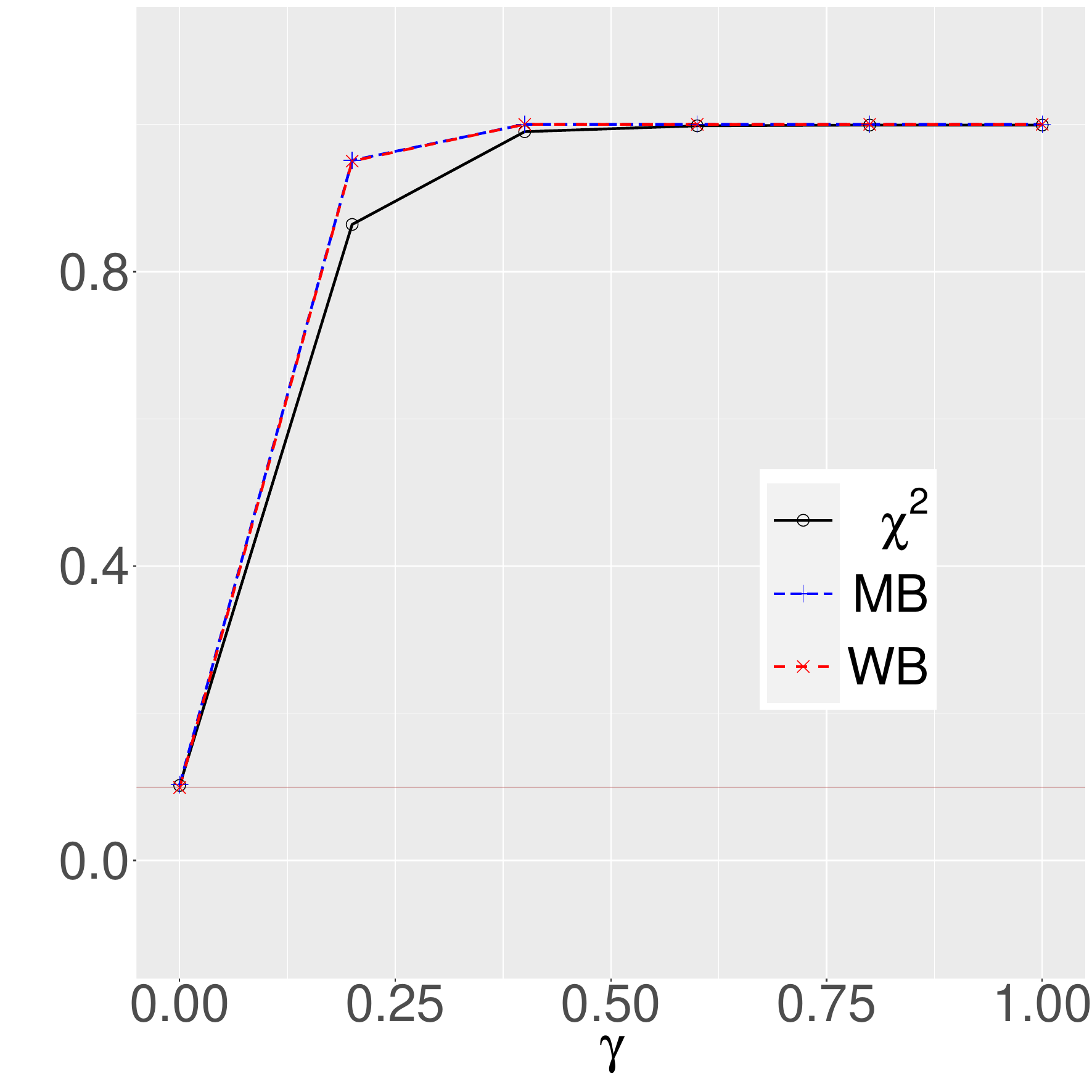}
\caption{10\%}
\end{subfigure}
\begin{subfigure}{0.32\textwidth}
\centering
\includegraphics[width=1\textwidth]{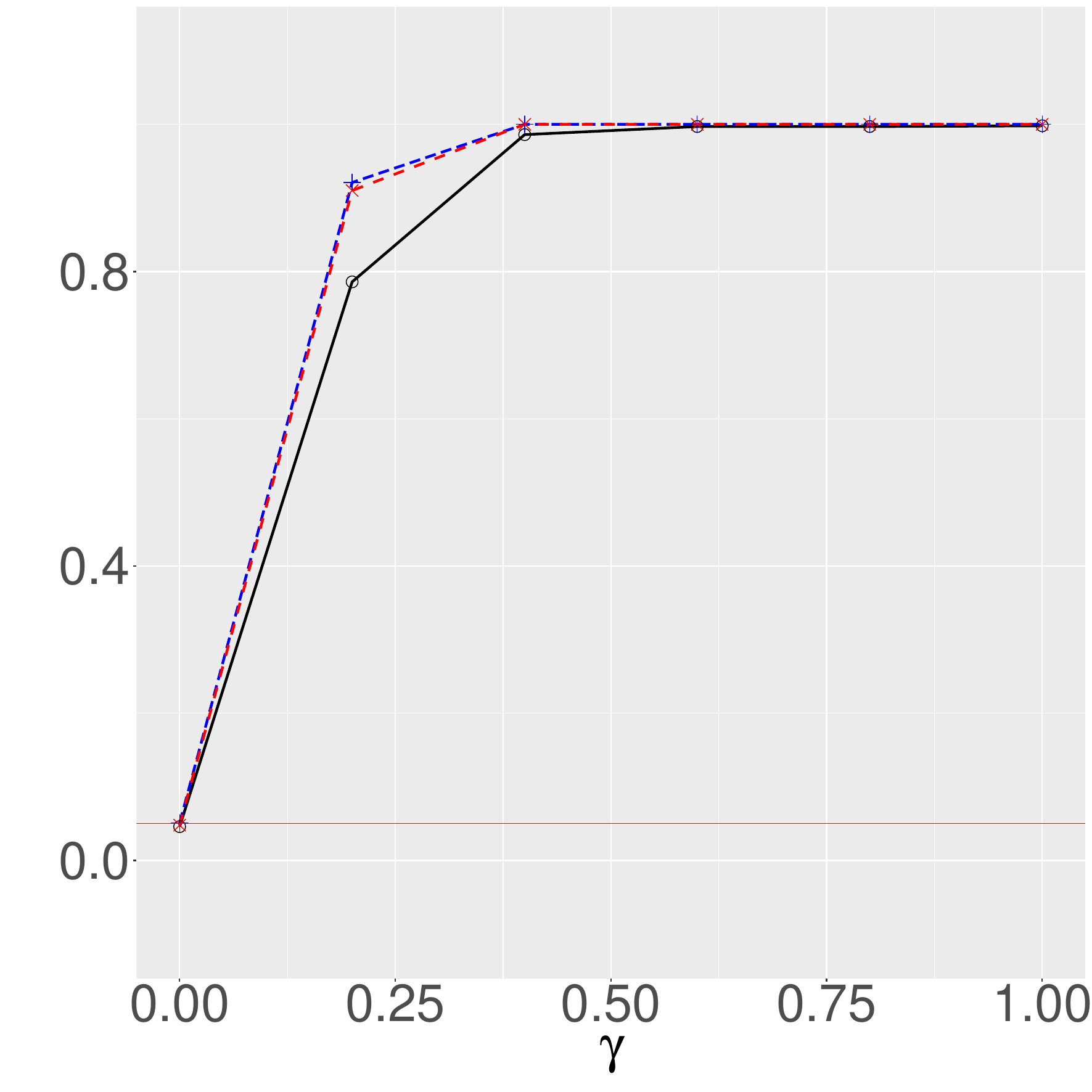}
\caption{5\%}
\end{subfigure}
\begin{subfigure}{0.32\textwidth}
\centering
\includegraphics[width=1\textwidth]{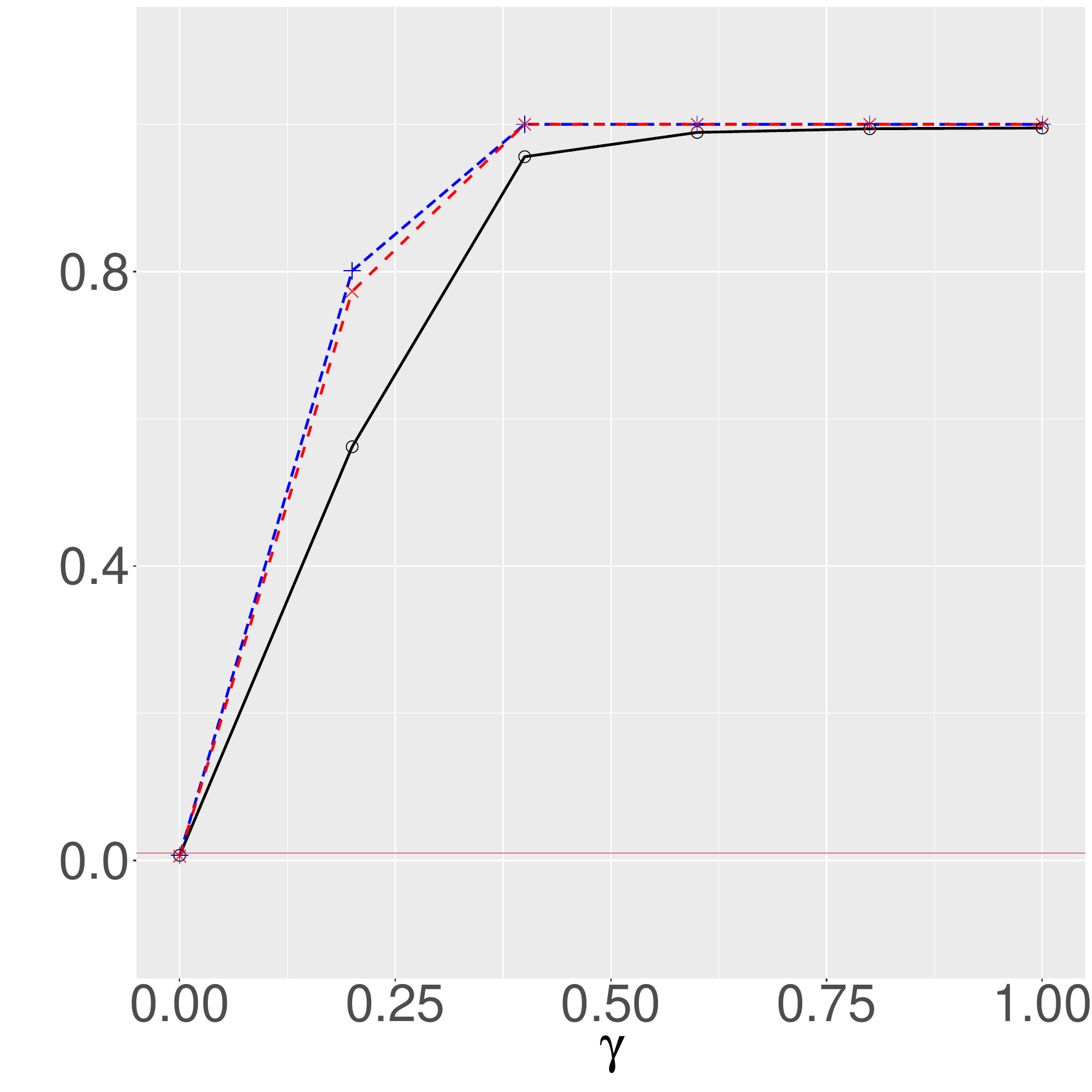}
\caption{1\%}
\end{subfigure}
\label{Fig:LS3_Gauss}
\end{figure}

\begin{figure}[H]
\centering 
\caption{DGP LS3 -- Negative Euclidean -- $n=400$.}
\begin{subfigure}{0.32\textwidth}
\centering
\includegraphics[width=1\textwidth]{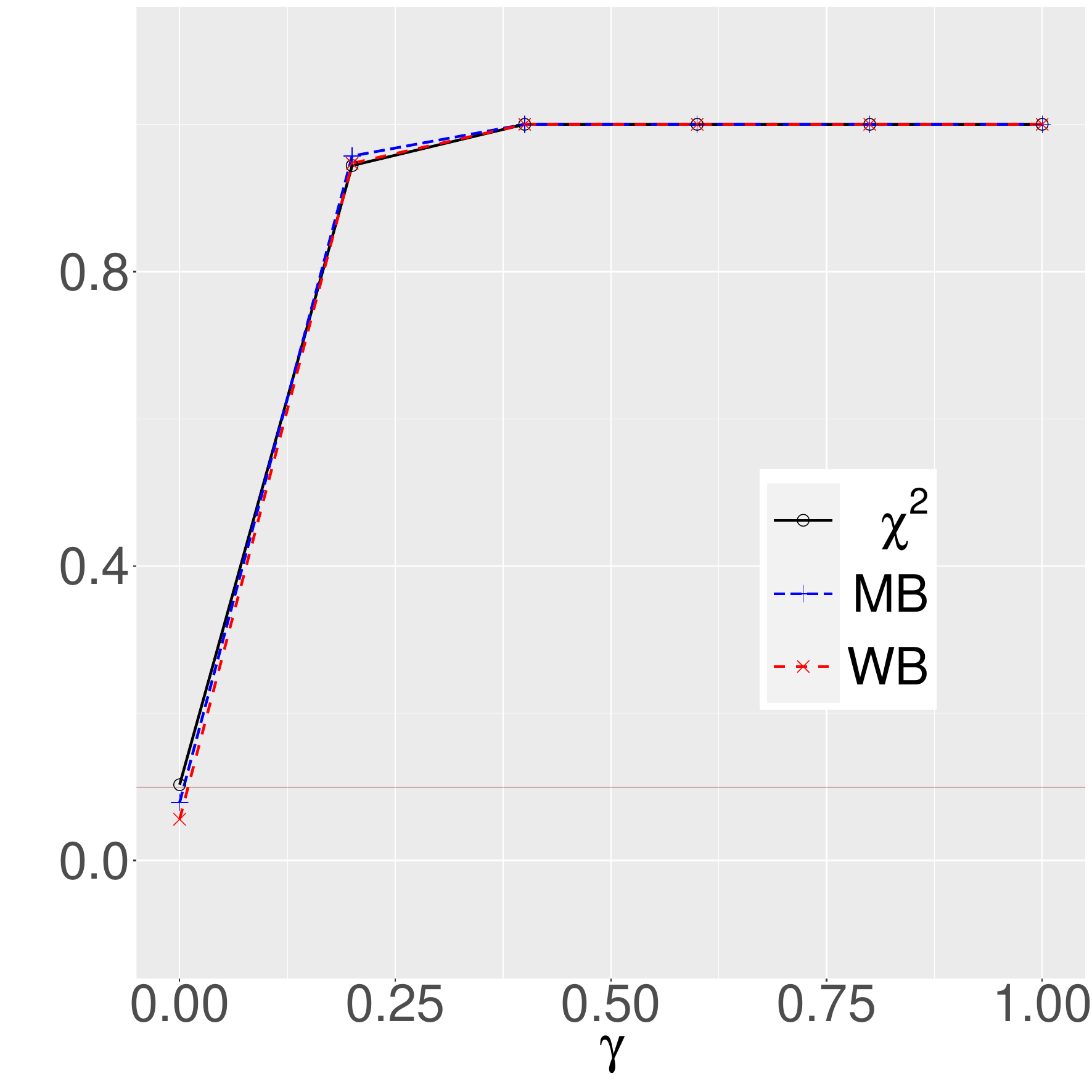}
\caption{10\%}
\end{subfigure}
\begin{subfigure}{0.32\textwidth}
\centering
\includegraphics[width=1\textwidth]{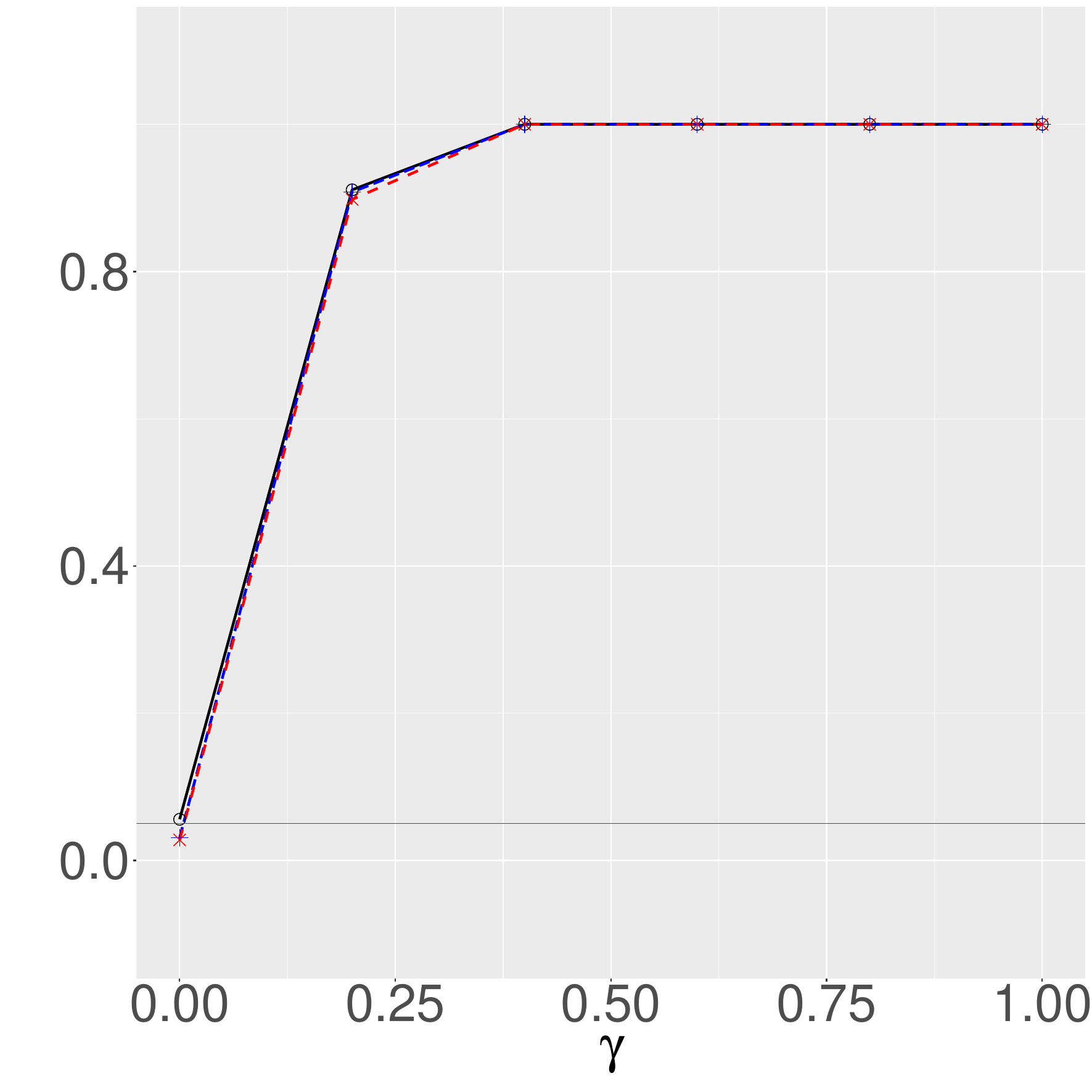}
\caption{5\%}
\end{subfigure}
\begin{subfigure}{0.32\textwidth}
\centering
\includegraphics[width=1\textwidth]{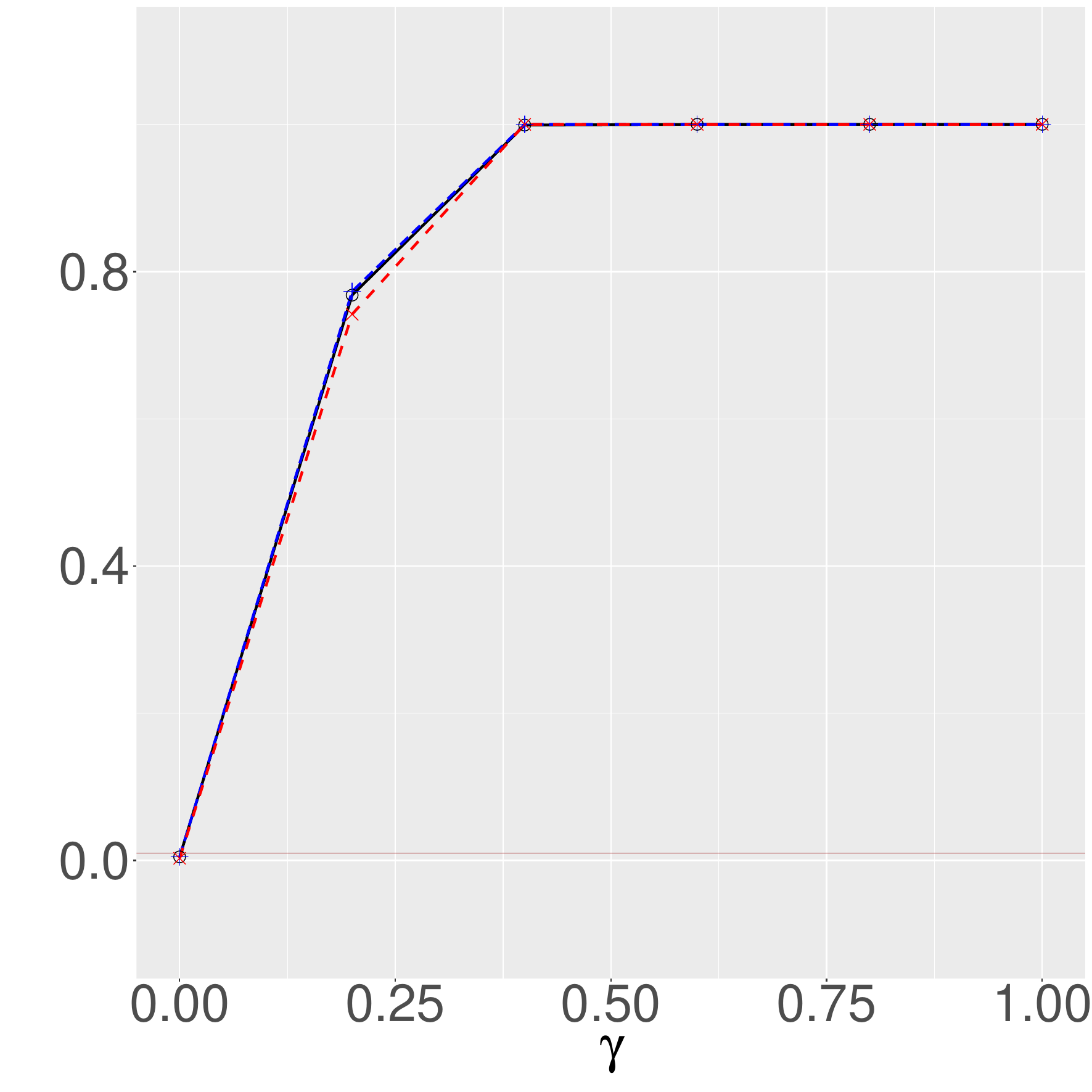}
\caption{1\%}
\end{subfigure}
\label{Fig:LS3_Euclid}
\end{figure}

\begin{figure}[H]
\centering 
\caption{DGP LS4 -- Gaussian Kernel -- $n=400$.}
\begin{subfigure}{0.32\textwidth}
\centering
\includegraphics[width=1\textwidth]{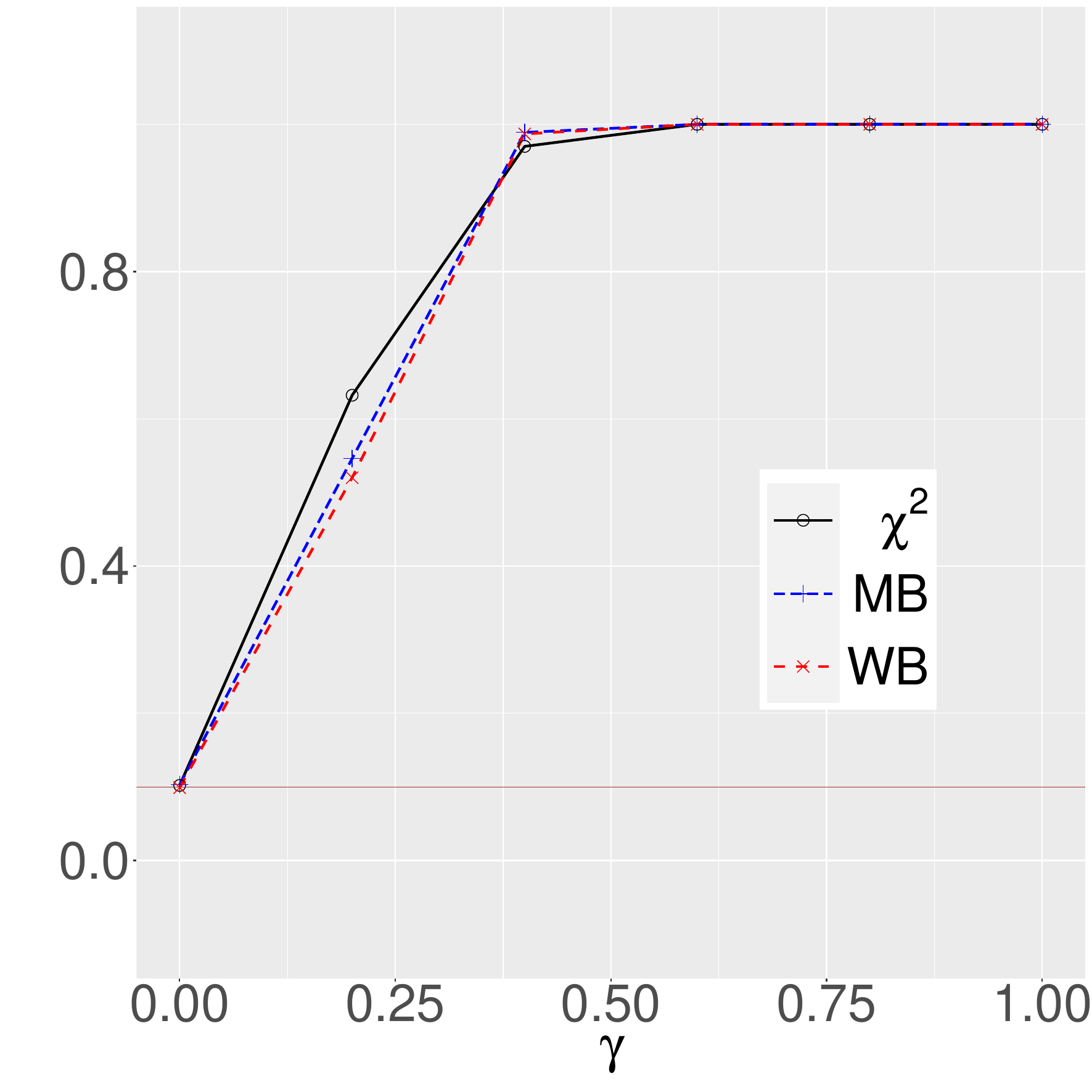}
\caption{10\%}
\end{subfigure}
\begin{subfigure}{0.32\textwidth}
\centering
\includegraphics[width=1\textwidth]{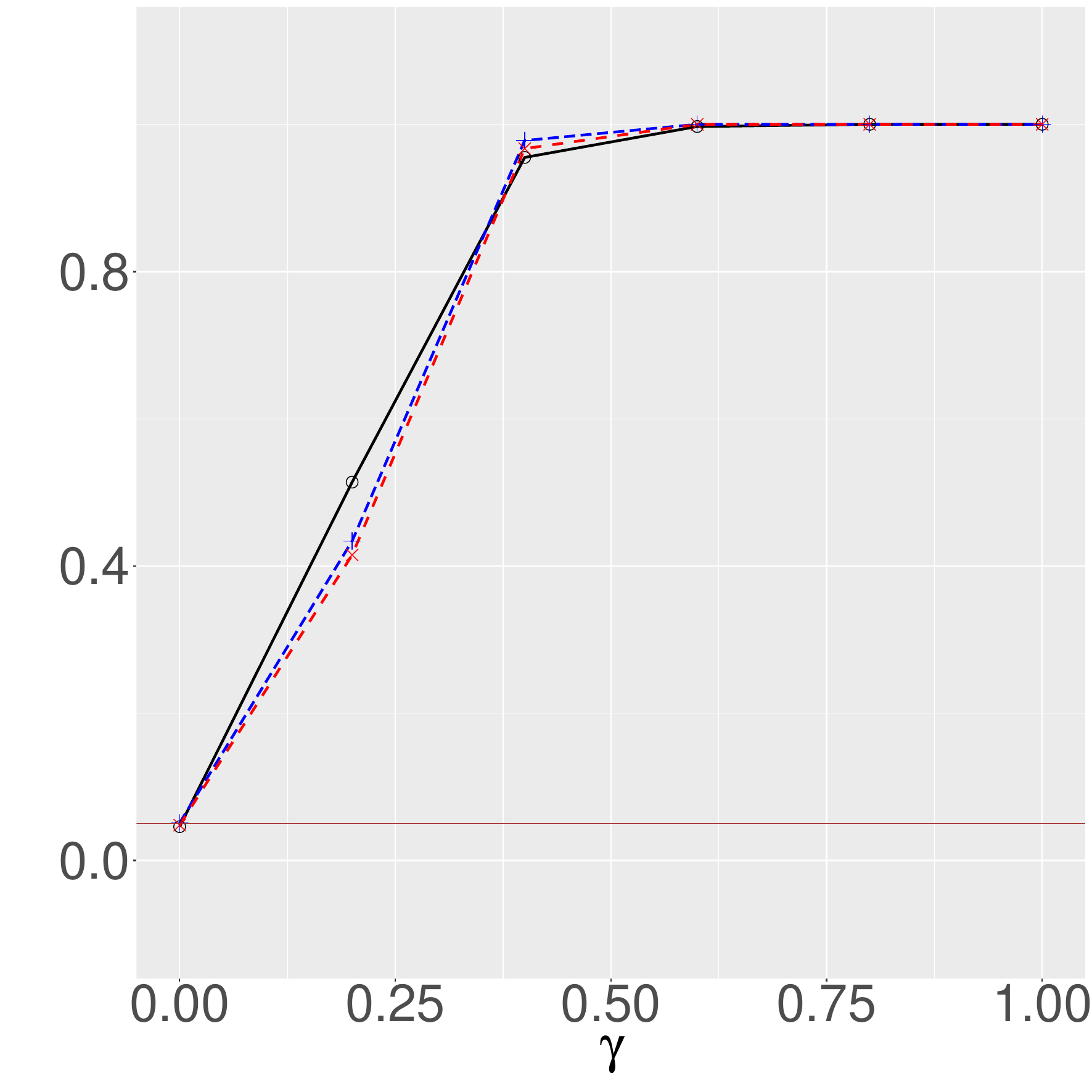}
\caption{5\%}
\end{subfigure}
\begin{subfigure}{0.32\textwidth}
\centering
\includegraphics[width=1\textwidth]{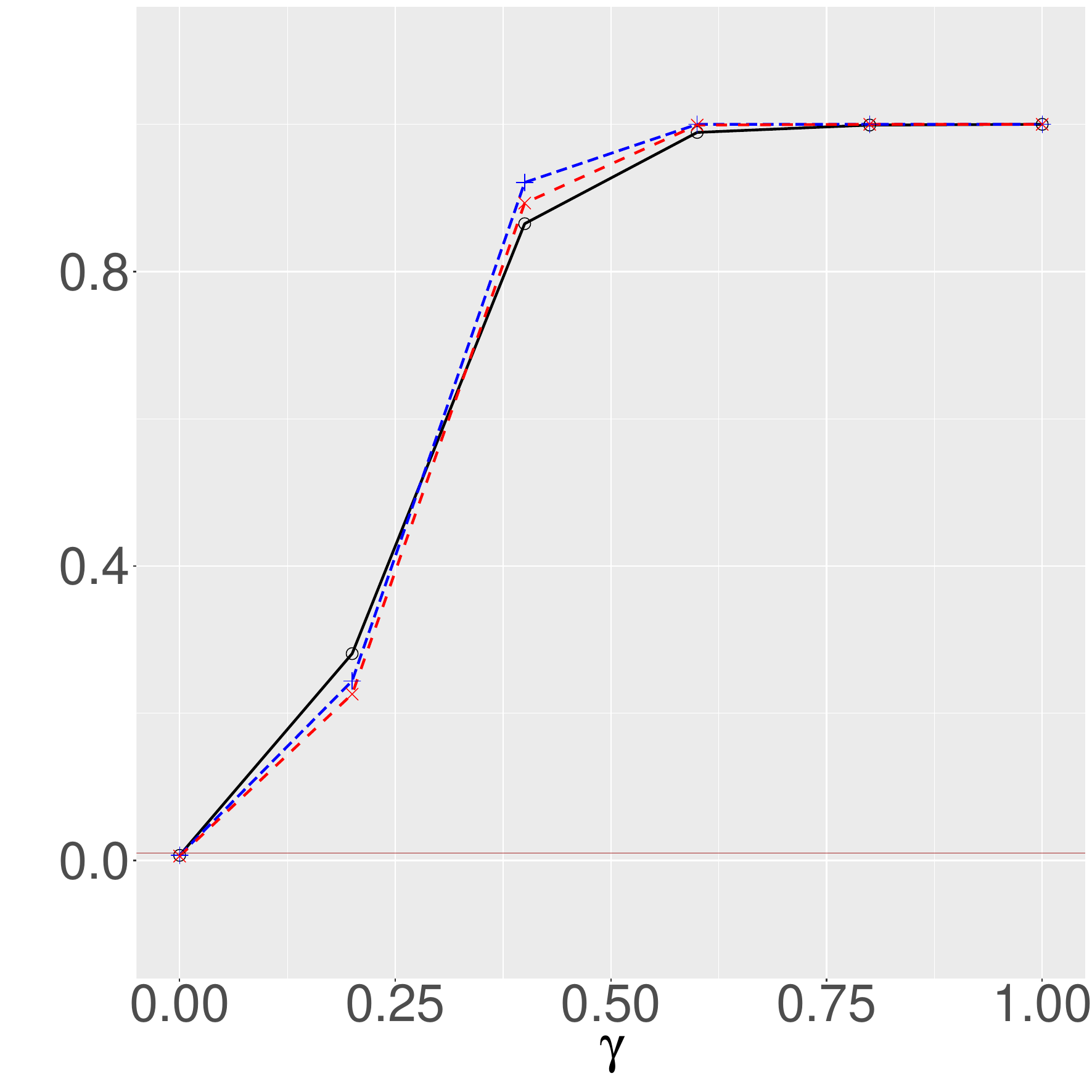}
\caption{1\%}
\end{subfigure}
\label{Fig:LS4_Gauss}
\end{figure}

\begin{figure}[H]
\centering 
\caption{DGP LS4 -- Negative Euclidean -- $n=400$.}
\begin{subfigure}{0.32\textwidth}
\centering
\includegraphics[width=1\textwidth]{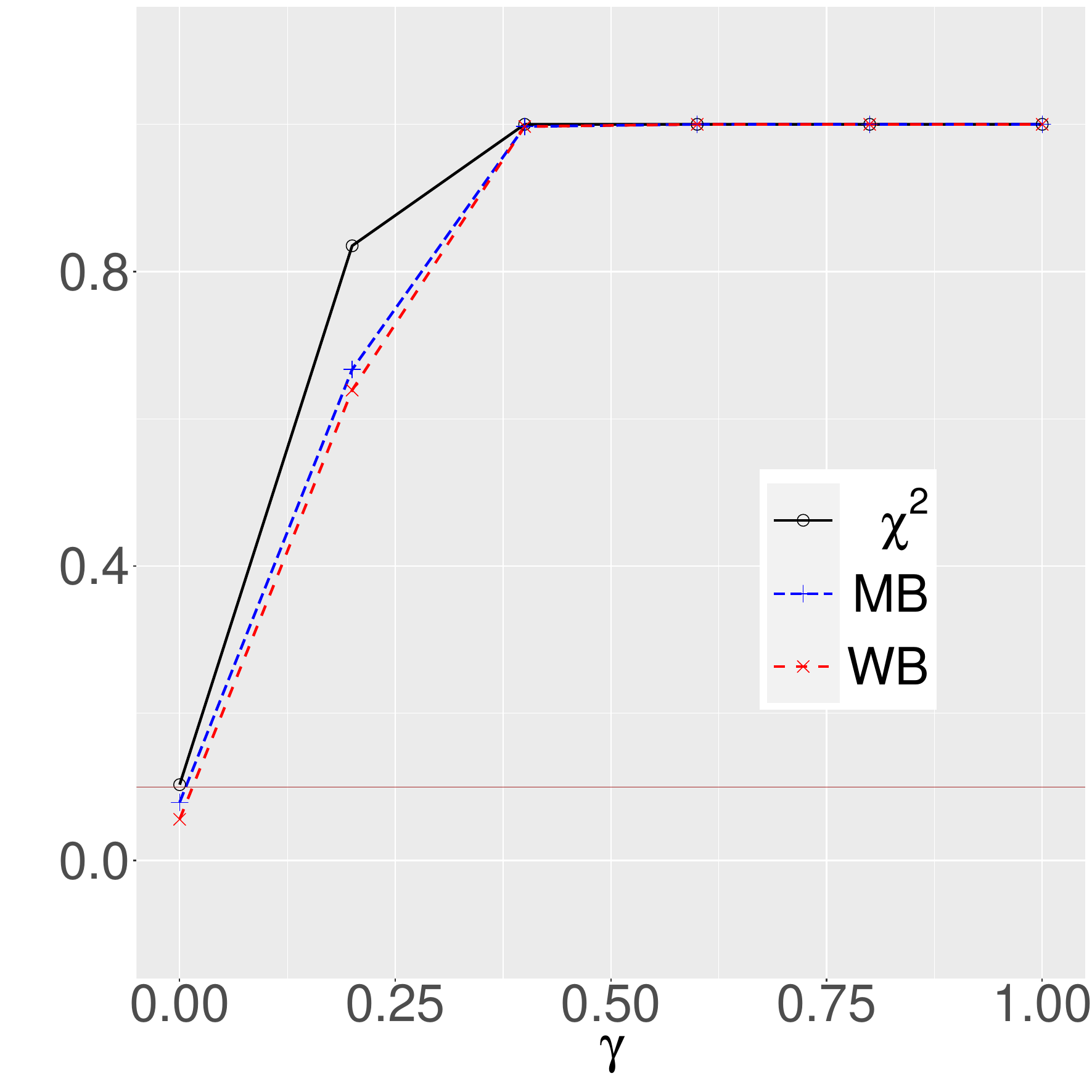}
\caption{10\%}
\end{subfigure}
\begin{subfigure}{0.32\textwidth}
\centering
\includegraphics[width=1\textwidth]{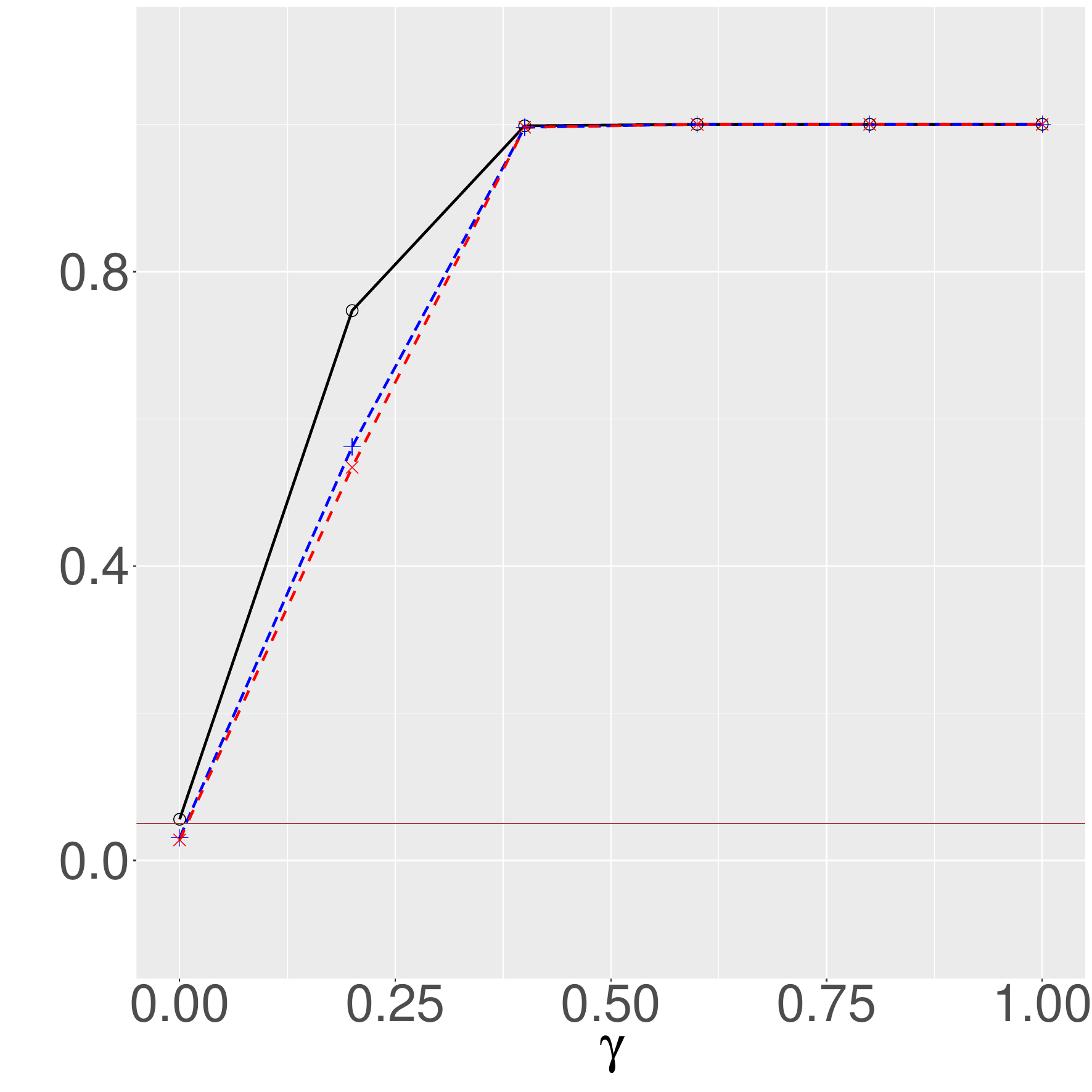}
\caption{5\%}
\end{subfigure}
\begin{subfigure}{0.32\textwidth}
\centering
\includegraphics[width=1\textwidth]{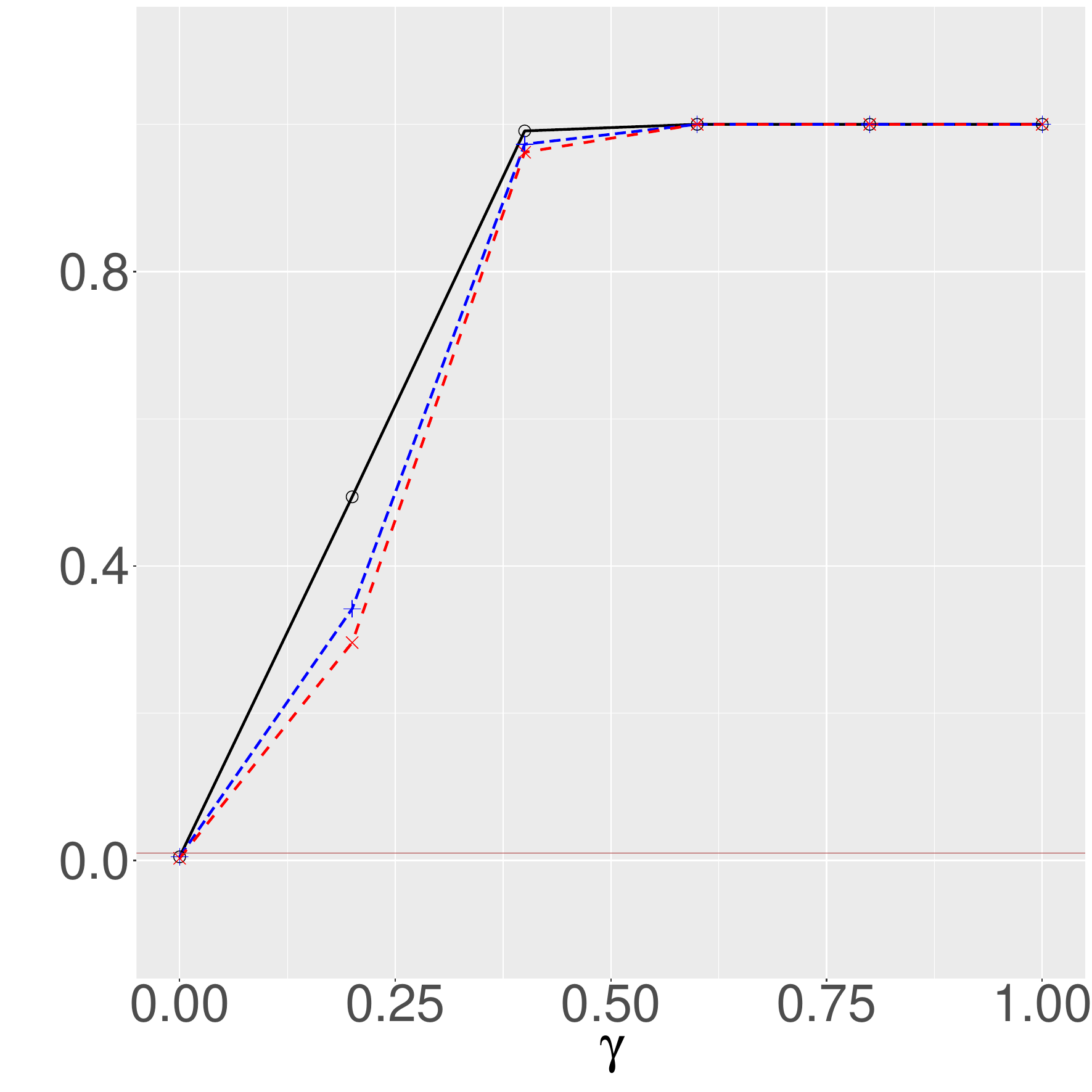}
\caption{1\%}
\end{subfigure}
\label{Fig:LS4_Euclid}
\end{figure}

To ensure that the good size control of the $\chi^2$-test in \Cref{Tab:Sim_Spec_LM12} is not achieved at the expense of power under the alternative, we consider its performance under both local and fixed alternatives. The results, presented in the third panel of \Cref{Tab:Sim_Spec_LM12}, indicate that the local power of the proposed $\chi^2$-test, along with the bootstrap-based procedures, is non-trivial. \Cref{Fig:LS2_Gauss,Fig:LS2_Euclid,Fig:LS3_Gauss,Fig:LS3_Euclid,Fig:LS4_Gauss,Fig:LS4_Euclid} present power curves corresponding to DGPs LS2 through LS4 using the Gaussian and Negative Euclidean kernels.  One observes that all tests demonstrate non-trivial power against the fixed alternative for variations of $\gamma$ away from zero. The proposed test exhibits highly competitive power performance across all three DGPs and both kernels considered. Notably, the relative power of the $\chi^2$-test compared to the bootstrap-based procedures demonstrates variability depending on the specific DGP and kernel, as highlighted in \Cref{thm_bahadur}.

The $\chi^2$-test proves useful when researchers seek a measure of correct model specification or mean independence without resorting to a formal hypothesis test, owing to the pivotality of its test statistic. In summary, our simulations demonstrate that the proposed $\chi^2$-test, in addition to its enhanced interpretability compared to bootstrap-based ICM specification tests, exhibits good size control and comparable power performance.

\subsection{Running Time}\label{subsec:run}
One key advantage of a pivotal test is its computational efficiency relative to bootstrap-based procedures, which becomes particularly critical in large samples. Accordingly, a comparison of the running times of competing tests is in order. \Cref{Tab:Sim_Time_LM0} presents a comparison between the proposed pivotal $\chi^2$-test and the bootstrap-based ICM tests in terms of computational time for fixed ICM kernels.\footnote{Computations were performed on a 2.8 GHz Quad-Core Intel Core i7 with 16 GB of RAM, running on a MacBook Pro.}

\begin{table}[!htbp]
    \centering
    \setlength{\tabcolsep}{4pt}
    \caption{Running Time - Specification Test - DGP LS1}
    \begin{tabular}{lcccccc}
    \toprule \toprule
        & & \multicolumn{3}{c}{Average Running Time (seconds)} & \multicolumn{2}{c}{Median Relative Time}  \\
        \cmidrule(lr){3-5} \cmidrule(lr){6-7}
        n & Kernel & $\chi^2$ & MB & WB & MB & WB \\ \midrule
    & Gauss  & 0.004 & 0.114 & 0.965 & 29.000 & 248.000 \\
200 &        & (0.002) & (0.021) & (0.063) & (16.833) & (142.208) \\
    & Euclid & 0.004 & 0.111 & 0.993 & 35.000 & 315.667 \\
    &        & (0.001) & (0.006) & (0.043) & (15.425) & (127.733) \\ \midrule
    & Gauss  & 0.011 & 0.440 & 1.100 & 47.222 & 116.422 \\
400 &        & (0.003) & (0.022) & (0.054) & (18.856) & (41.898) \\
    & Euclid & 0.009 & 0.435 & 1.095 & 58.786 & 145.571 \\
    &        & (0.002) & (0.016) & (0.055) & (23.858) & (55.071) \\ \midrule
    & Gauss  & 0.021 & 1.137 & 1.254 & 59.833 & 65.526 \\
600 &        & (0.004) & (0.035) & (0.048) & (18.407) & (18.479) \\
    & Euclid & 0.016 & 1.125 & 1.247 & 76.857 & 85.000 \\
    &        & (0.003) & (0.040) & (0.051) & (25.869) & (27.041) \\ \midrule
    & Gauss  & 0.038 & 2.427 & 1.552 & 65.859 & 41.789 \\
800 &        & (0.005) & (0.057) & (0.048) & (13.833) & (7.420) \\
    & Euclid & 0.030 & 2.471 & 1.584 & 87.554 & 55.930 \\
    &        & (0.004) & (0.053) & (0.040) & (22.298) & (13.505) \\ \bottomrule
    \end{tabular}
    \label{Tab:Sim_Time_LM0}
    \footnotesize
            
        \textit{Notes: The second row (in parentheses) for each sample size and kernel includes standard deviations and inter-quartile ranges for the average running times and relative times  the $\chi^2$-test as the benchmark), respectively.}
\end{table}
  
\Cref{Tab:Sim_Time_LM0} illustrates the substantial computational advantage of employing the $\chi^2$ specification test. The average running times for the $\chi^2$-test are negligible compared to those of the bootstrap-based procedures across all considered sample sizes. A striking difference emerges in terms of median relative computational time (with the $\chi^2$-test serving as a benchmark). While the multiplier bootstrap procedure is generally faster than the wild bootstrap procedure, our findings indicate that the median relative computational time of the multiplier bootstrap tends to increase with the sample size, whereas that of the wild bootstrap appears to decrease. At particularly large sample sizes ($n=600$ and $n=800$), the computational gain offered by the $\chi^2$-test becomes considerable.

\section{Conclusion}\label{Sect:Conclusion}
Despite the 40-year history of ICM tests, dating back to \textcite{bierens1982consistent} and encompassing numerous interesting contributions, a \emph{bona fide} pivotalized ICM specification test remains lacking. This paper achieves the objective of proposing an omnibus $\chi^2$-test of specification and mean independence based on ICM metrics.
		
The proposed $\chi^2$-test complements existing ICM tests by overcoming certain limitations of the latter. The test statistic can be constructed with functional forms that boost power in the direction of alternatives the researcher may have in mind. The test is computationally more efficient than commonly used bootstrap-based ICM tests and remains computationally viable even in large samples. In addition to providing a reliable pivotal test that derives its omnibus property from ICM metrics, the test statistic offers an easily interpretable metric of model specification and mean dependence, thereby obviating formal hypothesis tests.

In conclusion, we highlight several potential extensions to the current work. While this paper offers a viable solution for pivotalizing ICM tests, it is noteworthy that this approach necessitates regularization under the null hypothesis to address the ill-posed inverse problem. Future research could explore alternative formulations of $V$ that may yield solutions with more desirable theoretical properties. Furthermore, extending our analysis from the current fixed-dimensional setting to a high-dimensional framework presents an interesting avenue for research. Moreover, leveraging the projection approach proposed by \citet{escanciano2009simple,escanciano-2024-gaussian}, which eliminates the estimation effect, could be advantageous in dealing with estimators whose limiting distributions are difficult to characterize, for example, the LASSO. Finally, extending our methodology to encompass time series data, clustered data, and multiple-equation models with (non)-smooth objective functions warrants future exploration.


\printbibliography
\end{refsection}

\newpage
\setcounter{page}{1}
\begin{refsection}
\appendix
\renewcommand{\thetable}{S.\arabic{table}}
\renewcommand{\thefigure}{S.\arabic{figure}}
\renewcommand{\thesection}{S.\arabic{section}}
\renewcommand{\theequation}{\thesection\arabic{equation}}

\begin{center}
    \Large{\bf Supplementary Material}
\end{center}
\begin{center}
    Feiyu Jiang  ~~~~~~~~Emmanuel Selorm Tsyawo
\end{center}

The supplementary material contains all technical proofs and additional simulation results. Section \ref{Sect:Supporting_Lemma} provides lemmata supporting proofs in Section \ref{Sect_Appendix:Technical_Proofs}. Section \ref{sec:deltastar} studies the asymptotic properties of the $\chi^2$-test based on $\delta_h^*$. Section \ref{Sect_Appendix:Bahadur} provides numerical comparisons of the Bahadur slopes discussed in \Cref{SubSect:Bahadur}. Section \ref{Sect_Appendix:MC} contains simulation results on mean independence tests, {Section \ref{Sect_Suppl:MC_NLM} extends simulation results to specification testing of the non-linear logit model,} and Section \ref{Sect_Appendix:CM_Tests} concludes by discussing the relation between the proposed $\chi^2$-test and the family of CM tests.

\section{Supporting Lemmata}\label{Sect:Supporting_Lemma}

By the analog and plug-in principles, the estimator of $ \mathrm{ICM}(U \mid Z) $ is given by 
\[
\mathrm{ICM}_n(U \mid Z) := \frac{1}{n(n-1)}\sum_{i=1}^n\sum_{j\neq i}K(Z_i,Z_j)U(X_i;\widehat{\theta}_n)U(X_j;\widehat{\theta}_n).
\]

\begin{lemma}\label{lem:ICM_lin_asymp}
    Under \Cref{ass:bound_psiD,ass:Sampling_iid,ass:diff_Utheta,ass:theta_asymp_lin}, $\mathrm{ICM}_n(U \mid Z)$ has the following asymptotically linear representation:
    \[
    \sqrt{n}\big(\mathrm{ICM}_n(U \mid Z)-\mathrm{ICM}(U \mid Z)\big) = \frac{2}{\sqrt{n}}\sum_{i=1}^{n}\left\{\psi_U^{(1)}(D_i)-\mathrm{ICM}(U \mid Z) - H_2^\top \xi_{\theta,i} \right\} + o_p(1).
    \]
\end{lemma}

\textbf{Proof}

Under \Cref{ass:diff_Utheta} and the symmetry of the kernel $K(\cdot,\cdot)$,
\begin{flalign*}
\begin{split}
    \mathrm{ICM}_n(U \mid Z):=&\frac{1}{n(n-1)}\sum_{i=1}^n\sum_{j\neq i}K(Z_i,Z_j)U(X_i;\widehat{\theta}_n)U(X_j;\widehat{\theta}_n)\\
    =&\frac{1}{n(n-1)}\sum_{i=1}^n\sum_{j\neq i}K(Z_i,Z_j)\big(\widetilde{U}_i - G(X_i;\bar{\theta}_n)^\top(\widehat{\theta}_n - \theta_o) \big)\big(\widetilde{U}_j - G(X_j;\bar{\theta}_n)^\top(\widehat{\theta}_n - \theta_o) \big) \\ 
    =&\frac{1}{n(n-1)}\sum_{i\neq j}^{n}K(Z_i,Z_j)\widetilde{U}_i\widetilde U_j \\
    &+ (\widehat{\theta}_n - \theta_o)^\top\Bigg\{\frac{1}{n(n-1)}\sum_{i\neq j}^{n}K(Z_i,Z_j) G(X_i;\bar{\theta}_n) G(X_j;\bar{\theta}_n)^\top\Bigg\} (\widehat{\theta}_n - \theta_o)\\
    &- \frac{2}{n(n-1)}\sum_{i\neq j}^{n}K(Z_i,Z_j)\widetilde{U}_jG(X_i;\bar{\theta}_n)^\top(\widehat{\theta}_n - \theta_o)
    \\
    :=&\sum_{k=1}^3 G_k.
\end{split}
\end{flalign*}

The first summand is a second-order $U$-statistic. Recall $\psi_U(D_i,D_j) := K(Z_i,Z_j)\widetilde{U}_i\widetilde U_j$, and $\psi_U^{(1)}(D_i) = \E\big[\psi_U(D_i,D_j)|D_i\big]$. By \Cref{ass:Sampling_iid}, \Cref{ass:bound_psiD}, and the Hoeffding decomposition (e.g., Theorem 3 in section 1.3 of \citet{lee1990u}), 
\begin{flalign}\label{hoeffding1}
	\begin{split}
	 G_1= &\frac{1}{n(n-1)}\sum_{i\neq j}^{n}K(Z_i,Z_j)\widetilde{U}_i\widetilde U_j\\=  
     &{n\choose 2}^{-1}\sum_{i<j}^{n} K(Z_i,Z_j)\widetilde{U}_i\widetilde U_j\\
     =&\mathrm{ICM}(U \mid Z) + \frac{2}{n}\sum_{i=1}^{n}\big[ \psi_U^{(1)}(D_i)-\mathrm{ICM}(U \mid Z) \big]\\
     &+ {n\choose 2}^{-1}\sum_{i<j}^{n}\big[\psi_U(D_i,D_j)-\psi_U^{(1)}(D_i)-\psi_U^{(1)}(D_j)+\mathrm{ICM}(U \mid Z)\big].
     \\
     =&\mathrm{ICM}(U \mid Z)+ \frac{2}{n}\sum_{i=1}^{n}\big[\psi_U^{(1)}(D_i)-\mathrm{ICM}(U \mid Z)\big] + o_p(n^{-1/2}).
	\end{split}
\end{flalign}

\noindent By \Cref{ass:Sampling_iid,ass:theta_asymp_lin}, $ G_2 = O_p(n^{-1})$. For the third summand, note that by the law of large numbers for $U$-statistics, the continuous mapping theorem and \Cref{ass:Sampling_iid,ass:diff_Utheta,ass:bound_psiD},
\begin{flalign*}
    \frac{1}{n(n-1)}\sum_{i\neq j}^{n}K(Z_i,Z_j)\widetilde{U}_jG(X_i;\bar{\theta}_n)^\top \xrightarrow{p} \ \E[K(Z,Z^\dagger)\widetilde{U}^\dagger G(X;\theta_o)]^\top:= H_2^\top.
\end{flalign*}
Hence, $ \displaystyle
 G_3= -2H_2^\top \frac{1}{n}\sum_{i=1}^n \xi_{\theta,i} +o_p(n^{-1/2})$.

Combining all the terms above proves the assertion as claimed.

\qed

\section{Proofs of results in the main text}\label{Sect_Appendix:Technical_Proofs}

\subsection{Proof of \Cref{lem:key_nullity}}

\begin{align*}
    \delta_h^* &= \left[ \E[\widetilde{U}^\dagger \widetilde{h}(Z) K(Z,Z^{\dagger})],\   \mathrm{ICM}(U \mid Z) - \E[\widetilde{U}^\dagger \widetilde{h}(Z)K(Z,Z^{\dagger})]\right]^{\top}\\
    &+ \left[ \E[h(Z)]\E U ,\ (\E U)^2 - \E[h(Z)]\E U \right]^{\top} \E|K(Z,Z^{\dagger})|\\
    &= \Big[ \E[\widetilde{U}^\dagger \widetilde{h}(Z) K(Z,Z^{\dagger})] + \E|K(Z,Z^{\dagger})|\E[h(Z)]\E U , \\  
    &\mathrm{ICM}(U \mid Z) - \E[\widetilde{U}^\dagger \widetilde{h}(Z)K(Z,Z^{\dagger})] + \E|K(Z,Z^{\dagger})|\big((\E U)^2 - \E[h(Z)]\E U\big) \Big]^{\top}\\
    &:= [\delta_h^{*(1)},\ \delta_h^{*(2)}]^{\top}.
\end{align*}
Thus, 
$$
\delta_h^{*(1)}+\delta_h^{*(2)} = \mathrm{ICM}(U \mid Z) + (\E U)^2\E|K(Z,Z^{\dagger})|.
$$
\noindent If $\E[U|Z]=0 \ a.s. $, then  $\E U=0$ by the LIE. This implies that $\delta_h^*=0$. Conversely, if $\delta_h^*=0$, then we must have $\delta_h^{*(1)} + \delta_h^{*(2)} = 0$. Note that both $\mathrm{ICM}(U \mid Z)$ and $(\E U)^2\E|K(Z,Z^{\dagger})|$ are non-negative, thus we must have $\mathrm{ICM}(U \mid Z)=0$  and $(\E U)^2\E|K(Z,Z^{\dagger})|=0$, corresponding to $\E[U|Z]=\E U \ a.s. $ (by the omnibus property \eqref{eqn:Prop_Omnibus}) and $\E U=0$, respectively.  Hence $\E[U|Z]=0. \ a.s. $

\qed

\subsection{Proof of \Cref{lem:del_asymp_lin}}

The estimand $\delta_h$ can be expressed as
\begin{align*}
    \delta_h = \begin{bmatrix}
        \delta_h^{(1)}\\
        \mathrm{ICM}(U \mid Z) - \delta_h^{(1)}
    \end{bmatrix}
    =\begin{bmatrix}
        1 & 0 \\ -1 & 1
    \end{bmatrix}
    \begin{bmatrix}
        \delta_h^{(1)} \\ \mathrm{ICM}(U \mid Z)
    \end{bmatrix}
\end{align*}where $\delta_h^{(1)}:= \E\big[ K(Z,Z^\dagger)(h(Z^\dagger)-\E[h(Z)])(U - \E U) \big] $. 

Estimation of $\delta_h$ follows by applying the analog and plug-in principles: 

\begin{align*}
    \widehat{\delta}_h
    =\begin{bmatrix}
        1 & 0 \\ -1 & 1
    \end{bmatrix}
    \begin{bmatrix}
        \widehat{\delta}_h^{(1)} \\ \mathrm{ICM}_n(U \mid Z)
    \end{bmatrix}
\end{align*}where 
\begin{align*}
    \widehat{\delta}_h^{(1)} &= \frac{1}{n(n-1)}\sum_{i=1}^n \sum_{j\neq i} K(Z_i,Z_j)(h(Z_j)-\E_n[h(Z)])U(X_i; \widehat{\theta}_n) \text{ and} \\ 
    \mathrm{ICM}_n(U \mid Z) &= \frac{1}{n(n-1)}\sum_{i=1}^n\sum_{j\neq i}K(Z_i,Z_j)U(X_i; \widehat{\theta}_n)U(X_j; \widehat{\theta}_n).
\end{align*}

Thanks to \Cref{ass:diff_Utheta}, \Cref{ass:theta_asymp_lin}, and the continuous mapping theorem,
\begin{align*}
    \widehat{\delta}_h^{(1)} &= \frac{1}{n(n-1)}\sum_{i=1}^n \sum_{j\neq i} K(Z_i,Z_j)\widetilde{h}(Z_j)\widetilde{U}_i - \frac{1}{n(n-1)}\sum_{i=1}^n \sum_{j\neq i} K(Z_i,Z_j)\widetilde{h}(Z_j)G(X_i;\bar{\theta}_n)^\top(\widehat{\theta}_n - \theta_o)\\
    &- \Big(\frac{1}{n(n-1)}\sum_{i=1}^n \sum_{j\neq i} K(Z_i,Z_j)\widetilde{U}_i\Big) \E_n[\widetilde{h}(Z)] + \Big(\frac{1}{n(n-1)}\sum_{i=1}^n \sum_{j\neq i} K(Z_i,Z_j)G(X_i;\bar{\theta}_n)\Big) ^\top(\widehat{\theta}_n - \theta_o)\E_n[\widetilde{h}(Z)]\\
    &= \frac{1}{n(n-1)}\sum_{i=1}^n \sum_{j\neq i} K(Z_i,Z_j)\big( \widetilde{h}(Z_j)\widetilde{U}_i + \widetilde{h}(Z_i)\widetilde{U}_j \big)/2\\
    &- \begin{bmatrix} \E \big[K(Z, Z^\dagger)\widetilde{h}(Z^\dagger)G(X;\theta_o)\big]
        \\
        \E\big[K(Z,Z^\dagger)\widetilde{U}\big]
    \end{bmatrix}^\top \begin{bmatrix}
        \widehat{\theta}_n - \theta_o\\
        \E_n[h(Z)] - \E[h(Z)]
    \end{bmatrix} + o_p(n^{-1/2})\\
    &= \frac{1}{n(n-1)}\sum_{i=1}^n \sum_{j\neq i} \psi_h(D_i,D_j) - 2H_1^\top\big( (\widehat{\theta}_n - \theta_o)^\top, (\E_n[h(Z)] - \E[h(Z)])\big)^\top  + o_p(n^{-1/2})
\end{align*}where $\bar{\theta}_n$ satisfies $||\bar{\theta}_n - \theta_o|| \leq || \widehat{\theta}_n - \theta_o || $ and $ H_1:= (1/2)\begin{bmatrix} \E \big[K(Z, Z^\dagger)\widetilde{h}(Z^\dagger)G(X;\theta_o)\big]
        \\
        \E\big[K(Z,Z^\dagger)\widetilde{U}\big]
    \end{bmatrix} $.

The first summand in the above decomposition is a second-order $U$-statistic. Recall $\psi_h^{(1)}(D_i) =\E[\psi_h(D_i,D_j)|D_i]$, then by \Cref{ass:Sampling_iid}, \Cref{ass:bound_psiD}, and the Hoeffding decomposition (e.g., Theorem 3 in section 1.3 of \citet{lee1990u}), 
\begin{flalign*}
	\begin{split}
	&\frac{1}{n(n-1)}\sum_{i\neq j}^{n} \psi_h(D_i,D_j) =  {n\choose 2}^{-1}\sum_{i<j}^{n} \psi_h(D_i,D_j)\\
     =&\delta_h^{(1)} + \frac{2}{n}\sum_{i=1}^{n}[\psi_h^{(1)}(D_i) - \delta_h^{(1)}] + {n\choose 2}^{-1}\sum_{i<j}^{n}[\psi_h(D_i,D_j)-\psi_h^{(1)}(D_i) - \psi_h^{(1)}(D_j) + \delta_h^{(1)}]\\
     =&\delta_h^{(1)} + \frac{2}{n}\sum_{i=1}^{n}[\psi_h^{(1)}(D_i) - \delta_h^{(1)}] + o_p(n^{-1/2}).
	\end{split}
\end{flalign*}

\noindent In addition to \Cref{ass:theta_asymp_lin}, $ \displaystyle \sqrt{n}\big(\widehat{\delta}_h^{(1)} - \delta_h^{(1)} \big) = \frac{2}{\sqrt{n}}\sum_{i=1}^{n}\big[\psi_h^{(1)}(D_i) - \delta_h^{(1)} - H_1^\top[\xi_{\theta,i}^\top, \widetilde{h}(Z_i) ]^\top\big] + o_p(1) $.

Under the conditions of \Cref{lem:ICM_lin_asymp},
$$
\sqrt{n}[\mathrm{ICM}_n(U \mid Z)-\mathrm{ICM}(U \mid Z)]=\frac{2}{\sqrt{n}}\sum_{i=1}^{n}\left\{\psi_U^{(1)}(D_i)-\mathrm{ICM}(U \mid Z) - H_2^\top \xi_{\theta,i}\right\} + o_p(1).
$$

Thus, as claimed,
\begin{align*}
    \sqrt{n}\big(\widehat{\delta}_h - \delta_h\big)
    =\begin{bmatrix}
        1 & 0 \\ -1 & 1
    \end{bmatrix}\frac{2}{\sqrt{n}}\sum_{i=1}^{n}
    \begin{bmatrix}
        \psi_h^{(1)}(D_i) - \delta_h^{(1)} - H_1^\top[\xi_{\theta,i}^\top, \widetilde{h}(Z_i) ]^\top \\
        \psi_U^{(1)}(D_i)-\mathrm{ICM}(U \mid Z) - H_2^\top \xi_{\theta,i}
    \end{bmatrix} + o_p(1).
\end{align*} 

\qed

\subsection{Proof of \Cref{thm_all}}

Under the conditions of \Cref{lem:del_asymp_lin},
\begin{align*}
   \sqrt{n}(\widehat{\delta}_h - \delta_h) = \frac{1}{\sqrt{n}} \sum_{i=1}^n \xi_h(D_i) + o_p(1) = \frac{2}{\sqrt{n}} \sum_{i=1}^n \begin{bmatrix}
       \xi_{h,1}(D_i)\\ \xi_{h,2}(D_i) - \xi_{h,1}(D_i)
   \end{bmatrix} + o_p(1).
\end{align*}

Thanks to the LIE, we have the following:
\begin{equation}\label{eqn:del_LIE_comp}
    \begin{split}
    \psi_h^{(1)}(D_i) &= \frac{1}{2}\E[K(Z_i,Z_j)\widetilde{h}(Z_j)|Z_i]\widetilde{U}_i + \frac{1}{2}\E\big[K(Z_i,Z_j)\E[\widetilde{U}_j\mid Z_j]\mid Z_i\big]\widetilde{h}(Z_i);\\
    \psi_U^{(1)}(D_i)  &= \E\big[K(Z_i,Z_j)\E[\widetilde{U}_j\mid Z_j]|Z_i\big]\widetilde{U}_i;\\
    \mathrm{ICM}(U\mid Z)&= \E \big[ K(Z,Z^\dagger)\E[\widetilde{U}^\dagger\mid Z^\dagger]\E[\widetilde{U}\mid Z]\big]; \text{ and}\\
    H_2 &= \E \big[ K(Z,Z^\dagger)\E[\widetilde{U}^\dagger\mid Z^\dagger]G(X;\theta_o)\big].
\end{split}
\end{equation}

(i) Under $\mathbb{H}_o$, $\E[\widetilde{U}\mid Z]=0 \ a.s.$ and we have $\delta_h=0$. From \eqref{eqn:del_LIE_comp}, 
$
\xi_{h,2}(D_i):= \psi_U^{(1)}(D_i)-\mathrm{ICM}(U \mid Z) - H_2^\top \xi_{\theta,i} = 0 \ a.s.
$ as each summand of $\xi_{h,2}(D_i)$ is zero $a.s.$ under $\mathbb{H}_o$.  This implies that under $\mathbb{H}_o$,
\begin{flalign*}
   \sqrt{n}\widehat{\delta}_h= [1, \ -1]^\top
   \frac{2}{\sqrt{n}}\sum_{i=1}^{n}\xi_{h,1}(D_i) + o_p(1).
\end{flalign*}
Asymptotic normality thus follows from \Cref{ass:Sampling_iid}, \Cref{ass:bound_psiD}, \Cref{ass:theta_asymp_lin}, and the Lindberg-L\'evy Central Limit Theorem. 

    (ii) Under $ \mathbb{H}_{an} $, $ \sqrt{n}\delta_h^{(1)} = \E\big[\widetilde{h}(Z)a(Z^{\dagger})K(Z,Z^{\dagger})\big] $ and $\sqrt{n}\mathrm{ICM}(U \mid Z) = \E\big[a(Z)a(Z^{\dagger})K(Z,Z^{\dagger})\big]/\sqrt{n}$. Thus, under \Cref{ass:bound_psiD}, the $\mathrm{ICM}(U \mid Z)$ part does not contribute to the local power since $\displaystyle \lim_{n\rightarrow \infty } \sqrt{n}\mathrm{ICM}(U \mid Z) = 0 $ under $ \mathbb{H}_{an} $. Thus, as $n\rightarrow\infty$ under $\mathbb{H}_{an}$,
\begin{align*}
    \sqrt{n}\delta_h &= \sqrt{n}\begin{bmatrix}
        \delta_h^{(1)}\\
        \mathrm{ICM}(U \mid Z) - \delta_h^{(1)}
    \end{bmatrix}
    =\begin{bmatrix}
        1 & 0 \\ -1 & 1
    \end{bmatrix}
    \begin{bmatrix}
        \sqrt{n}\delta_h^{(1)} \\ \sqrt{n}\mathrm{ICM}(U \mid Z)
    \end{bmatrix}
    =\begin{bmatrix}
        1 & 0 \\ -1 & 1
    \end{bmatrix}
    \begin{bmatrix}
        \E\big[ K(Z,Z^\dagger)h(Z^\dagger)a(Z) \big] \\ \E\big[a(Z)a(Z^{\dagger})K(Z,Z^{\dagger})\big]/\sqrt{n}
    \end{bmatrix}\\
    &\rightarrow \E\big[ K(Z,Z^\dagger)h(Z^\dagger)a(Z) \big] \begin{bmatrix}
        1 \\ -1
    \end{bmatrix} := a_o.
\end{align*}

Furthermore, by the LIE and \Cref{ass:theta_asymp_lin}
\begin{align*}
    \E[\xi_{h,1}(D_i)] &= \E\big[\psi_h^{(1)}(D_i) - \delta_h^{(1)}\big] - H_1^\top\E[[\xi_{\theta,i}^\top, \widetilde{h}(Z_i) ]^\top] = 0 \text{ and} \\
    \E[\xi_{h,2}(D_i)] &= \E\big[\psi_U^{(1)}(D_i)-\mathrm{ICM}(U \mid Z)\big] - H_2^\top \E[\xi_{\theta,i}] = 0,
\end{align*} i.e., $\E[\xi_h(D_i)] = 2\E\big[\xi_{h,1}(D_i), \ \xi_{h,2}(D_i) - \xi_{h,1}(D_i)\big]^\top = 0 $.

From the foregoing, we obtain that 
$$
 \sqrt{n}\widehat{\delta}_h = \sqrt{n}\delta_h + \frac{1}{\sqrt{n}}\sum_{i=1}^{n}\xi_h(D_i) + o_p(1) =  a_o + \frac{1}{\sqrt{n}}\sum_{i=1}^{n}\xi_h(D_i) + o_p(1).
$$
Under $\mathbb{H}_{an}$, \Cref{ass:Sampling_iid}, and \Cref{ass:bound_psiD}, asymptotic normality follows. Since $ \E[\widetilde{U}|Z]= n^{-1/2}a(Z) $ under $\mathbb{H}_{an}$ it follows from \eqref{eqn:del_LIE_comp} that
$$
\operatorname{Var}(\xi_h(D_i)) - \Omega_{h,o}=O(n^{-1/2}),
$$
and the continuous mapping theorem applies. 

(iii) Under the local alternative $\mathbb{H}_{an}': 
\E[\widetilde{U}|Z]=n^{-1/4}a(Z)$, we have $\sqrt{n}\mathrm{ICM}(U \mid Z) = \E\big[a(Z)a(Z^{\dagger})K(Z,Z^{\dagger})\big] > 0$ and $\sqrt{n}\delta_h^{(1)} = n^{1/4}\E\big[\widetilde{h}(Z)a(Z^{\dagger})K(Z,Z^{\dagger})\big]=0$ if $\E\big[a(Z^{\dagger})K(Z,Z^{\dagger})\mid Z\big]$ is orthogonal to $h(Z)$. By arguments akin to part (ii) above for $ \E[\widetilde{U}|Z]= n^{-1/4}a(Z) $ under $\mathbb{H}_{an}'$, it follows from \eqref{eqn:del_LIE_comp} that
$$
\operatorname{Var}(\xi_h(D_i)) - \Omega_{h,o}=O(n^{-1/4}).
$$ Invoking the continuous mapping theorem completes this part of the proof.

(iv) $\delta_h\neq0$ under $\mathbb{H}_a$. Thanks to the asymptotically linear expression in \Cref{lem:del_asymp_lin}, the result follows from \Cref{ass:Sampling_iid,ass:bound_psiD,ass:theta_asymp_lin,ass:diff_Utheta}. 

\qed

\subsection{Proof of \Cref{thm_wald}}
By Theorem 2, equation (3.3) in \citet{maesono1998asymptotic}, we know that $$
\widetilde{\Omega}_{h,n}-\Omega_h=O_p(n^{-1/2}), 
$$
for both $\mathbb{H}_o$ and $\mathbb{H}_a$.  Furthermore, under $\mathbb{H}_{an}$, we know that $\Omega_h-\Omega_{h,o}=O(n^{-1/2})$.
Hence, for any small $\iota\in(0,1/2)$,  $n^{1/2-\iota}(\widetilde{\Omega}_{h,n}-{\Omega}_{h,o})=o_p(1)$, under $\mathbb{H}_o$ and $\mathbb{H}_{an}$.  
This implies that Assumption 2.2 in \citet{dufour2016rank} holds by setting $b_n=n^{1/2}$ and $c_n=Cn^{-1/2+\iota}$ for some constant $C>0$. Furthermore,  by Proposition 9.1 in \citet{dufour2016rank}, we know that $\widehat{\Omega}_{h,n}^{-}\overset{p}{\to}\Omega_{h,o}^-$ under $\mathbb{H}_o$ and $\mathbb{H}_{an}$.   Similarly, $\widehat{\Omega}_{h,n}^{-}\overset{p}{\to}\Omega_{h,a}^-$ under $\mathbb{H}_a$. 

(i). By Corollary 9.3 in \citet{dufour2016rank}, $T_{h,n}\overset{d}{\to} \chi^2_1$.

(ii). By the continuous mapping theorem and \Cref{thm_all} (ii), 
$$
T_{h,n}\overset{d}{\to} \chi^2_1(b_o).
$$

(iii) By the continuous mapping theorem and \Cref{thm_all} (iii), 
$$
T_{h,n}\overset{d}{\to} \chi^2_1(b_o').
$$

(iv). When $\delta_h\not\in \mathcal{M}_0$, we have that $\displaystyle
\lim_{n\to\infty} T_{h,n}=\lim_{n\to\infty} n \widehat{\delta}_h^{\top} \Omega_{h,a}^{-}\widehat{\delta}_h = \infty,
$ in probability, and the result follows. 

\qed

\subsection{Proof of the Result in \Cref{rem_m0}}
\begin{lemma}\label{lem_del_null_space}
$\delta_h\not\in\mathcal{M}_0$ for $ V_h $ in \eqref{eqn:V_h}.
\end{lemma}
\begin{proof}

Under $\mathbb{H}_a$, $a(Z)=\E[U|Z]-\E[U]$ is non-degenerate with $\E a(Z)=0.$ Recall $\widetilde{h}(Z): = h(Z)-\E[h(Z)]$, and define
\begin{flalign*}
    &m_{\widetilde{V}}(Z) := \left[\mathbb{E}\{K(Z,Z^{\dagger})\widetilde{h}(Z^{\dagger})|Z\},\ \mathbb{E}\{K(Z,Z^{\dagger})[a(Z^{\dagger})-\widetilde{h}(Z^{\dagger})]|Z\}\right]^{\top} \text{ and}  \\
    &m_{\widetilde{U}}(Z) := \mathbb{E}\big[K(Z,Z^{\dagger})a(Z^{\dagger})|Z\big]. 
\end{flalign*}

First, when $\Omega_a$ is of rank 2, the claim holds because $\delta_h\neq \bm{0}^{\top}$ under $\mathbb{H}_a$.  

Second,  recall that $\Omega_a = \mathrm{Var}\{\xi_h(D)\}$ using the expression $ \xi_h(D) =[m_{\widetilde{U}}(Z)-\E m_{\widetilde{U}}(Z)](V_h - \E V_h)+ [m_{\widetilde{V}}(Z)-\E m_{\widetilde{V}}(Z)](U-\E U)\text{ and } \Omega_h = \mathrm{Var}\left\{\xi_h(D)\right\}.$ By the Cauchy-Schwarz inequality, we know that 
$\Omega_a$ is of rank one if and only if for some constant $C\neq -1$, $$\xi_h(D)^{\top}[-C,1]=0\ a.s. $$
Here we rule out $C=-1$ as it indicates that $\mathrm{ICM}(a(Z)|Z)=0$, a contradiction of $\mathbb{H}_a$.

Let $f(Z^{\dagger}):={a}(Z^{\dagger})-(C+1)\widetilde{h}(Z^{\dagger})$, this implies that 
\begin{equation}\label{contradiction}
   [m_{\widetilde{U}}(Z)-\E m_{\widetilde{U}}(Z)][\widetilde{U}-(C+1)\widetilde{h}(Z)] + [m_{\widetilde{f}}(Z)-\E m_{\widetilde{f}}(Z)]\widetilde{U}=0 \ a.s. 
\end{equation}
where $m_{\widetilde{f}}(Z)=\E\{K(Z,Z^{\dagger})f(Z^{\dagger})|Z\}$.

Recall that $\E [\widetilde{U}|Z]=a(Z)$, by taking conditional expectation of \eqref{contradiction} w.r.t. $Z$, we have    
\begin{equation}\label{restrict_a}
        [m_{\widetilde{U}}(Z)-\E m_{\widetilde{U}}(Z)]f(Z)+[m_{\widetilde{f}}(Z)-\E m_{\widetilde{f}}(Z)]a(Z)=0\ a.s.
    \end{equation}
Therefore, by  taking the difference, we have 
    $$\left\{m_{\widetilde{U}}(Z)+m_{\widetilde{f}}(Z)-\E m_{\widetilde{U}}(Z)+\E m_{\widetilde{f}}(Z)\right\}[\widetilde{U}-a(Z)]=0\ a.s.$$
In light of the foregoing, either one of the following holds:
\begin{enumerate}[(1)]
    \item $\widetilde{U}=a(Z)\ a.s.$
    \item  $\widetilde{U}\neq a(Z)$, so that the coefficient on $\widetilde{U}$ in \eqref{contradiction} is zero, i.e. 
    $$\left\{m_{\widetilde{U}}(Z)+m_{\widetilde{f}}(Z)-\E m_{\widetilde{U}}(Z)+\E m_{\widetilde{f}}(Z)\right\}=0\ a.s.$$ which further implies that $[m_{\widetilde{U}}(Z)-\E m_{\widetilde{U}}(Z)]\widetilde{h}(Z)=0\ a.s.$ in combination of \eqref{contradiction}.
\end{enumerate}
\noindent We proceed using proof by contradiction.

If (1) holds, we have  $$\Omega_a
=\mathrm{Var}\left\{[m_{\widetilde{U}}(Z)-\E m_{\widetilde{U}}(Z)]\widetilde{h}(Z)+[m_{\widetilde{h}}(Z)-\E m_{\widetilde{h}}(Z)]a(Z)|D]\right\}\begin{bmatrix}
				1 & C\\ 
				C & C^2
			\end{bmatrix}
$$
which implies  $\mathcal{M}_0=\{(x,y)^{\top}|x=-Cy, y\in\mathbb{R}\}$. Note in this case, by taking the expectation of \eqref{restrict_a}, we can show that $\E\{K(Z,Z^{\dagger})a(Z^{\dagger})[a(Z)-\widetilde{h}(Z)]\}=C\E[K(Z,Z^{\dagger})a(Z^{\dagger})\widetilde{h}(Z)]$, so  that $\delta_h=\E[K(Z,Z^{\dagger})a(Z^{\dagger})\widetilde{h}(Z)][1,C]^{\top} \not \in \mathcal{M}_0 $.

If (2) holds, then we have $m_{\widetilde{U}}(Z)-\E m_{\widetilde{U}}(Z)=0 \ a.s.$ since $\widetilde{h}(Z)$ is non-degenerate. This implies that $\E\{[m_{\widetilde{U}}(Z)-\E m_{\widetilde{U}}(Z)]a(Z)\}=\mathrm{ICM}(U|Z)=0$, a contradiction. 

\end{proof}

\subsection{Proof of \Cref{thm_bahadur}}
For two functions $f_a(x)$ and $f_b(x)$, write $f_a(x)\sim f_b(x)$ if and only if $f_a(x)/f_b(x)\to 1$ as $x\to\infty$. By \citet{zolotarev1961concerning},  we know that 
 $$
 \log \mathbb{P}\Big(\sum_{k=1}^{\infty}\lambda_k G_k^2>x \Big) \sim  -x/(2\lambda_1), \quad \mbox{as } x\to\infty.
 $$
Clearly, under $\mathbb{H}_a,$ we have $n\mathrm{ICM}_n(U \mid Z)\to\infty$ in probability. Thus,
\begin{flalign*}
    c_G=\plim_{n\to\infty}\frac{n\mathrm{ICM}_n(U \mid Z)}{n\lambda_1}= \frac{\mathrm{ICM}(U \mid Z)}{\lambda_1},
\end{flalign*}
where we note that $\mathrm{ICM}_n(U \mid Z)\to_{a.s.}\mathrm{ICM}(U \mid Z)$ under \Cref{ass:bound_psiD}.

Next, by the large deviation result for the chi-squared distribution,  
$$
\log\mathbb{P}\big(\chi^2_1 > x\big) \sim -x/2\quad \mbox{as } x\to\infty.
$$
Also, $\displaystyle \plim_{n\to\infty} T_{h,n} = \plim_{n\to\infty} n\widehat{\delta}_h^{\top}\widehat{\Omega}_{h,n}^{-}\widehat{\delta}_h  = \lim_{n\to\infty} n\delta_a^{\top}\Omega_{h,a}^{-}\delta_a = \infty$ under $\mathbb{H}_a$, thus 
\begin{flalign*}
    c_T = \lim_{n\to\infty} \frac{2}{n} \frac{n\delta_a^{\top}\Omega_{h,a}^{-}\delta_a}{2} = \delta_a^{\top}\Omega_{h,a}^{-}\delta_a.
\end{flalign*}

\qed

\section{Asymptotic properties of $\widehat{\delta}_h^*$.} \label{sec:deltastar}

A natural empirical estimator for $\delta_h^*$ is given by 
$
\widehat{\delta}_h^* = \widehat{\delta}_h + \widehat{\eta}_h^*
$ where
\[
    \widehat{\eta}_h^*:= \frac{1}{n(n-1)}\sum_{i=1}^n\sum_{j\neq i}|K(Z_i,Z_j)|\E_n[U]\big[\E_n[h(Z)], \ \E_n[U] - \E_n[h(Z)] \big]^\top.
\] Accordingly, define $ \eta_h^*:= \E\big[|K(Z,Z^{\dagger})|\big]\E U \E\big[h(Z), \ U - h(Z) \big]^\top $. We consider the simpler case of mean independence with nullity -- specification with nullity testing follows straightforwardly. For the asymptotic analysis of $\widehat{\delta}_h^*$, we consider the following sequence of local alternatives:
\[
\mathbb{H}_{an}^*: \E[U \mid Z] = n^{-1/2} a^*(Z),
\]
where $a^*(Z)$ is a nonzero, possibly degenerate function of $Z$.

Thanks to the Hoeffding decomposition
\begin{align*}
    \sqrt{n}(\widehat{\eta}_h^* - \eta_h^*) 
    &= \frac{1}{\sqrt{n}(n-1)} \sum_{i\neq j} 
    \begin{bmatrix}
     \E[U]\E[h(Z)] \\ (\E[U])^2 - \E[U]\E[h(Z)]
    \end{bmatrix} \big(|K(Z_i, Z_j)| - \E\big[|K(Z, Z^\dagger)|\big]\big)\\
    &+ \frac{1}{\sqrt{n}} \sum_{i=1}^n \begin{bmatrix}
     \E[h(Z)] \\ \E[U] - \E[h(Z)]
    \end{bmatrix}\E\big[|K(Z,Z^\dagger)|\big]\big( U_i - \E[U] \big)\\
    &+ \frac{1}{\sqrt{n}} \sum_{i=1}^n \E\big[|K(Z,Z^\dagger)|\big]\E[U]\begin{bmatrix}
        h(Z_i) - \E[h(Z_i)] \\ \big( U_i - \E[U] \big) - \big(h(Z_i) - \E[h(Z)]\big)
    \end{bmatrix}\\
    &= \frac{1}{\sqrt{n}} \sum_{i=1}^n\Bigg\{ 
    \begin{bmatrix}
     \E[U]\E[h(Z)] \\ (\E[U])^2 - \E[U]\E[h(Z)]
    \end{bmatrix} 2\big( \E\big[(|K(Z_i, Z_j)|)\mid Z_i\big] - \E\big[|K(Z, Z^\dagger)|\big]\big)\\
    &\quad \quad \quad \quad + \begin{bmatrix}
     \E[h(Z)] \\ \E[U] - \E[h(Z)]
    \end{bmatrix}\E\big[|K(Z, Z^\dagger)|\big]\big( U_i - \E[U] \big)\\
    &\quad \quad \quad \quad + \E\big[|K(Z,Z^\dagger)|\big]\E[U]\begin{bmatrix}
        h(Z_i) - \E[h(Z)] \\ \big( U_i - \E[U] \big) - \big(h(Z_i) - \E[h(Z)]\big)
    \end{bmatrix} \Bigg\} + o_p(1)\\
    =&: \frac{1}{\sqrt{n}} \sum_{i=1}^n \phi_h^*(D_i) + o_p(1).
\end{align*}

In addition to \Cref{lem:del_asymp_lin}, the following is the asymptotically linear representation of $\widehat{\delta}_h^*$:

\begin{align*}
    \sqrt{n}(\widehat{\delta}_h^* - \delta_h^*) = \sqrt{n}(\widehat{\delta}_h - \delta_h) + \sqrt{n}(\widehat{\eta}_h^* - \eta_h^*) =  \frac{1}{\sqrt{n}} \sum_{i=1}^n \big\{\xi_h(D_i) + \phi_h^*(D_i) \big\} + o_p(1).
\end{align*}

As $n\rightarrow \infty$ under $ \mathbb{H}_{an}^* $, 
  \[
  \sqrt{n}\delta_h^* = \sqrt{n}\delta_h + \sqrt{n}\eta_h^* \rightarrow \begin{bmatrix} 1 \\ -1 \end{bmatrix} \Big(\E\big[h(Z)\big(a^*(Z^{\dagger}) - \E[a^*(Z^{\dagger})] \big)K(Z,Z^{\dagger})\big] + \E\big[|K(Z,Z^\dagger)|\big]\E[a^*(Z)]\E[h(Z)] \Big):= a_o^*
  \]
   where $\displaystyle \lim_{n\rightarrow \infty}\sqrt{n}\eta_h^* = \E\big[|K(Z,Z^\dagger)|\big]\E[a^*(Z)]\E[h(Z)]\big[ 1, -1 \big]^\top $.

\begin{theorem}
\label{thm:star}
   Let the conditions of \Cref{thm_all} hold, then
    \begin{enumerate}[(i)]
    \item under $\mathbb{H}_o^*$,
 $
 \sqrt{n}\widehat{\delta}_h^*\overset{d}{\rightarrow}\mathcal{N}(0,\Omega_{h,o}^*) $;
\item under $\mathbb{H}_{an}^*$,  
 $
 \sqrt{n}\widehat{\delta}_h^*\overset{d}{\rightarrow}\mathcal{N}(a_o^*,\Omega_{h,o}^*); \text{ and}
 $
      \item under $\mathbb{H}_{a}^*$;
     $
 \sqrt{n}(\widehat{\delta}_h^*-\delta_h^*)\overset{d}{\rightarrow}\mathcal{N}(0,\Omega_{h,a}^*),$
\end{enumerate}
where $\Omega_{h,o}^*$ and $\Omega_{h,a}^*$ correspond to specific expressions of $\Omega_h^*=\operatorname{Var}\big[\xi_h(D) + \phi_h^*(D)\big]$ under $\mathbb{H}_o^*$  and $\mathbb{H}_a^*$, respectively.
\end{theorem}

\begin{proof}[\textbf{Proof}]
The proof follows that of \Cref{thm_all} and the discussion preceding the theorem. The details are therefore omitted.

\end{proof}

\section{Numerical Illustration and Computation of the Bahadur slopes} \label{Sect_Appendix:Bahadur}

This section presents additional details concerning the Bahadur slopes of the $\chi^2$-test and the bootstrap-based ICM specification tests. The first subsection offers concrete examples that illustrate \Cref{thm_bahadur}, while the second subsection describes the numerical integration procedure used in the first subsection.

\subsection{Numerical Illustration}

To make the \Cref{thm_bahadur} results concrete, we consider the simple design 
$$
U = \exp(-Z^2/3) - \sqrt{3/5} + \mathcal{E}, \quad \text{with  } Z \sim\mathcal{N}(0,1)\quad \text{and  } \mathcal{E} \sim \mathcal{U}[-\sqrt{3},\sqrt{3}], 
$$ where $\mathcal{E}$ is independent of $Z,$  hence $\E[U \mid Z]:=a(Z)=\exp(-Z^2/3) - \sqrt{3/5}$, and $\E U=0$. To examine the power behavior of the competing tests, we introduce variations in the construction of $V$. Using the Gaussian kernel $K(Z,Z^{\dagger}) = \exp\big(-0.5(Z-Z^{\dagger})^2\big) $ for both tests, \Cref{Tab:Bahadur} shows the Bahadur slopes of the bootstrap-based ICM test alongside the $\chi^2$-test with
\begin{enumerate}[(1)]
    \item $V_1 = [h_1(Z),\ U - h_1(Z)]^{\top}$, $h_1(Z)=\exp(Z)$ which is agnostic of $a(Z)$;
    \item $V_{1a} = [h_1(Z),\ U - h_1(Z),\ Z]^{\top}$;
    \item $V_2 = [h_2(Z),\ U - h_2(Z)]^{\top}$, $h_2(Z)=a(Z)$ which results in a singular covariance matrix under $\mathbb{H}_a$;
    \item $V_3 = [h_3(Z),\ U - h_3(Z)]^{\top}$, $h_3(Z) = \sqrt{3}\exp(-Z^2/2) $ which satisfies $\E[K(Z,Z^{\dagger})h(Z^{\dagger}) \mid Z] =\exp(-Z^2/3) - \sqrt{3/5} := a(Z) $; and
    \item $V_{3a} = [h_3(Z),\ U - h_3(Z),\ Z]^{\top}$.
\end{enumerate}
In scenarios (3) - (5), we use prior knowledge of the alternative. In (3), it can be shown that $a_o = [\mathrm{ICM}(U \mid Z),0]^{\top}$, and 
$$
\Omega_{h,a}=  A\times \begin{bmatrix}
    1 & -1\\ -1 & 1
\end{bmatrix},
$$
where $A = \operatorname{Var}\left\{ \widetilde{\xi}_{h,2}(D)\right\}$ and $\widetilde{\xi}_{h,2}(D) = (U-\E U)\E[a(Z^\dagger)K(Z,Z^\dagger) \mid Z]$.  Clearly, $\E[\widetilde{\xi}_{h,2}(D)] = \mathrm{ICM}(U \mid Z)$, therefore $a_o^{\top}\Omega_{h,a}^{-}a_o= {\mathrm{ICM}^2(U \mid Z)}/[4\operatorname{Var}(\xi_{h,2}(D))].$ This is in fact a worst-case scenario under $\mathbb{H}_a$ for the $\chi^2$-test.  To fully make use of the information of $a(Z)$, the choice of 
$h(Z)$ in scenario (4) maximizes the linear dependence with $ \E\big[K(Z,Z^\dagger)(U^\dagger- \E U) \mid Z \big] $ in the first element of $\delta_h$ while the second term is not degenerate: $\delta_h = \E\big[a^2(Z), \ \mathrm{ICM}(U \mid Z) - a^2(Z) \big]^{\top}$. Therefore, power should be augmented. Scenarios (2) and (5) are $V$s of scenarios (1) and (4) augmented with $Z$.
\begin{table}[!htp]
\centering
\caption{Bahadur Slopes}
\begin{tabular}{@{}ccccccc@{}}
\toprule
              & Boot & \multicolumn{5}{c}{$\chi^2$} \\ \midrule
        &      &     $ V_1 $  & $V_{1a}$   &  $V_2$ &  $V_3$  & $V_{3a}$     \\ \cmidrule(lr){3-3}\cmidrule(lr){4-4}\cmidrule(lr){5-5}\cmidrule(lr){6-6}\cmidrule(lr){7-7}
Bahadur Slope &   0.0109 & 0.0214 & 0.0242 & 0.0056 & 0.0246 & 0.0246 \\ \bottomrule
\end{tabular}
\label{Tab:Bahadur}
\end{table}

\Cref{Tab:Bahadur} shows that the bootstrap-based test is an intermediate case between the agnostic choice with $h(Z)=\exp(Z)$ (scenario (1)) and the worst case with $h(Z)=a(Z)$ (scenario (3)).  In the case where the linear dependence is maximized with respect to either element of $V_h = [h(Z), \ U - h(Z)]^{\top}$ without degeneracy in the other  (scenario (4)), the $\chi^2$-test Bahadur slope is larger. Moreover, we note that when augmenting $V$, one observes a slight increase in the Bahadur slope for scenario (2) but none for scenario (5). The latter case is not surprising as $Z$ is orthogonal to the direction under alternative, namely, $\E[K(Z,Z^{\dagger})h(Z^{\dagger}) \mid Z] = a(Z) = \exp(-Z^2/3) - \sqrt{3/5} $ is orthogonal to $Z$ thus the augmentation with $Z$ does not augment power.

\subsection{Numerical Integration}
Here, we briefly provide details pertaining to the Monte Carlo numerical integration used to obtain the Bahadur slopes in the preceding sub-section. The approach proceeds by computing $\E[a(Z)a(Z^\dagger)K(Z,Z^\dagger)]$, $\lambda_1$, $a_o$, and $\Omega_{h,a}$ on a sample of $1\ 000$ random draws following the DGP in \Cref{SubSect:Bahadur}. $\lambda_1=0.6175762$ is computed using steps 1-3 of \citet[Algorithm 1]{seri-2022-computing} with $(i,j)$'th element $\exp(-0.5(Z_i-Z_j)^2)\mathcal{E}_i\mathcal{E}_j/(n-1)$. The numerical values of the Bahadur slopes in \Cref{Tab:Bahadur} are then obtained using averages of the quantities $\E[a(Z)a(Z^\dagger)K(Z,Z^\dagger)]$, $\lambda_1$, $a_o$, and $\Omega_{h,a}$ over $10\ 000$ replications.

\section{Monte Carlo Experiments - Mean Independence Test}\label{Sect_Appendix:MC}
This section examines the empirical size control and power performance of the test of mean independence via simulations. Section \ref{subsec:sim_mean} presents five DGPs; Section \ref{subsec:sizepowMI} examines the size control and power performance of the tests of mean independence; Section \ref{subsec:pv_more} examines the performance of the $\chi^2$-test with $V$ augmented to dimensions $p_v>2$; Section \ref{subsec:robustness_cn} conducts sensitivity analyses of the size and power performance of the test to variations of the tuning rule $c_n = \widetilde{\lambda}_1n^{-1/3}$ and other selection criteria used in the literature; and Section \ref{subsec:nullity} examines the test of nullity $\E[U \mid Z]=\E[U]=0 \ a.s. $ via simulations.

\subsection{Specifications}\label{subsec:sim_mean}
Five different DGPs with conditional heteroskedasticity are considered for the test of mean independence. The first four DGPs follow directly from LS1 through LS4 in the main text with $\theta_l=0, \, l\geq 1$. The fifth DGP examines alternatives that are binary and non-monotone in $Z$.

\begin{enumerate}[label=(\roman*)]
	\item[MI 1:] $\displaystyle U = 1 + \frac{\mathcal{E}}{\sqrt{1+Z_1^2}} $;
			
\item[MI 2:] $ \displaystyle U = 1 + \frac{ \gamma }{5\sqrt{2}} \sum_{l=1}^5 Z_l^2 + \frac{\mathcal{E}}{\sqrt{1+Z_1^2}} $; 
						
\item[MI 3:] $ \displaystyle U =  1 + \gamma \sum_{l=1}^5 \frac{\cos(2Z_l)}{\sqrt{2(1-\exp(-8))}} + \frac{\mathcal{E}}{\sqrt{1+Z_1^2}} $; 

\item[MI 4:] $ \displaystyle U =  1 + \gamma\sum_{l=1}^5 \big(\exp(-Z_l^2/3) - \sqrt{3/5}\big)  + \frac{\mathcal{E}}{\sqrt{1+Z_1^2}} $; and

\item[MI 5:] $ \displaystyle U = 1 + \frac{\gamma}{2}\sum_{l=1}^5\mathrm{I}\big( \mid Z_l|< -\Phi^{-1}(1/4)\big) + \frac{\mathcal{E}}{\sqrt{1+Z_1^2}} $.
\end{enumerate}

\subsection{Empirical Size and Power}\label{subsec:sizepowMI}

\begin{table}[!htbp]
\centering
\caption{Empirical Size \& Local Power}
\begin{tabular}{lccccccc}
\toprule
& & \multicolumn{2}{c}{$ 10\% $} & \multicolumn{2}{c}{$ 5\% $} & \multicolumn{2}{c}{$ 1\% $}  \\
\cmidrule(lr){3-4} \cmidrule(lr){5-6}\cmidrule(lr){7-8}
$n$ & Kernel & $\chi^2$ & MB & $\chi^2$ & MB & $\chi^2$ & MB \\ \midrule

\multicolumn{1}{c}{\textcolor{blue}{MI 1}} & & \multicolumn{6}{c}{ Empirical Size }\\ \cmidrule(lr){1-1}  \cmidrule(lr){3-8}

200   &Gauss &0.100   &0.095 &0.043 &0.048 &0.005 &0.008 \\ 
       &Euclid &0.078  &0.093  &0.023  &0.044  &0.004  &0.007  \\ \cmidrule[0.25pt](lr){2-8}
400   &Gauss &0.094 &0.089 &0.040  &0.041 &0.007 &0.007 \\ 
       &Euclid &0.098  &0.086  &0.044  &0.04   &0.006  &0.006  \\ 
       \cmidrule[0.25pt](lr){2-8}
600   &Gauss &0.077 &0.093 &0.042 &0.047 &0.009 &0.012 \\ 
       &Euclid &0.096  &0.098  &0.039  &0.045  &0.006  &0.012  \\ 
       \cmidrule[0.25pt](lr){2-8}
800   &Gauss &0.095 &0.089 &0.047 &0.041 &0.004 &0.005 \\ 
       &Euclid &0.105  &0.094  &0.042  &0.044  &0.007  &0.007  \\ \midrule
   
\multicolumn{1}{c}{\textcolor{blue}{MI 2}} & & \multicolumn{6}{c}{ Local Power: $\gamma = 5/\sqrt{n}$  }\\ \cmidrule(lr){1-1}  \cmidrule(lr){3-8}

200   &Gauss &0.439 &0.421 &0.291 &0.308 &0.095 &0.140  \\ 
       &Euclid &0.598  &0.243  &0.438  &0.113  &0.192  &0.020   \\ 
       \cmidrule[0.25pt](lr){2-8}
400   &Gauss &0.464 &0.435 &0.318 &0.306 &0.125 &0.111 \\ 
       &Euclid &0.632  &0.231  &0.501  &0.118  &0.245  &0.023  \\ 
       \cmidrule[0.25pt](lr){2-8}
600   &Gauss &0.471 &0.427 &0.327 &0.308 &0.12  &0.134 \\ 
       &Euclid &0.637  &0.244  &0.504  &0.136  &0.245  &0.022  \\ 
       \cmidrule[0.25pt](lr){2-8}
800   &Gauss &0.463 &0.438 &0.328 &0.325 &0.142 &0.131 \\ 
       &Euclid &0.628  &0.233  &0.504  &0.130   &0.259  &0.024  \\ 
\bottomrule
\end{tabular}    
\label{Tab:Sim_MI_1}
\end{table}

\Cref{Tab:Sim_MI_1} compares the empirical sizes of the proposed $\chi^2$-test for mean independence with those of the multiplier bootstrap procedure, using DGP MI 1. It also reports their power under local alternatives in DGP MI 2.\footnote{The multiplier and wild bootstrap procedures coincide for the test of mean independence since there is no model (re)-estimation.} As shown in \Cref{Tab:Sim_MI_1}, both tests exhibit accurate size control and non-trivial local power across all nominal levels and ICM kernels.

\begin{figure}[H]
\centering 
\caption{DGP MI 2 -- Gaussian Kernel -- $n=400$.}
\begin{subfigure}{0.32\textwidth}
\centering
\includegraphics[width=1\textwidth]{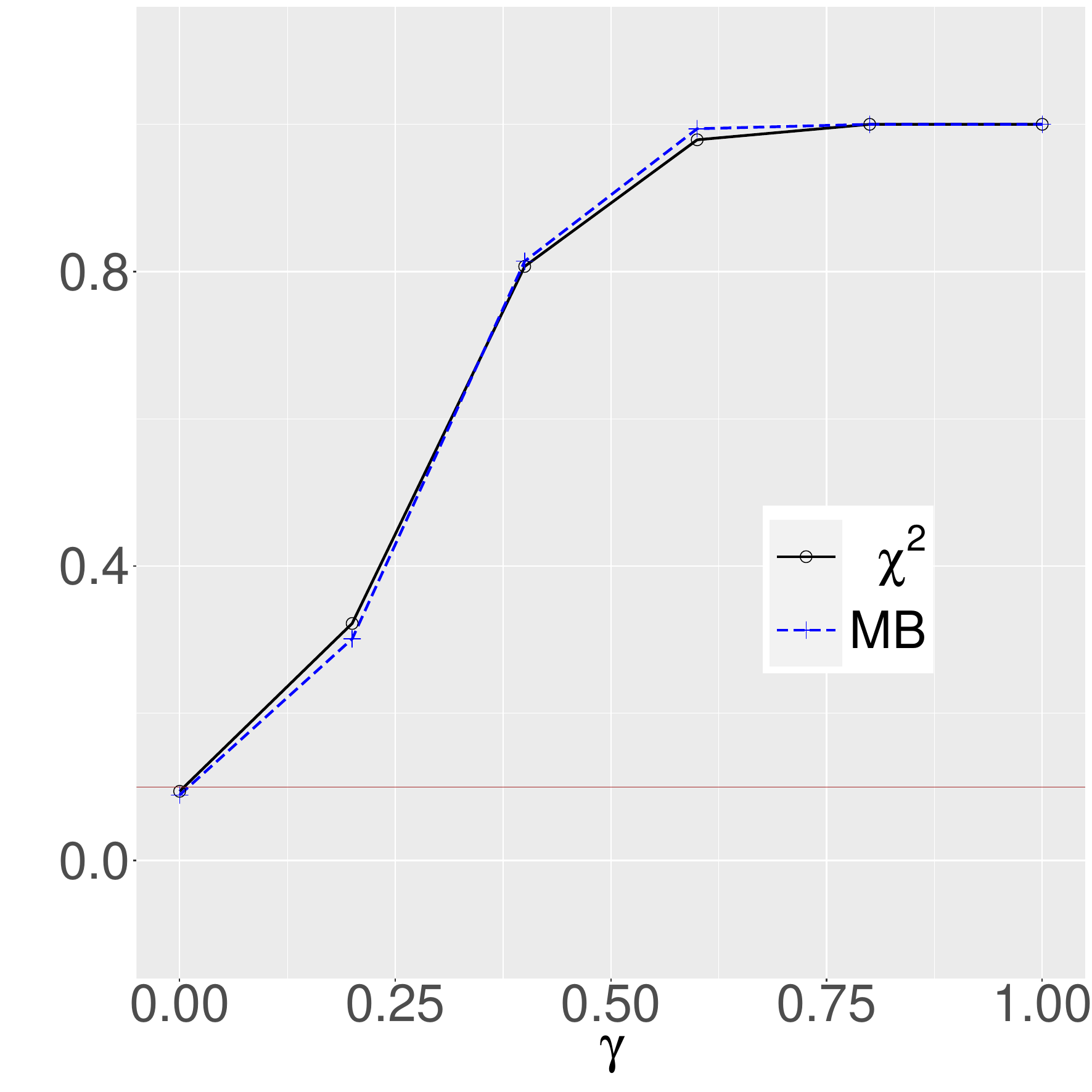}
\caption{10\%}
\end{subfigure}
\begin{subfigure}{0.32\textwidth}
\centering
\includegraphics[width=1\textwidth]{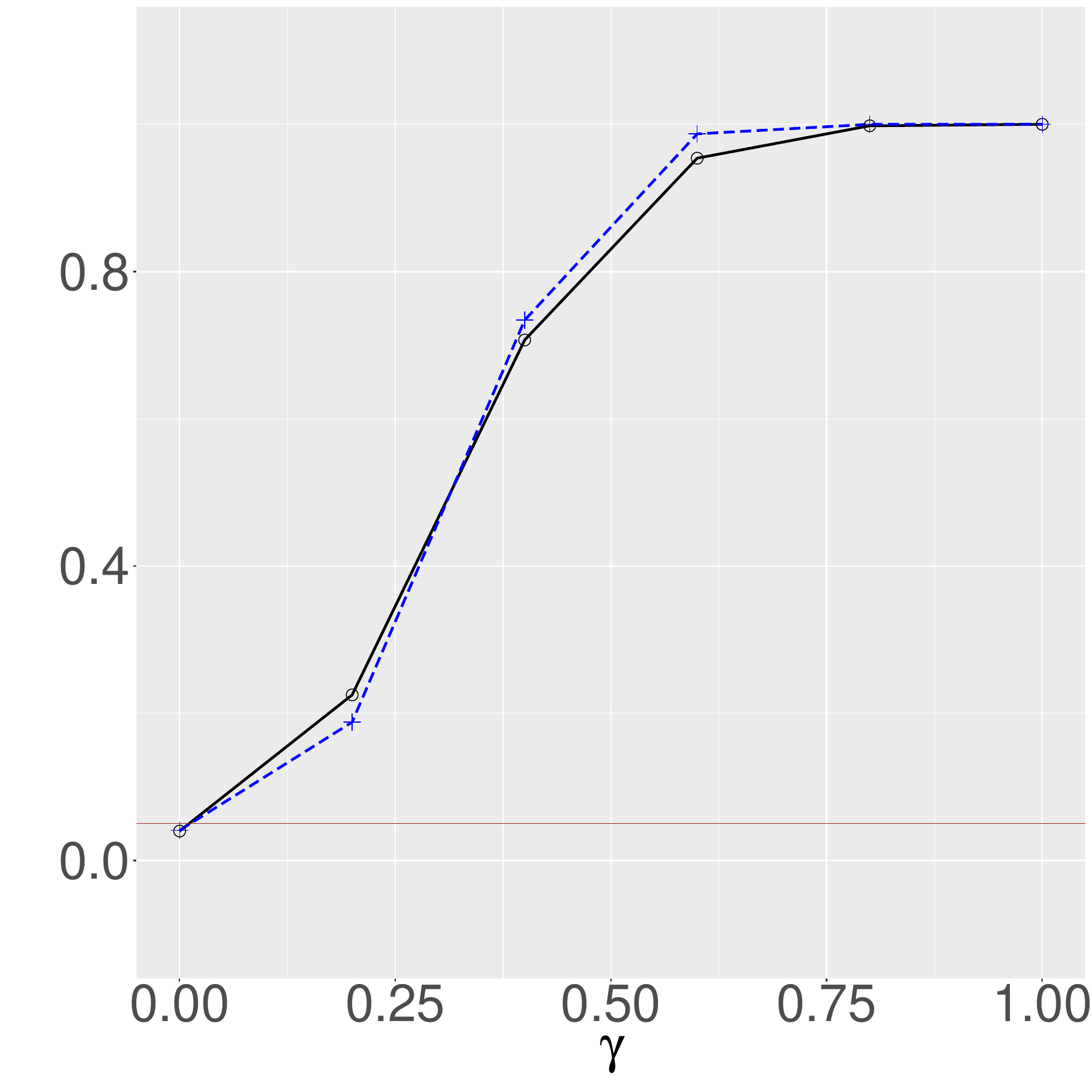}
\caption{5\%}
\end{subfigure}
\begin{subfigure}{0.32\textwidth}
\centering
\includegraphics[width=1\textwidth]{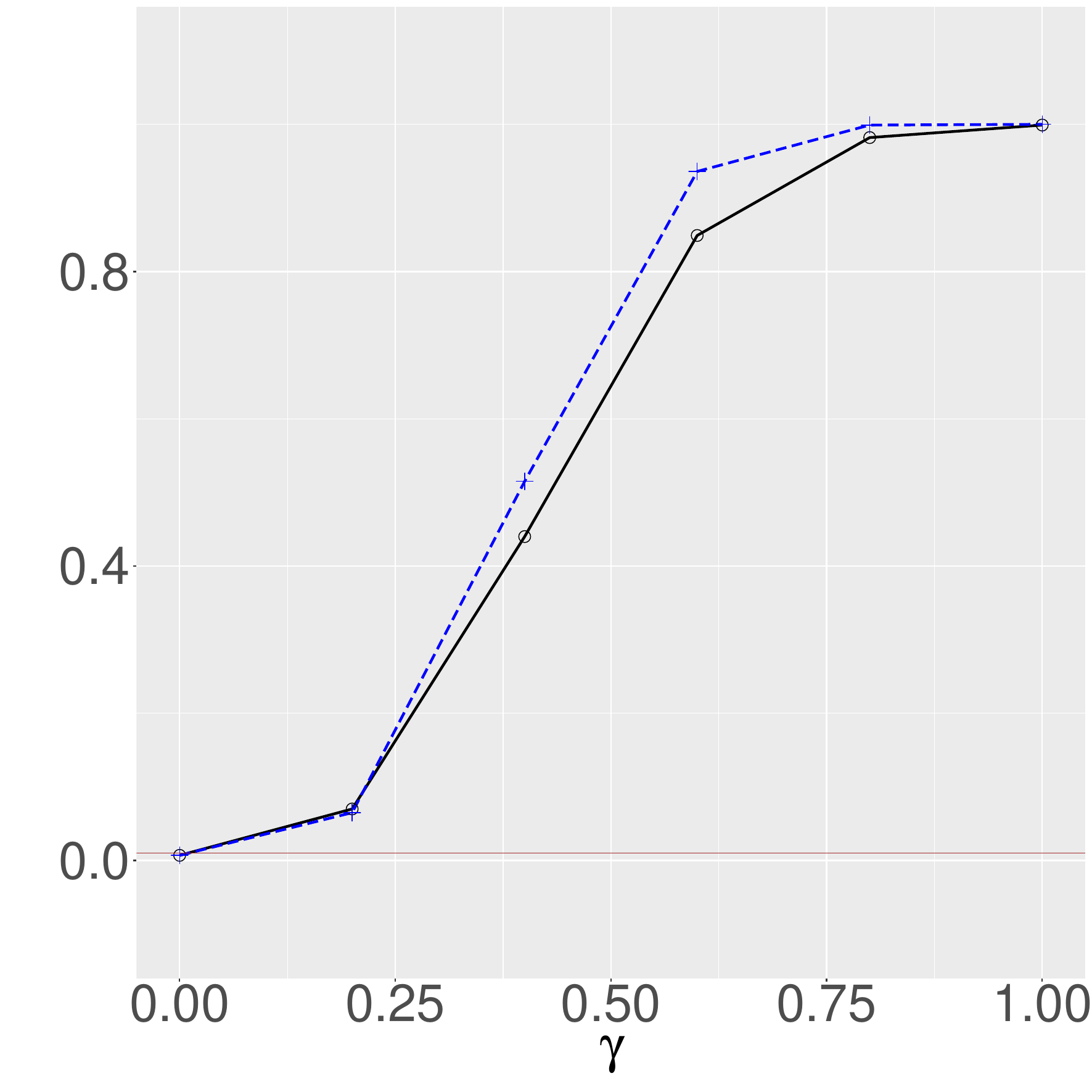}
\caption{1\%}
\end{subfigure}
\label{Fig:MI2_Gauss}
\end{figure}

\begin{figure}[H]
\centering 
\caption{DGP MI 2 -- Negative Euclidean -- $n=400$.}
\begin{subfigure}{0.32\textwidth}
\centering
\includegraphics[width=1\textwidth]{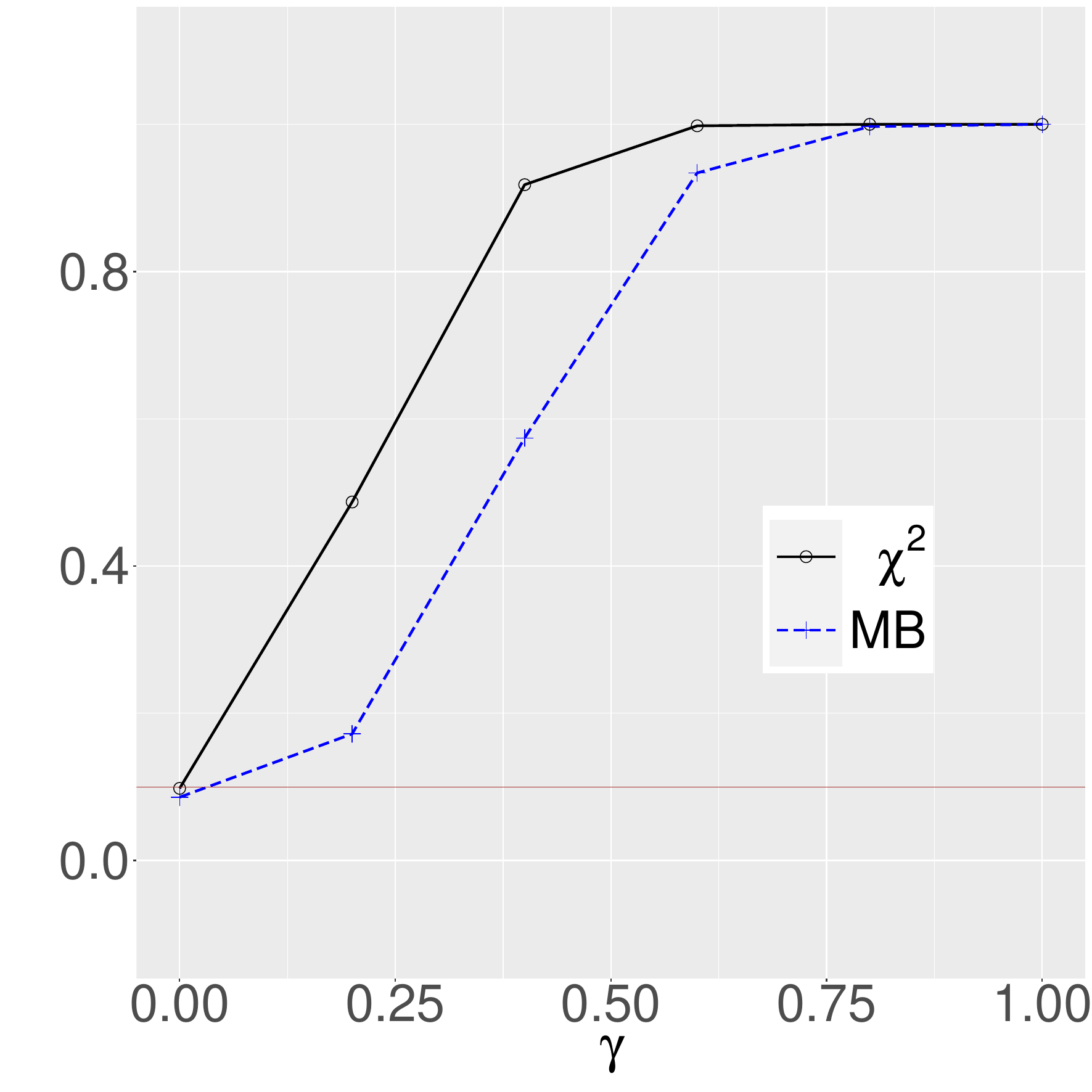}
\caption{10\%}
\end{subfigure}
\begin{subfigure}{0.32\textwidth}
\centering
\includegraphics[width=1\textwidth]{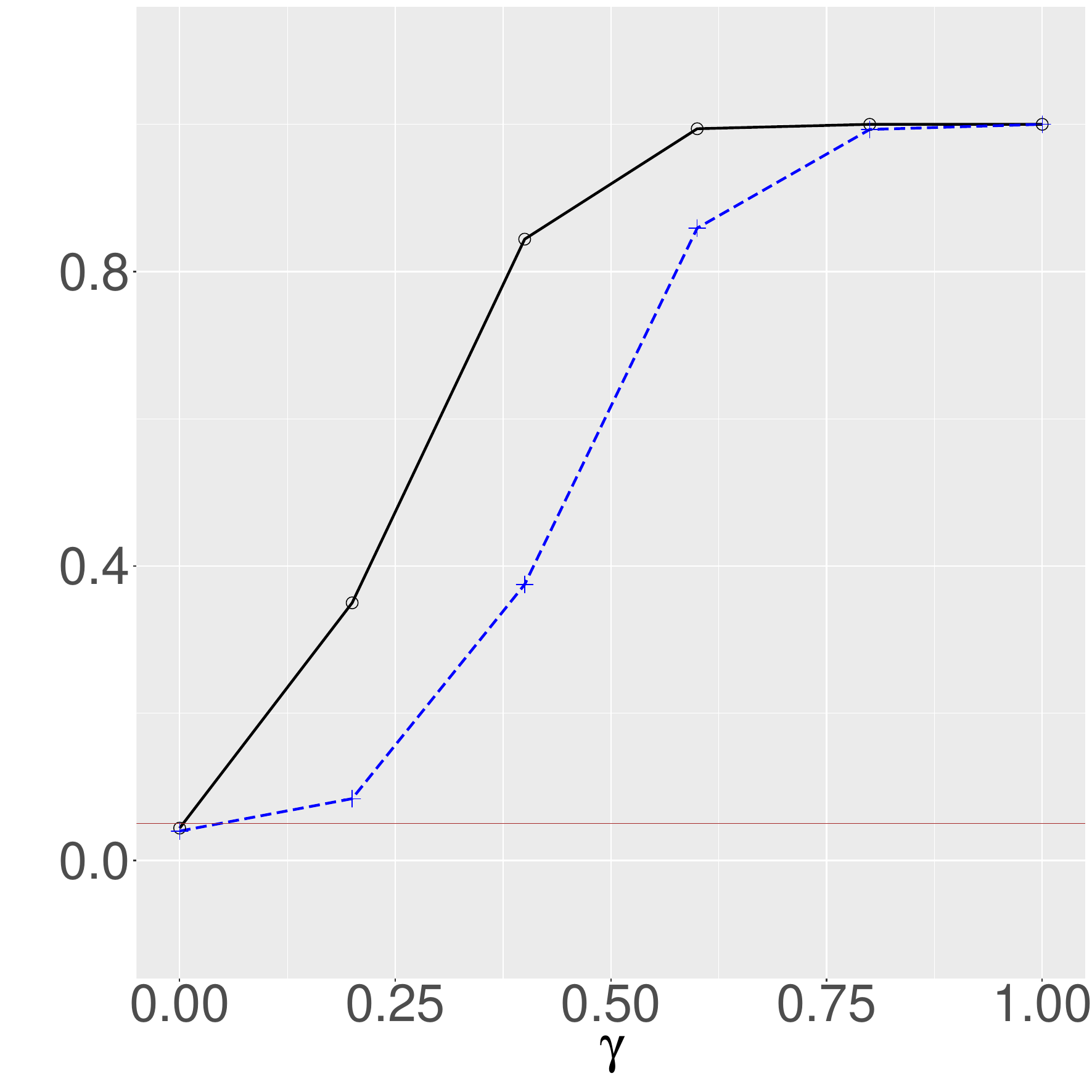}
\caption{5\%}
\end{subfigure}
\begin{subfigure}{0.32\textwidth}
\centering
\includegraphics[width=1\textwidth]{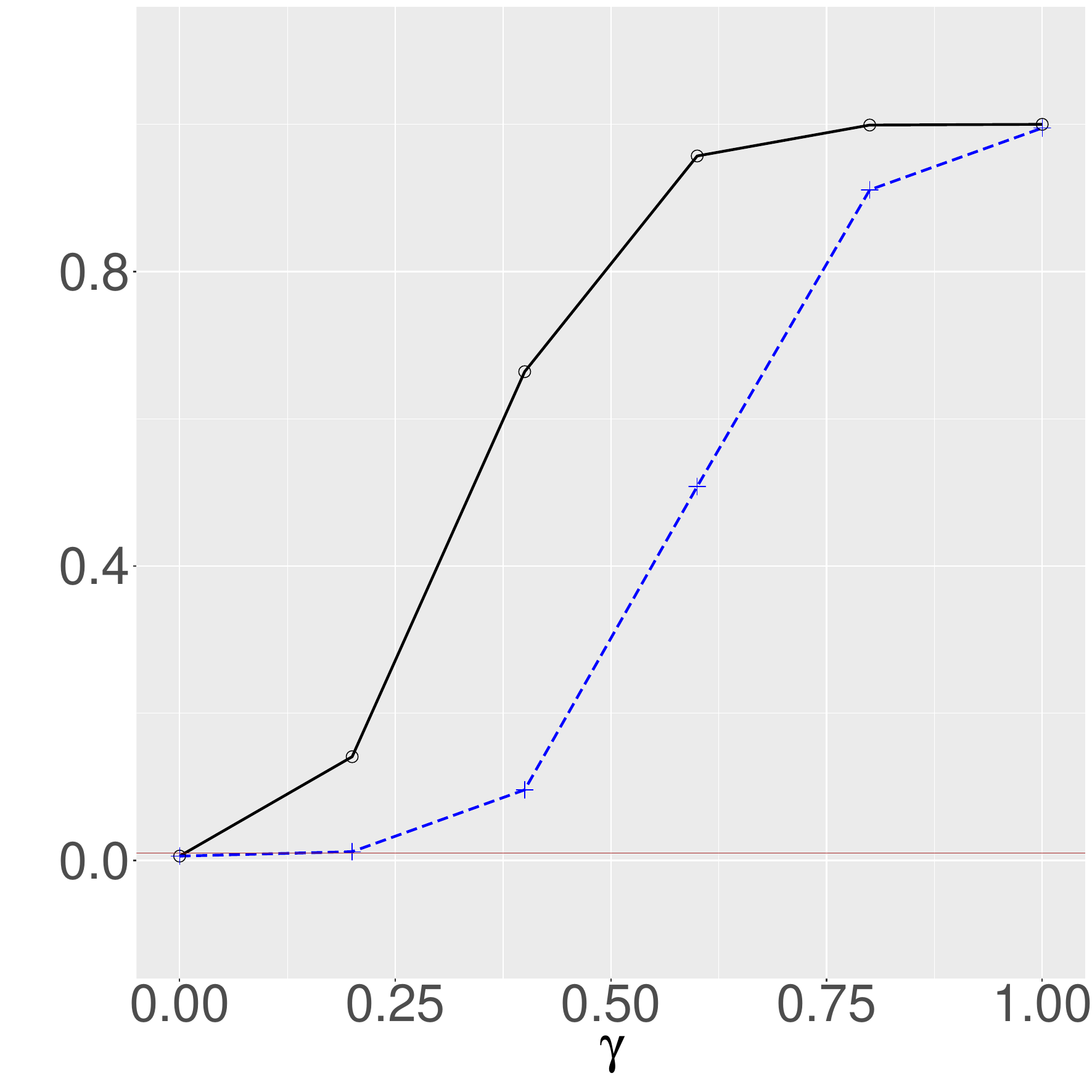}
\caption{1\%}
\end{subfigure}
\label{Fig:MI2_Euclid}
\end{figure}

\begin{figure}[H]
\centering 
\caption{DGP MI 3 -- Gaussian Kernel -- $n=400$.}
\begin{subfigure}{0.32\textwidth}
\centering
\includegraphics[width=1\textwidth]{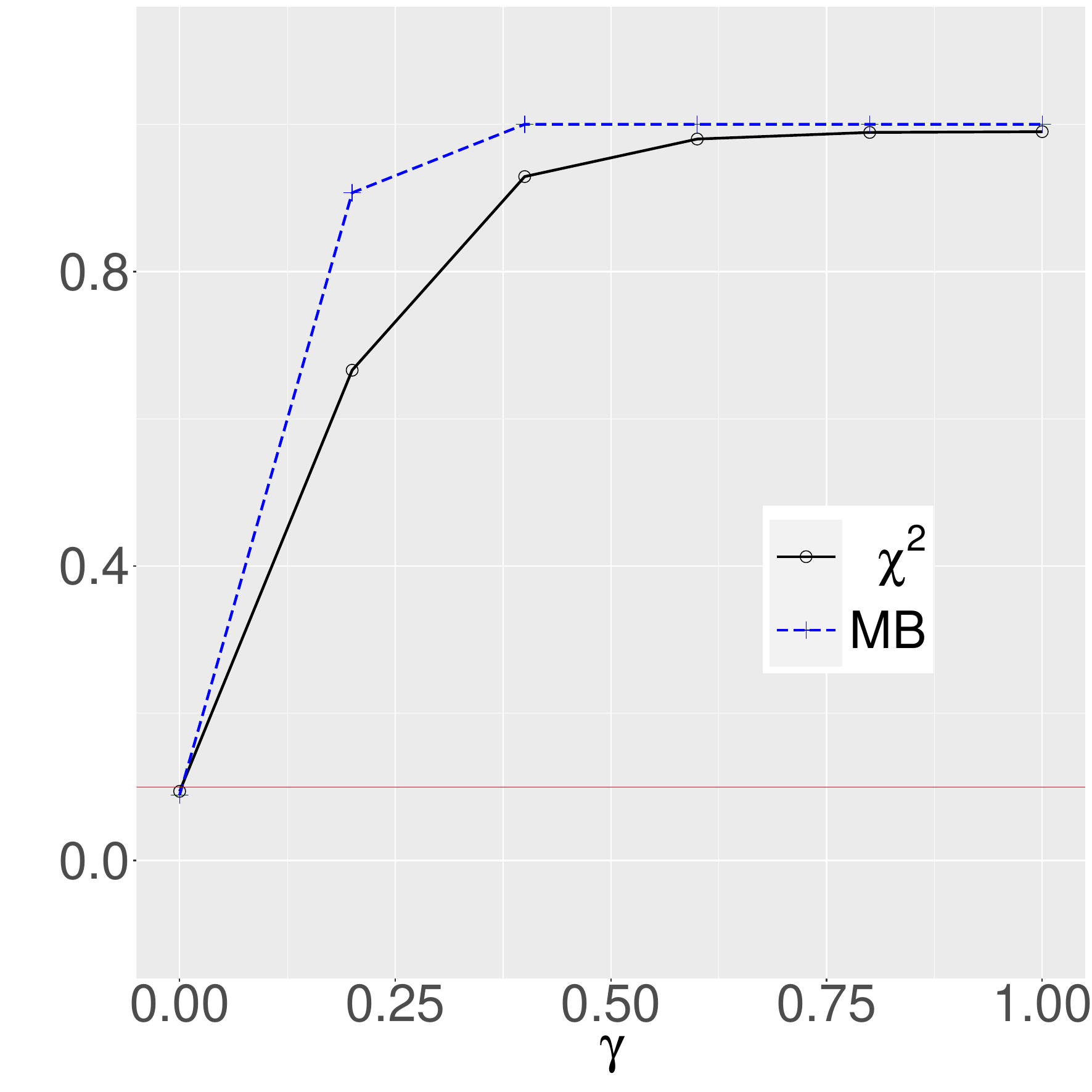}
\caption{10\%}
\end{subfigure}
\begin{subfigure}{0.32\textwidth}
\centering
\includegraphics[width=1\textwidth]{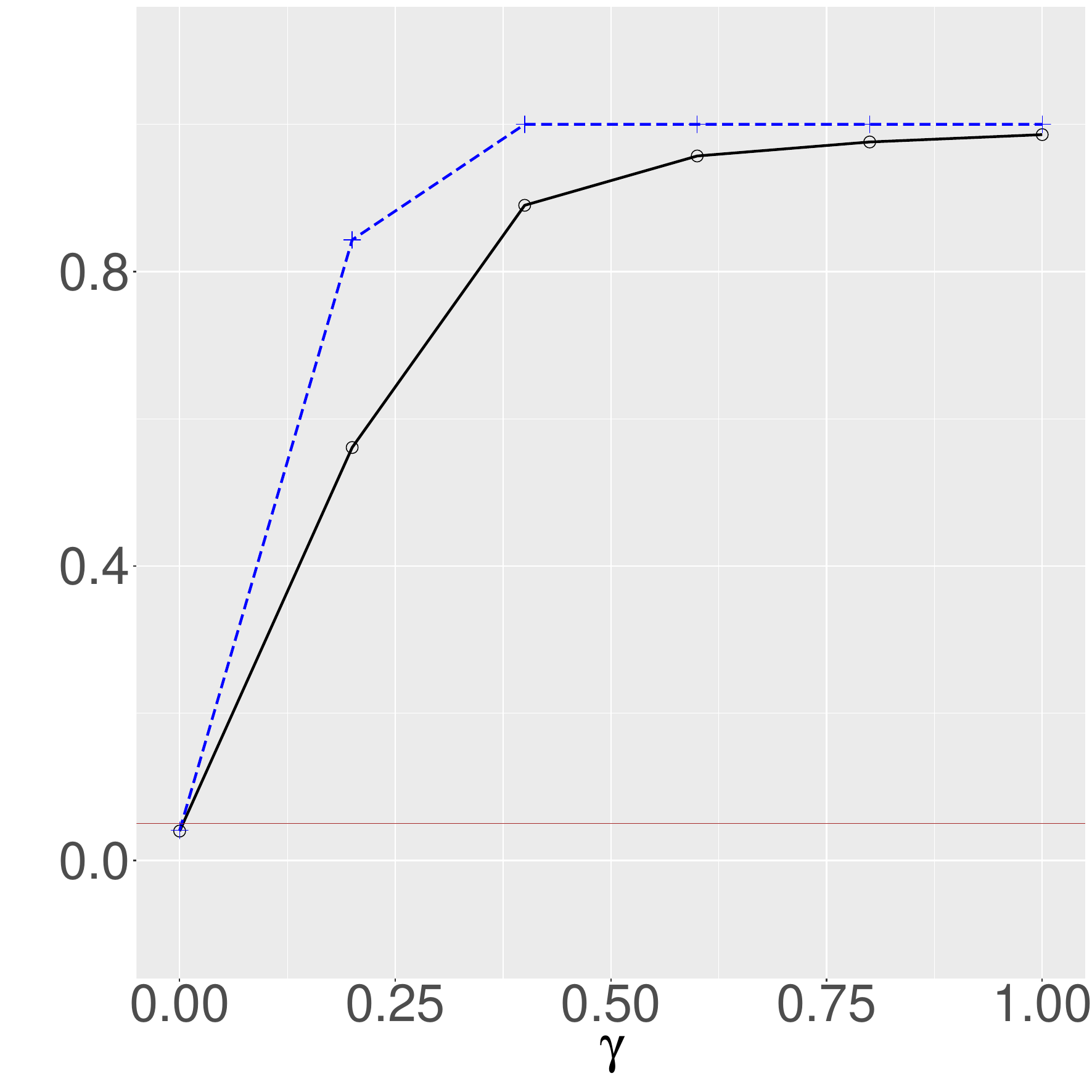}
\caption{5\%}
\end{subfigure}
\begin{subfigure}{0.32\textwidth}
\centering
\includegraphics[width=1\textwidth]{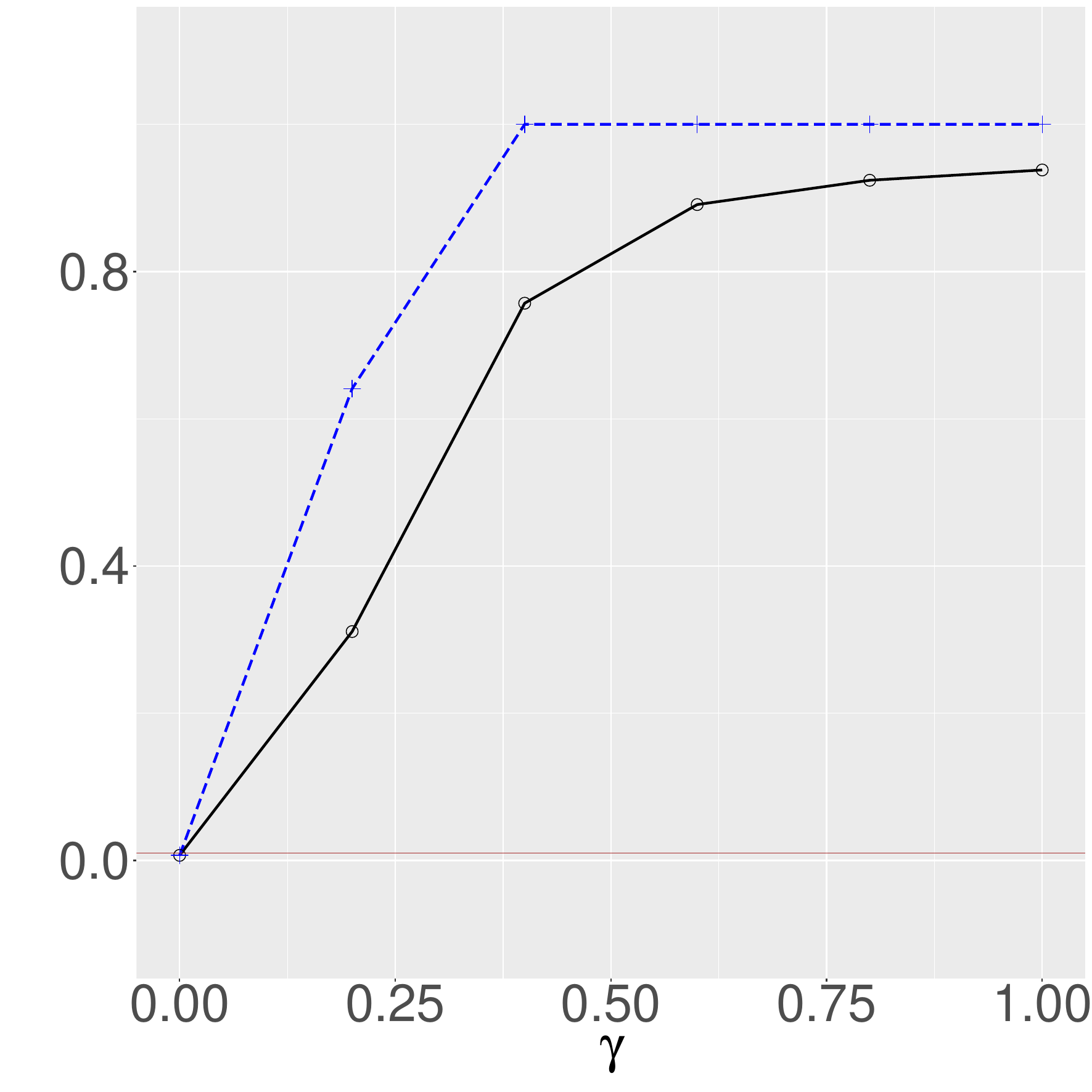}
\caption{1\%}
\end{subfigure}
\label{Fig:MI3_Gauss}
\end{figure}

\begin{figure}[H]
\centering 
\caption{DGP MI 3 -- Negative Euclidean -- $n=400$.}
\begin{subfigure}{0.32\textwidth}
\centering
\includegraphics[width=1\textwidth]{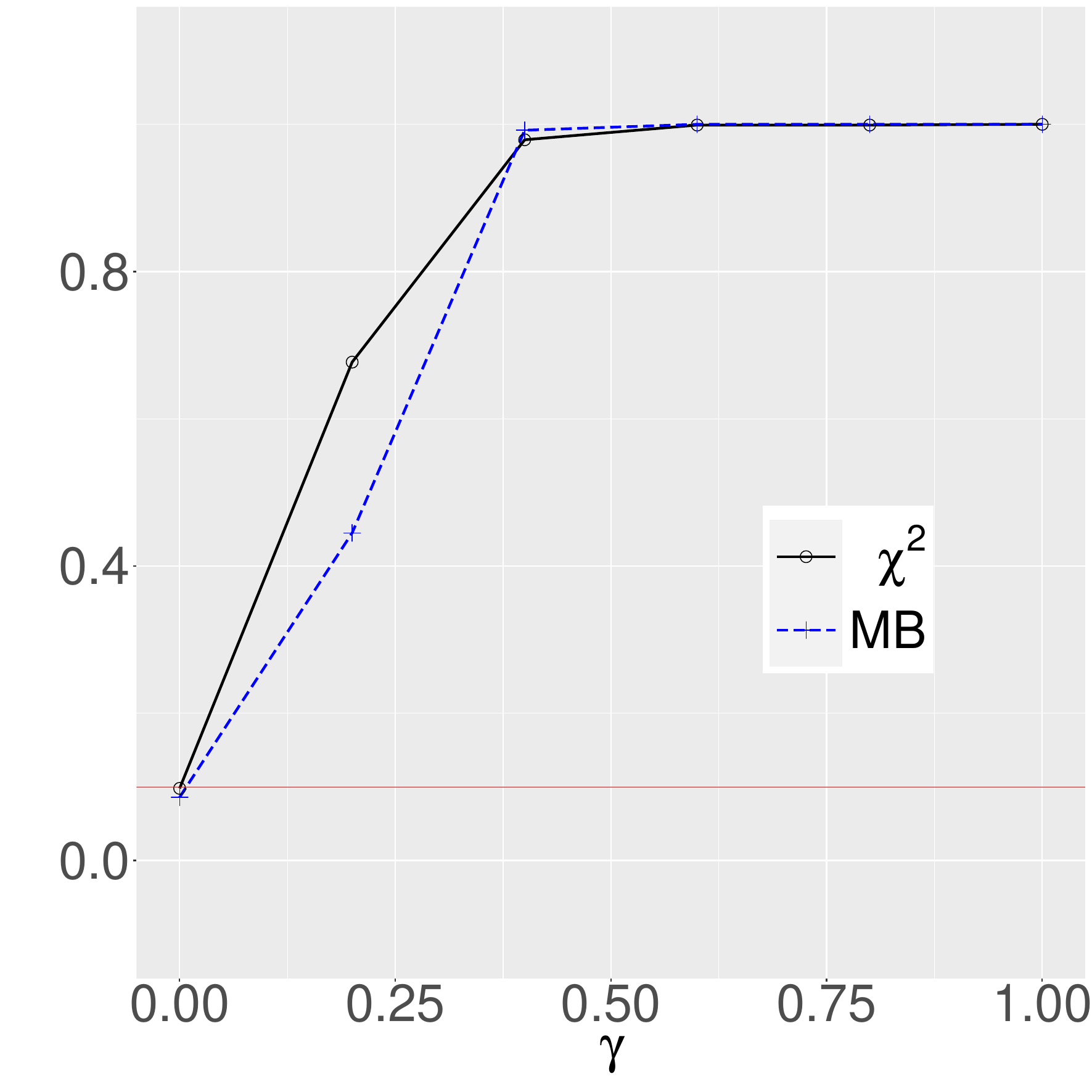}
\caption{10\%}
\end{subfigure}
\begin{subfigure}{0.32\textwidth}
\centering
\includegraphics[width=1\textwidth]{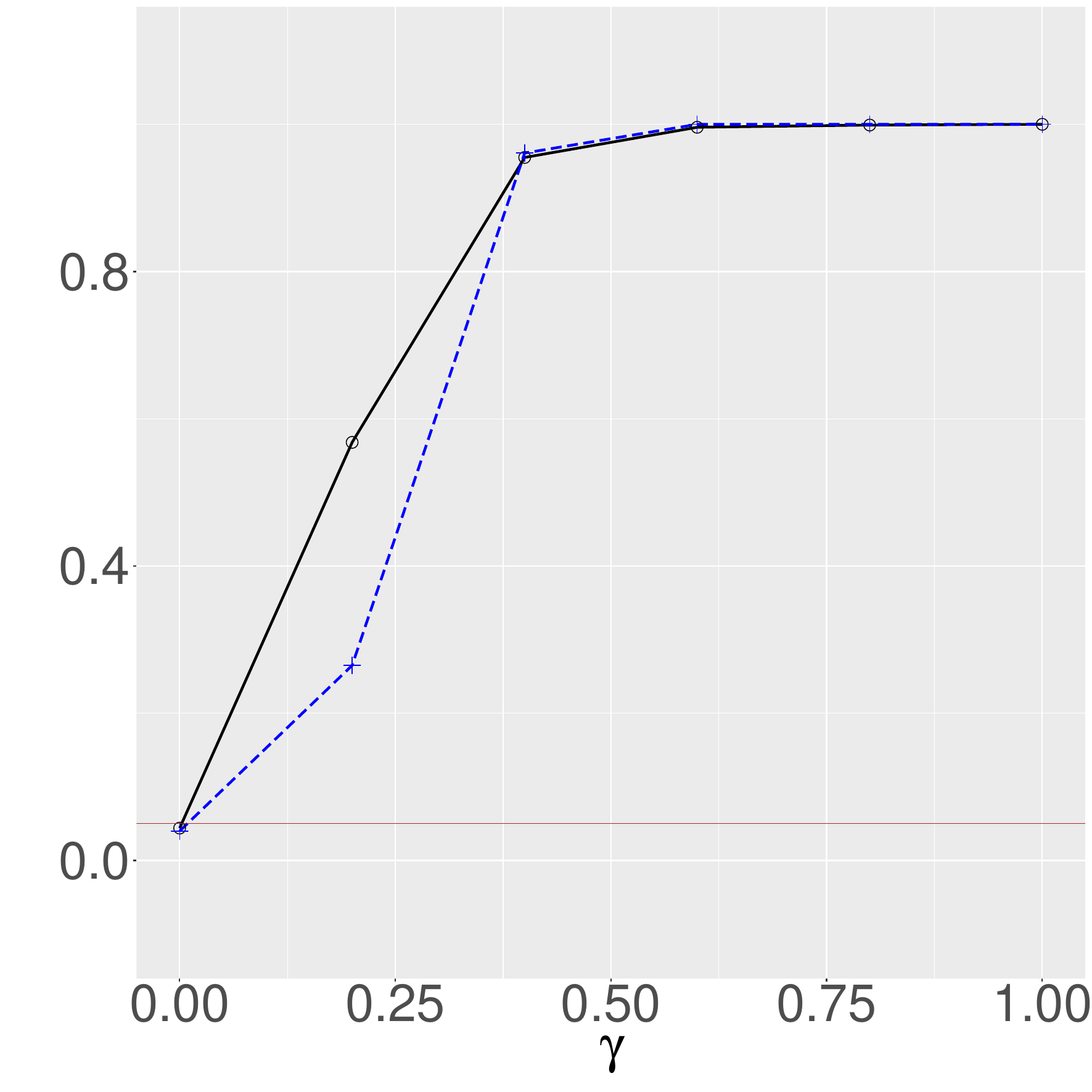}
\caption{5\%}
\end{subfigure}
\begin{subfigure}{0.32\textwidth}
\centering
\includegraphics[width=1\textwidth]{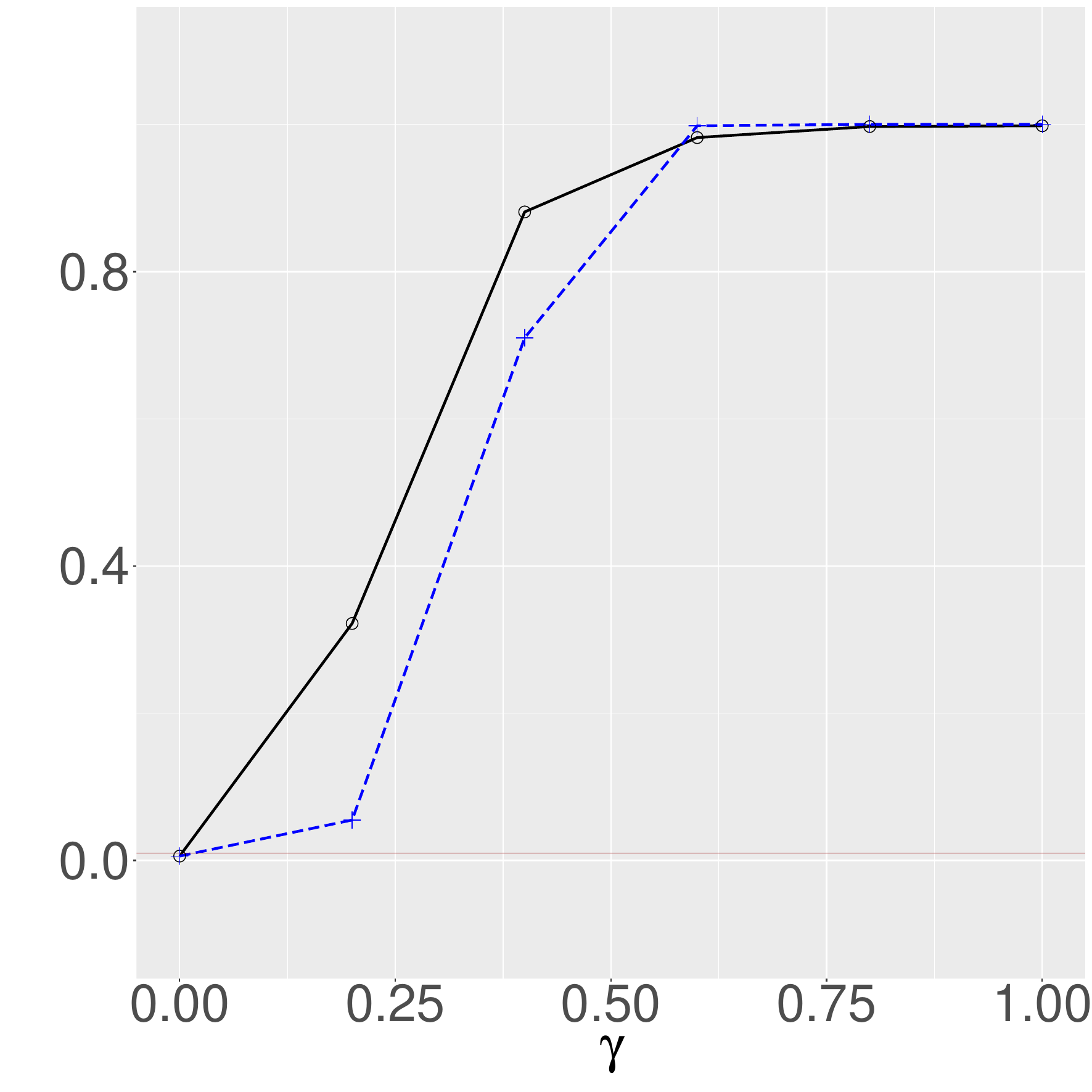}
\caption{1\%}
\end{subfigure}
\label{Fig:MI3_Euclid}
\end{figure}

\begin{figure}[H]
\centering 
\caption{DGP MI 4 -- Gaussian Kernel -- $n=400$.}
\begin{subfigure}{0.32\textwidth}
\centering
\includegraphics[width=1\textwidth]{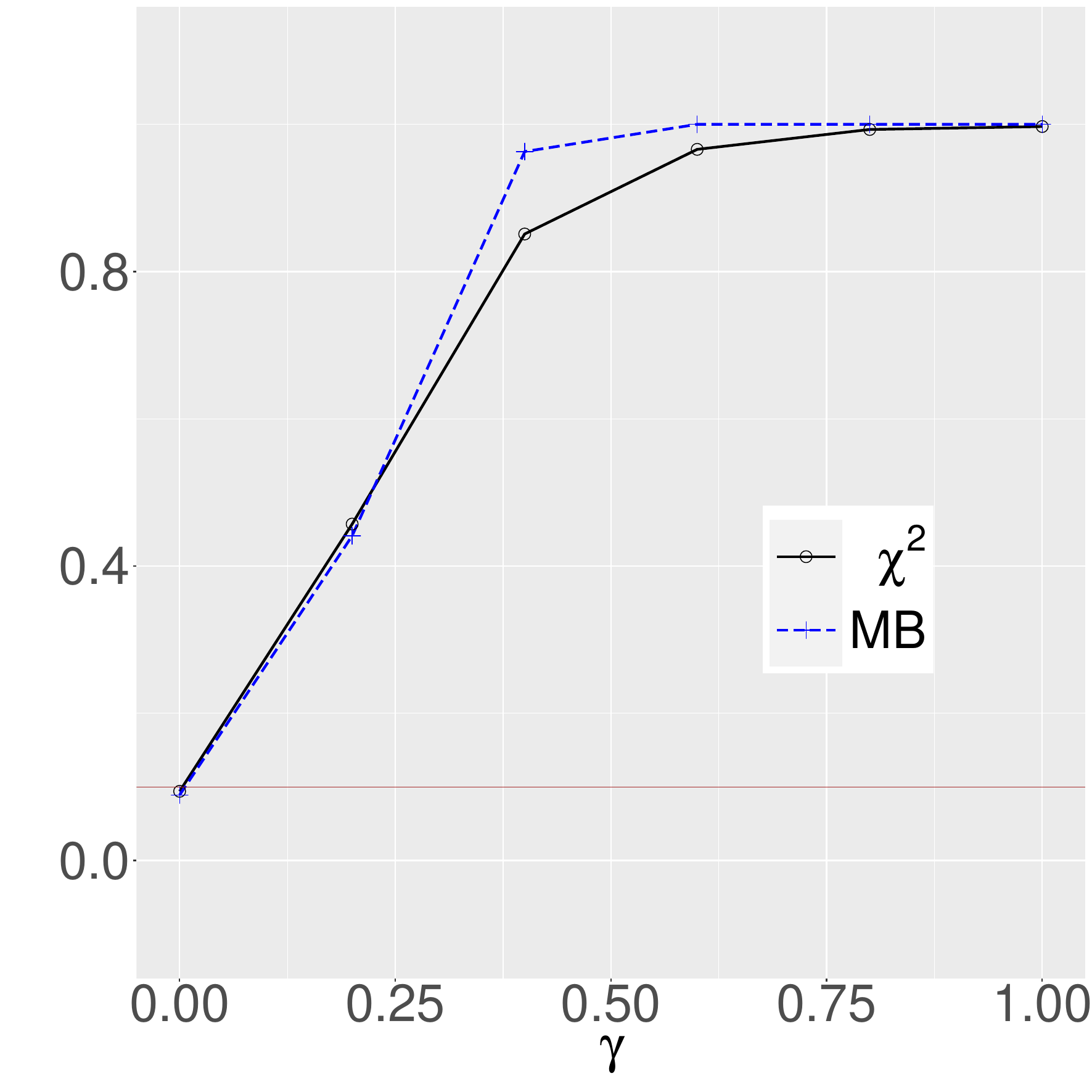}
\caption{10\%}
\end{subfigure}
\begin{subfigure}{0.32\textwidth}
\centering
\includegraphics[width=1\textwidth]{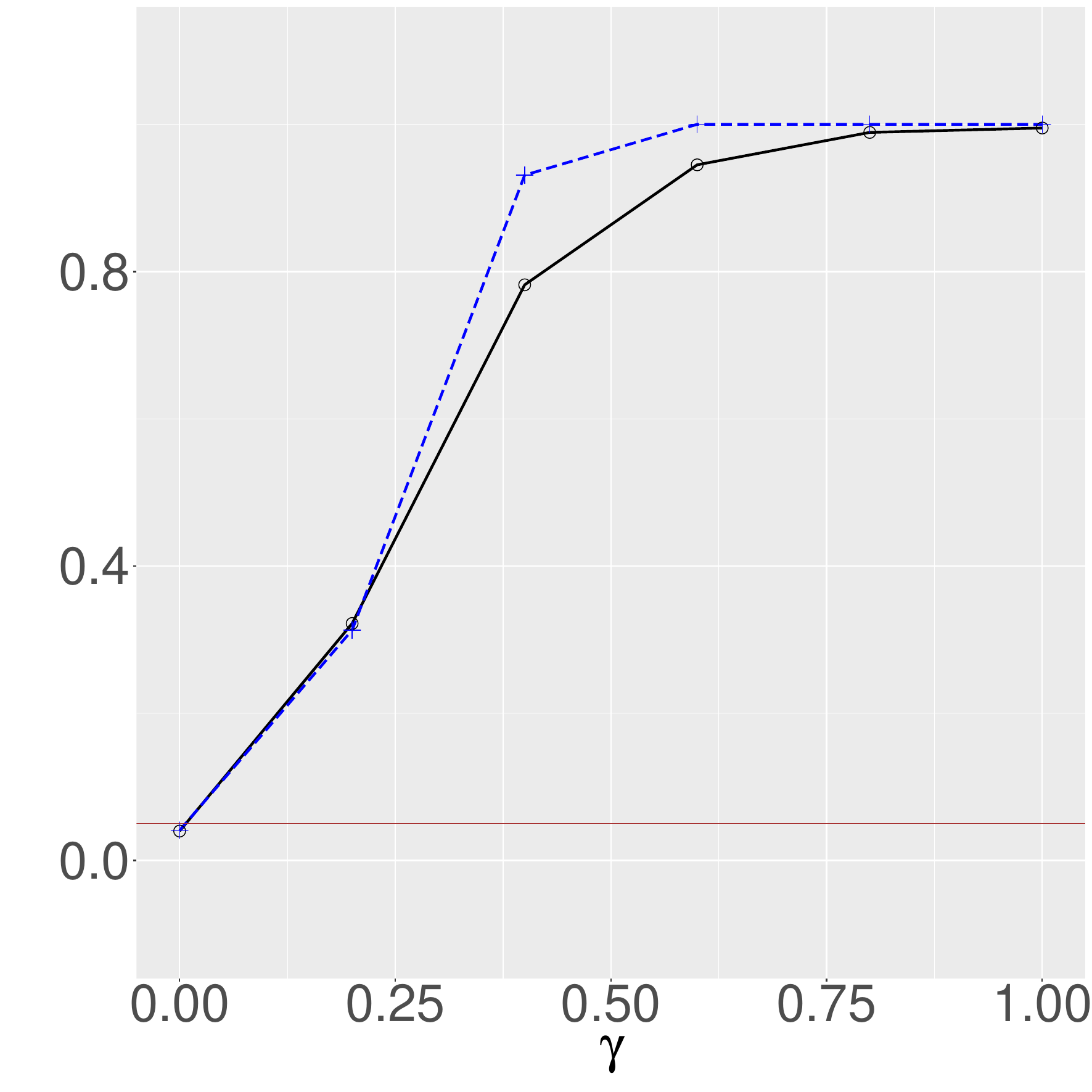}
\caption{5\%}
\end{subfigure}
\begin{subfigure}{0.32\textwidth}
\centering
\includegraphics[width=1\textwidth]{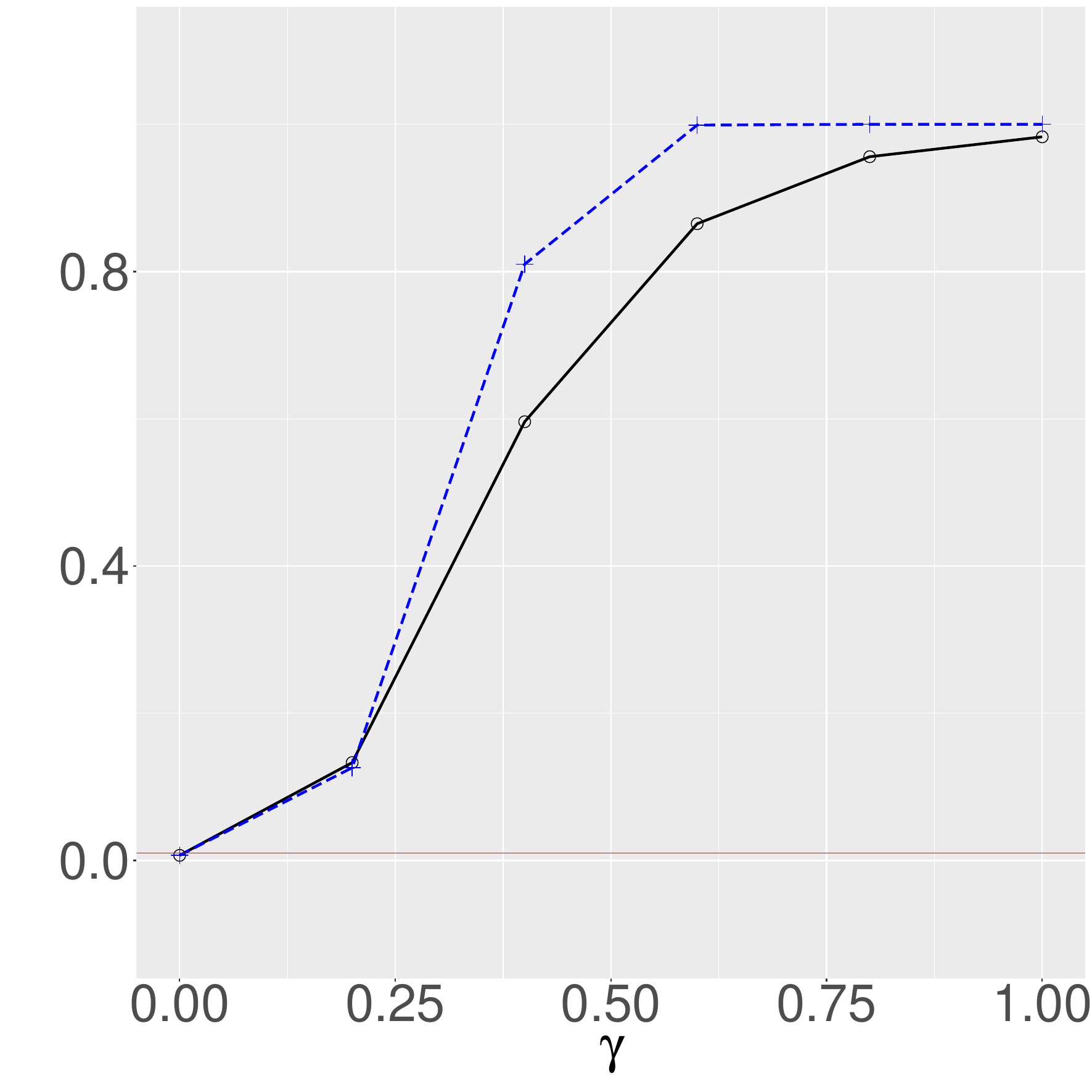}
\caption{1\%}
\end{subfigure}
\label{Fig:MI4_Gauss}
\end{figure}

\begin{figure}[H]
\centering 
\caption{DGP MI 4 -- Negative Euclidean -- $n=400$.}
\begin{subfigure}{0.32\textwidth}
\centering
\includegraphics[width=1\textwidth]{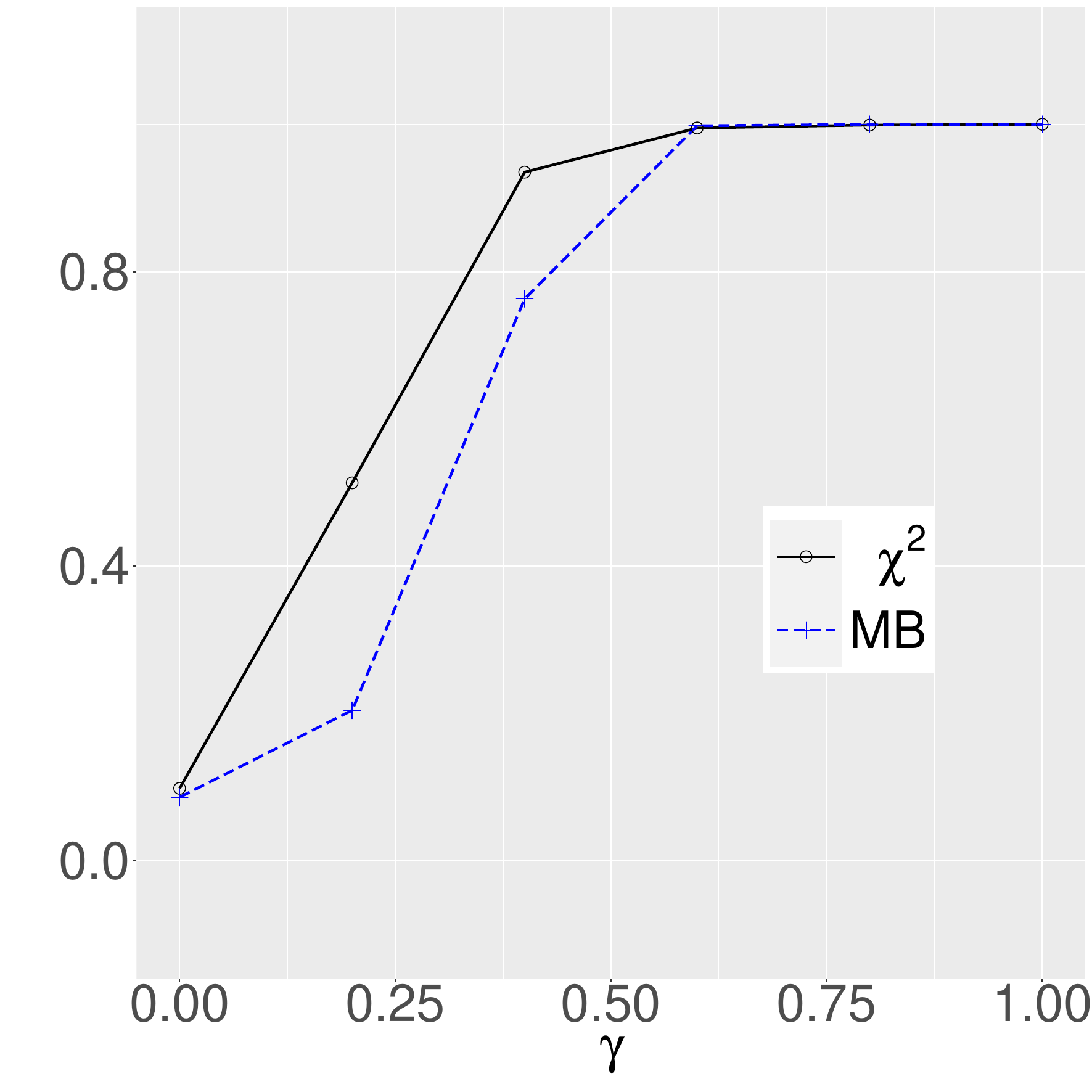}
\caption{10\%}
\end{subfigure}
\begin{subfigure}{0.32\textwidth}
\centering
\includegraphics[width=1\textwidth]{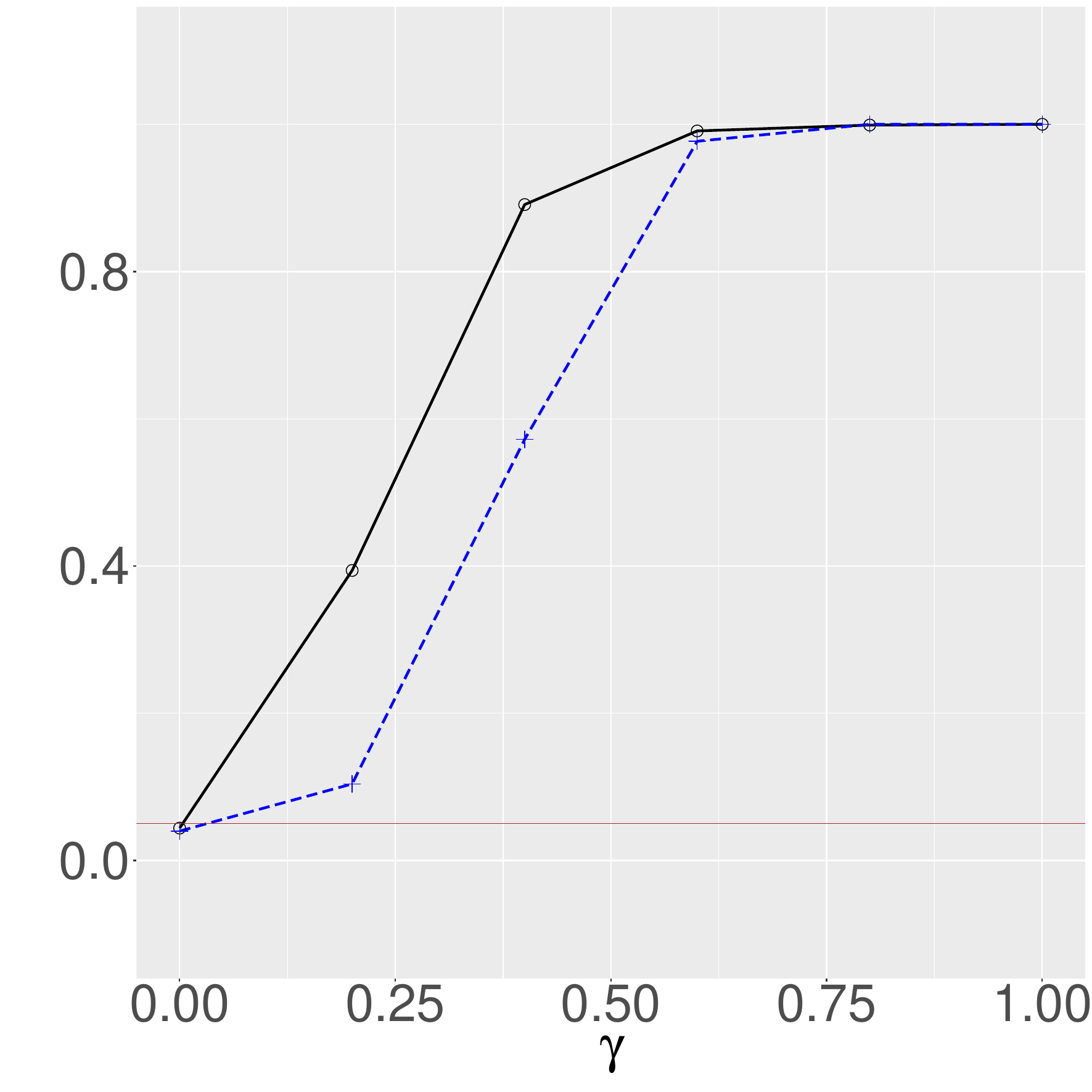}
\caption{5\%}
\end{subfigure}
\begin{subfigure}{0.32\textwidth}
\centering
\includegraphics[width=1\textwidth]{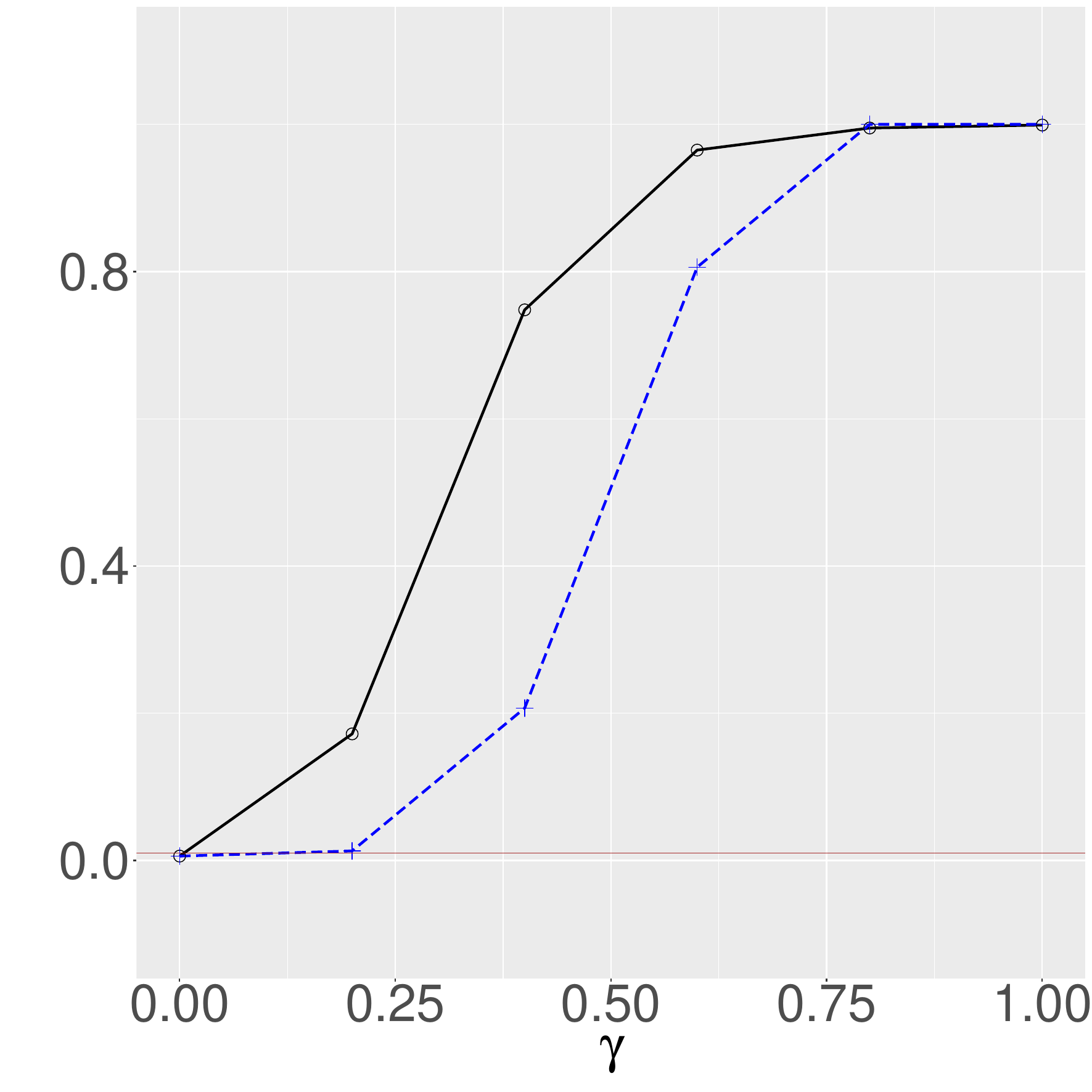}
\caption{1\%}
\end{subfigure}
\label{Fig:MI4_Euclid}
\end{figure}

\begin{figure}[H]
\centering 
\caption{DGP MI 5 -- Gaussian Kernel -- $n=400$.}
\begin{subfigure}{0.32\textwidth}
\centering
\includegraphics[width=1\textwidth]{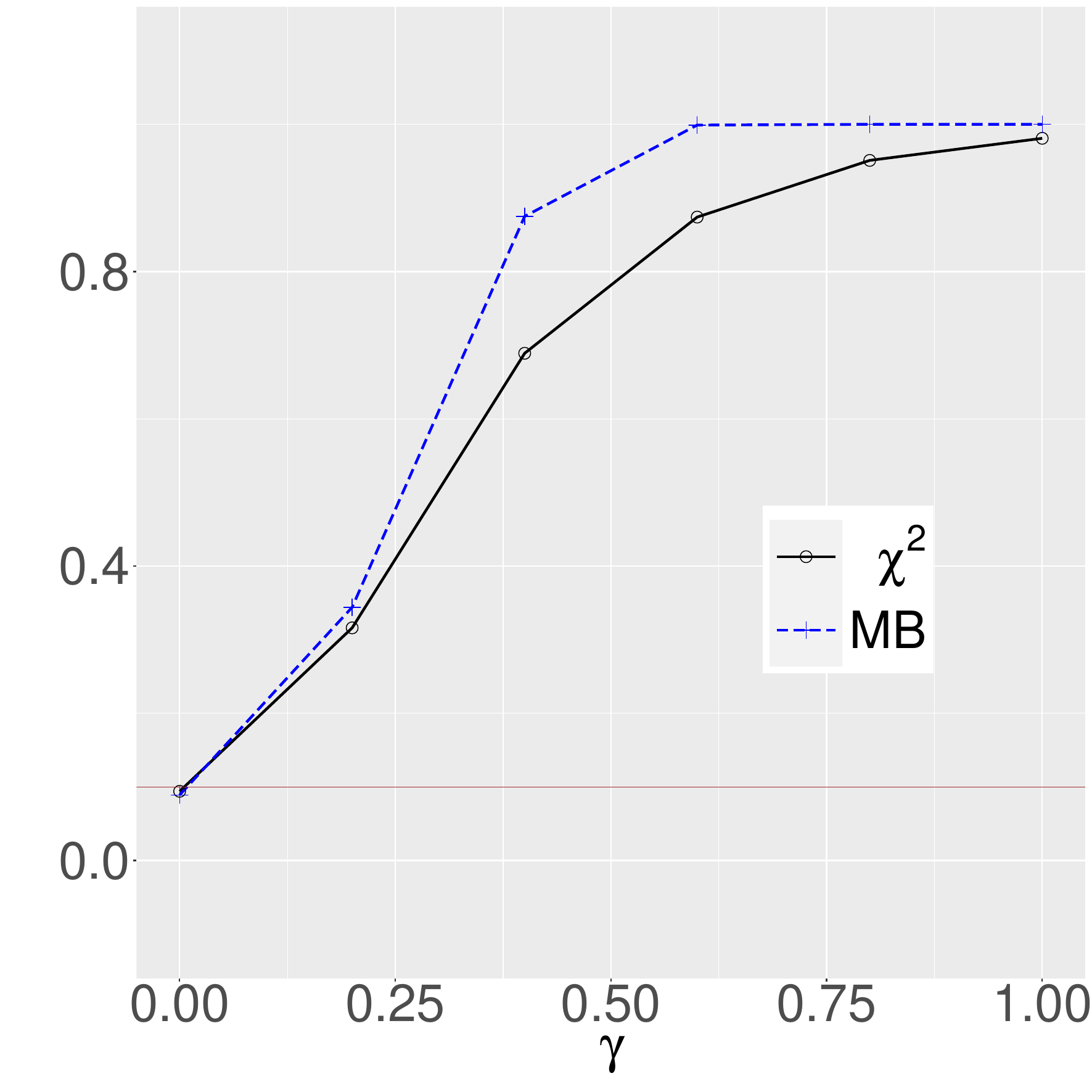}
\caption{10\%}
\end{subfigure}
\begin{subfigure}{0.32\textwidth}
\centering
\includegraphics[width=1\textwidth]{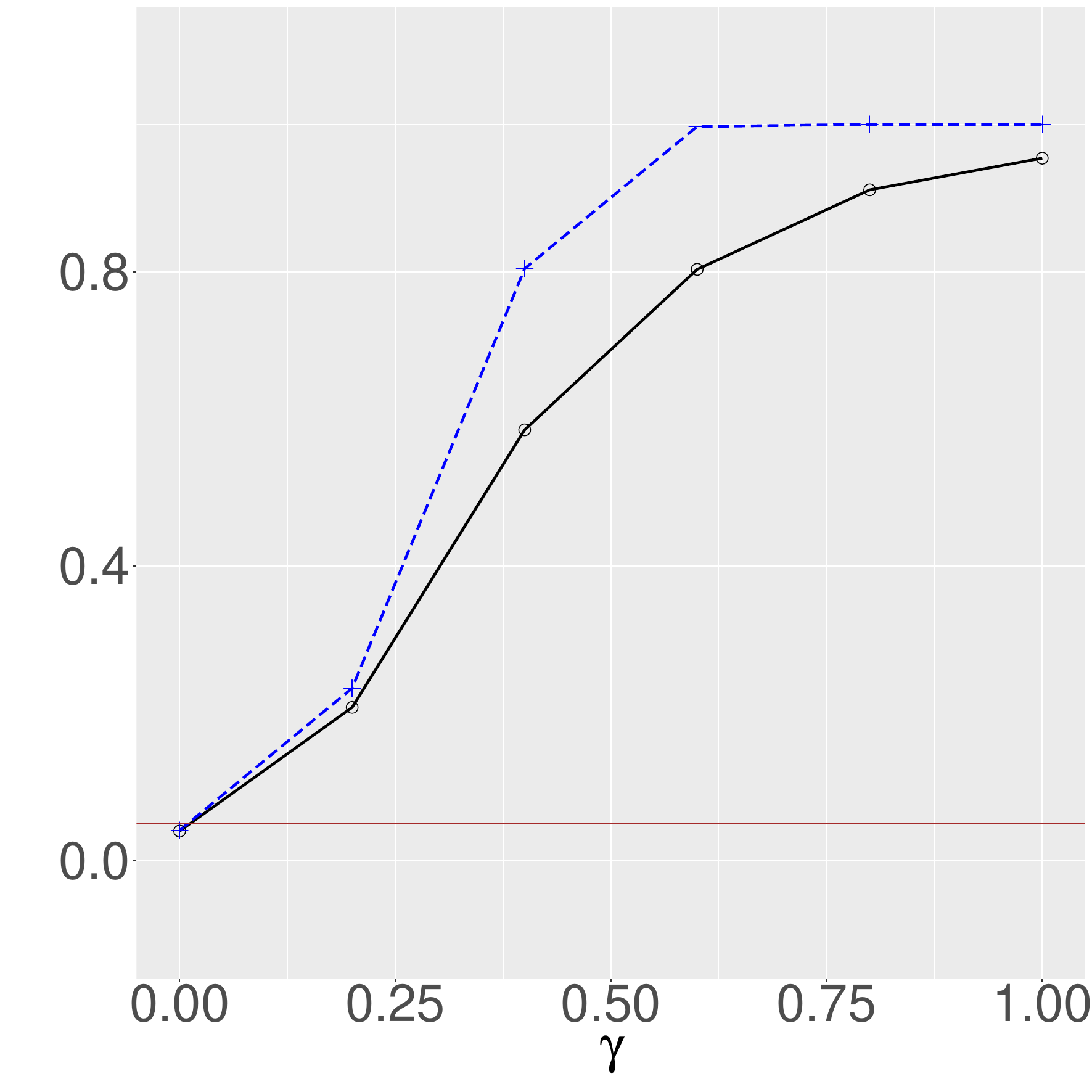}
\caption{5\%}
\end{subfigure}
\begin{subfigure}{0.32\textwidth}
\centering
\includegraphics[width=1\textwidth]{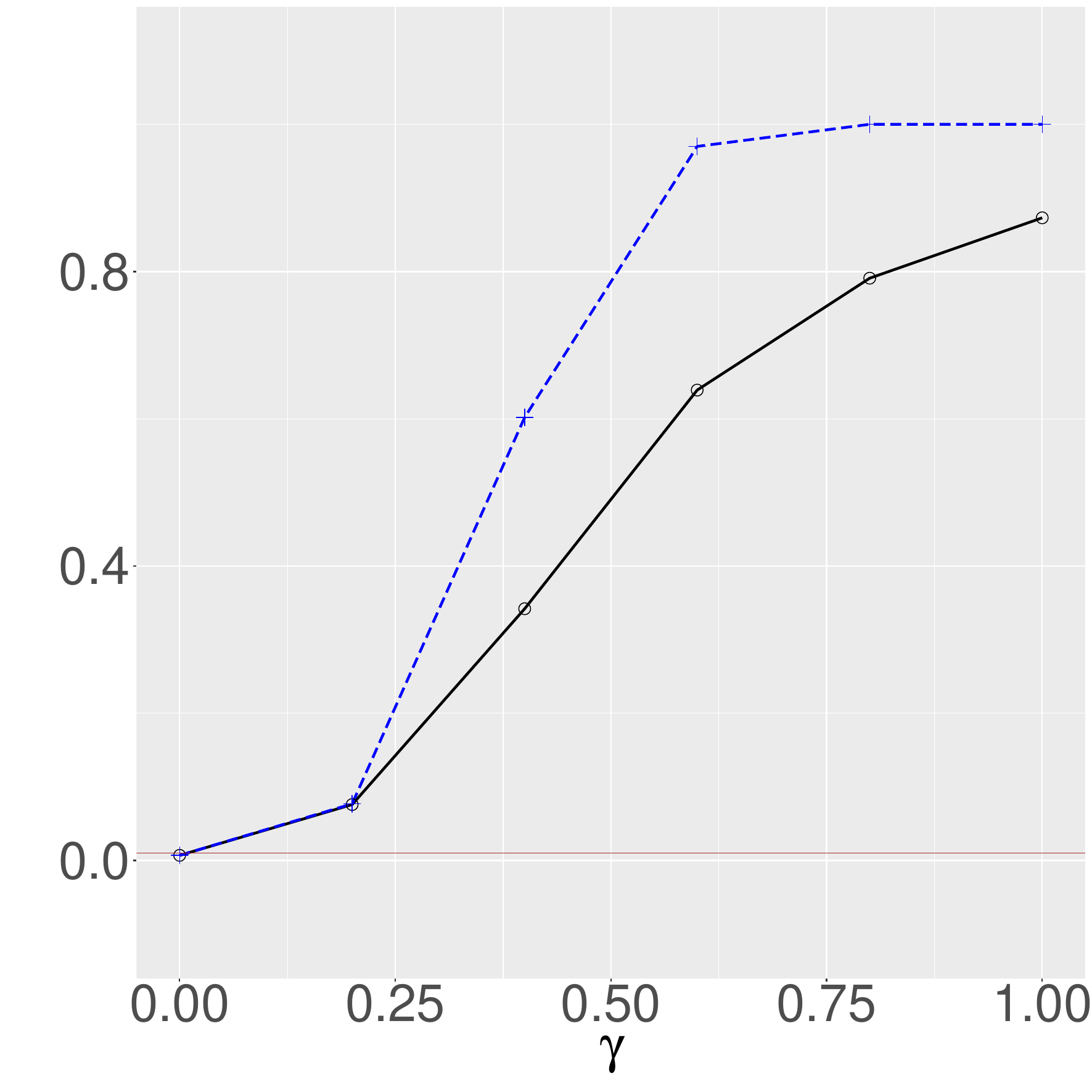}
\caption{1\%}
\end{subfigure}
\label{Fig:MI5_Gauss}
\end{figure}

\begin{figure}[H]
\centering 
\caption{DGP MI 5 -- Negative Euclidean -- $n=400$.}
\begin{subfigure}{0.32\textwidth}
\centering
\includegraphics[width=1\textwidth]{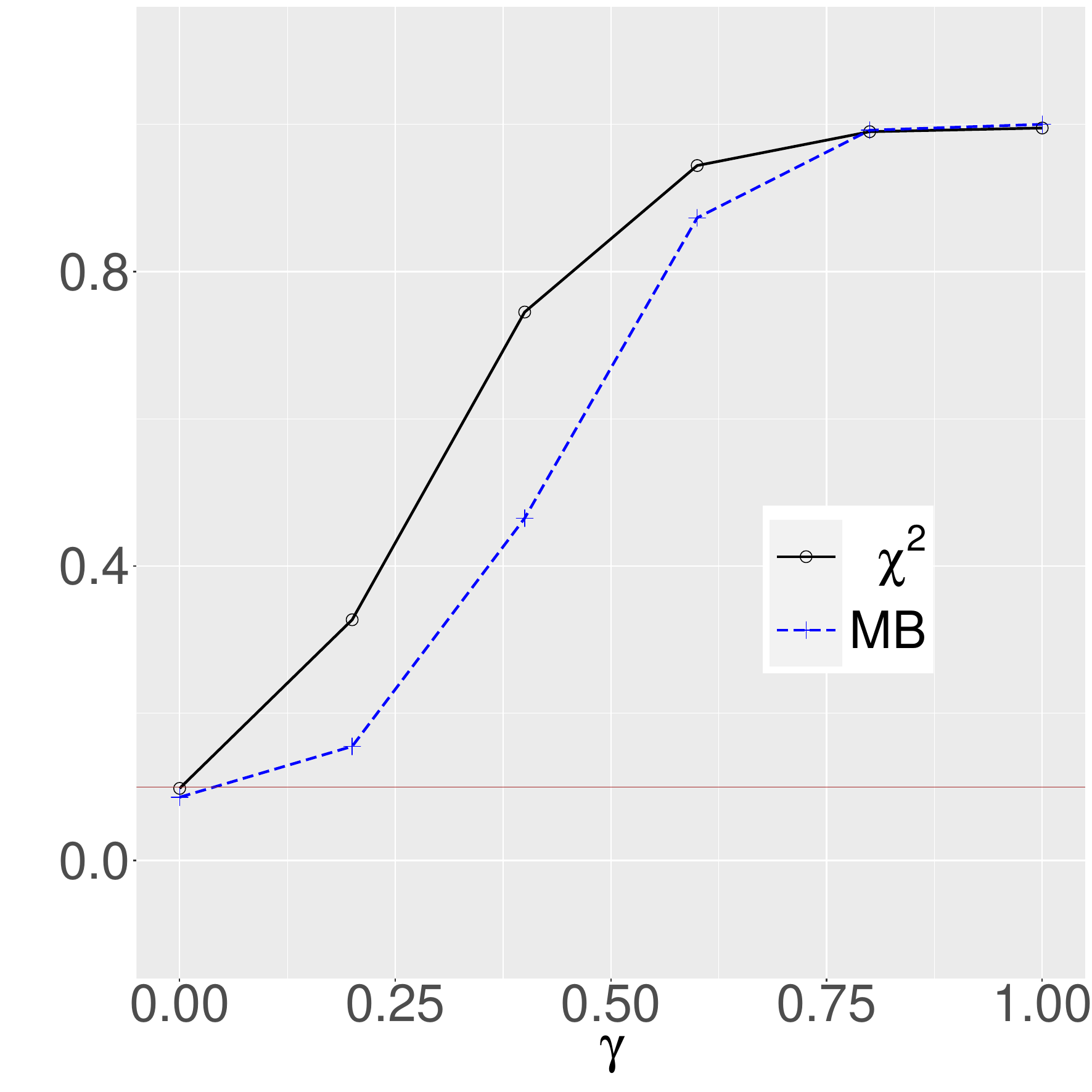}
\caption{10\%}
\end{subfigure}
\begin{subfigure}{0.32\textwidth}
\centering
\includegraphics[width=1\textwidth]{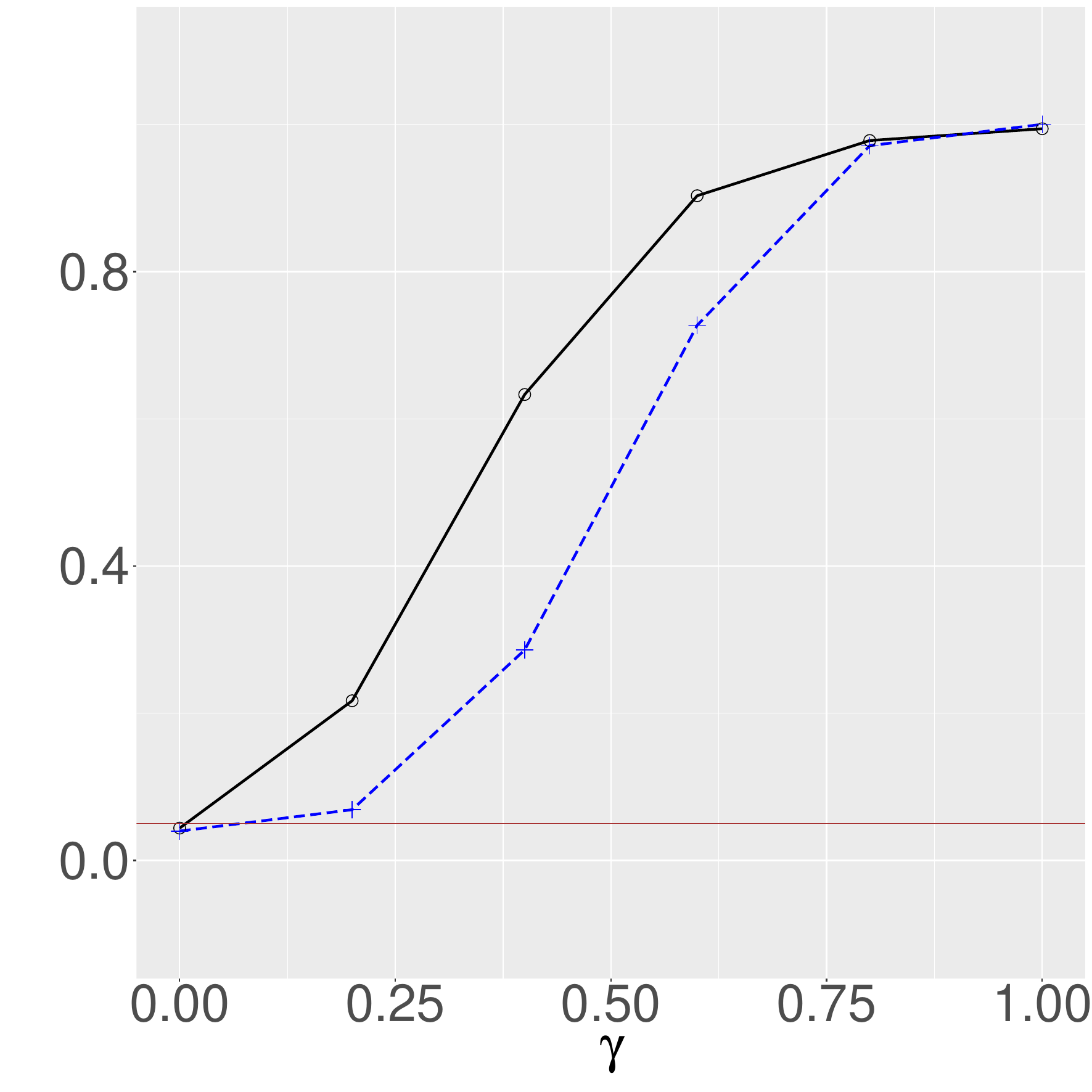}
\caption{5\%}
\end{subfigure}
\begin{subfigure}{0.32\textwidth}
\centering
\includegraphics[width=1\textwidth]{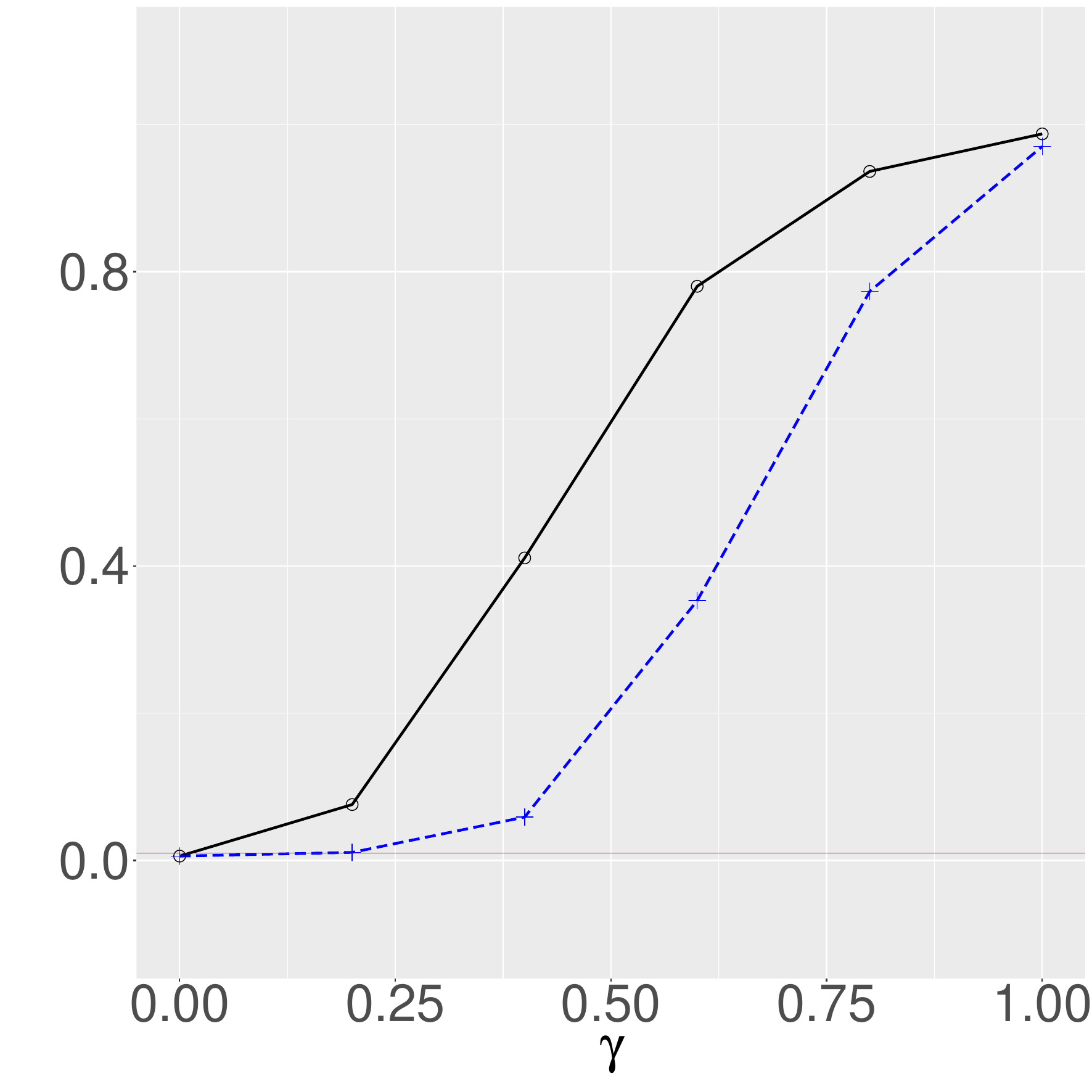}
\caption{1\%}
\end{subfigure}
\label{Fig:MI5_Euclid}
\end{figure}

\Cref{Fig:MI2_Gauss,Fig:MI2_Euclid,Fig:MI3_Gauss,Fig:MI3_Euclid,Fig:MI4_Gauss,Fig:MI4_Euclid,Fig:MI5_Gauss,Fig:MI5_Euclid} illustrate that all tests generally perform well in detecting violations of mean independence. The $\chi^2$-test shows competitive performance across all DGPs. As indicated by \Cref{thm_bahadur}, relative performance depends on, for example, the choice of kernel. Overall, no single test consistently outperforms the other, highlighting the complementary strengths of the $\chi^2$- and bootstrap-based ICM tests.

\subsection{$p_v>2$ }\label{subsec:pv_more}
The goal of this subsection is to study the sensitivity of the $\chi^2$-test to the dimension of $V$ alongside bootstrap-based ICM tests. The $\chi^2$-test is implemented in the rest of this section using the Gaussian kernel. Consider the following DGP:
\begin{enumerate}
    \item[MI 6:] $U = 1 + \frac{\gamma}{0.233}(\exp(-Z^2/3) - \sqrt{3/5}) + \mathcal{E} $, $Z \sim\mathcal{N}(0,1)$, and $\mathcal{E} \sim \mathcal{U}[-\sqrt{3},\sqrt{3}] $.
\end{enumerate}

\noindent Define the following: $h_1(Z):= \exp(Z) $; $h_2(Z):= 4\sqrt{3} \exp(-Z^2/2) $; $V_1:= [h_1(Z),\ U - h_1(Z) ]^{\top} $; $V_{1A}:= [h_1(Z),\ U - h_1(Z), \ h_2(Z) ]^{\top} $; $V_2:= [h_2(Z),\ U - h_2(Z) ]^{\top} $; and $V_{2A}:= [h_2(Z),\ U - h_2(Z), \ h_1(Z) ]^{\top} $. Let $g_p(Z) \in \mathbb{R}^p $ denote a vector of orthogonal polynomials of $Z$ with degrees $1$ through $p$. Then, we generate higher dimensional $V$ given by $V_{1B}$ and $V_{2B}$, respectively, which augment $V_{1A}$ and $V_{2A}$ using $g_2(Z)$, and $V_{1C}$ and $V_{2C}$, respectively, which augment $V_{1A}$ and $V_{2A}$ using $g_7(Z)$. When $p_v \geq 3$, the degrees of freedom of the $\chi^2$-test is set to the number of positive eigenvalues of $\widehat{\Omega}_{h,n}$. In contrast to $V_2$ where $h_2(Z)$ targets the alternative, $V_1$ is agnostic about the alternative.

\begin{table}[!htbp]
\centering
\setlength{\tabcolsep}{4pt}
\caption{DGP MI 6 - Sensitivity to $p_v$}
\begin{tabular}{@{}clccccccccccc@{}}
\toprule
                      $\gamma$ & Sig-Lev & \multicolumn{8}{c}{$ \chi^2 $-test} & \multicolumn{3}{c}{Wild Bootstrap} \\ \cmidrule(lr){1-1} \cmidrule(lr){2-2} \cmidrule(lr){3-10} \cmidrule(lr){11-13}
                &  & $V_1$ & $V_{1A}$ & $V_{1B}$ & $V_{1C}$ & $V_2$ & $V_{2A}$ & $V_{2B}$ & $V_{2C}$ & Gauss & MDD & Esc6 \\
                &  & $p_v=2$ & $p_v=3$ & $p_v=5$ & $p_v=10$ & $p_v=2$ & $p_v=3$ & $p_v=5$ & $p_v=10$ &  &  &  \\ \midrule
&10\%   &0.112 &0.112 &0.102 &0.100 &0.113 &0.116 &0.110 &0.107 &0.100 &0.100 &0.104 \\
0.0   &5\%    &0.051 &0.051 &0.050 &0.050 &0.058 &0.058 &0.056 &0.056 &0.049 &0.053 &0.053 \\
      &1\%    &0.007 &0.007 &0.007 &0.007 &0.008 &0.008 &0.009 &0.007 &0.009 &0.009 &0.007 \\
\midrule
      &10\%   &0.976 &0.976 &0.973 &0.973 &0.989 &0.989 &0.982 &0.977 &0.925 &0.818 &0.707 \\
0.2   &5\%    &0.953 &0.953 &0.946 &0.945 &0.980 &0.979 &0.967 &0.958 &0.854 &0.678 &0.463 \\
      &1\%    &0.838 &0.838 &0.822 &0.822 &0.912 &0.911 &0.863 &0.852 &0.622 &0.264 &0.115 \\
\midrule
      &10\%   &1.000 &1.000 &1.000 &1.000 &1.000 &1.000 &1.000 &1.000 &1.000 &1.000 &1.000 \\
0.4   &5\%    &1.000 &1.000 &1.000 &1.000 &1.000 &1.000 &1.000 &1.000 &1.000 &1.000 &0.999 \\
      &1\%    &1.000 &1.000 &1.000 &1.000 &1.000 &1.000 &1.000 &1.000 &1.000 &0.998 &0.965 \\
\midrule
      &10\%   &1.000 &1.000 &1.000 &1.000 &1.000 &1.000 &1.000 &1.000 &1.000 &1.000 &1.000 \\
0.6   &5\%    &1.000 &1.000 &1.000 &1.000 &1.000 &1.000 &1.000 &1.000 &1.000 &1.000 &1.000 \\
      &1\%    &1.000 &1.000 &1.000 &1.000 &1.000 &1.000 &1.000 &1.000 &1.000 &1.000 &1.000 \\
\midrule
      &10\%   &1.000 &1.000 &1.000 &1.000 &1.000 &1.000 &1.000 &1.000 &1.000 &1.000 &1.000 \\
0.8   &5\%    &1.000 &1.000 &1.000 &1.000 &1.000 &1.000 &1.000 &1.000 &1.000 &1.000 &1.000 \\
      &1\%    &1.000 &1.000 &1.000 &1.000 &1.000 &1.000 &1.000 &1.000 &1.000 &1.000 &1.000 \\
\midrule
      &10\%   &1.000 &1.000 &1.000 &1.000 &1.000 &1.000 &1.000 &1.000 &1.000 &1.000 &1.000 \\
1.0   &5\%    &1.000 &1.000 &1.000 &1.000 &1.000 &1.000 &1.000 &1.000 &1.000 &1.000 &1.000 \\
      &1\%    &1.000 &1.000 &1.000 &1.000 &1.000 &1.000 &1.000 &1.000 &1.000 &1.000 &1.000 \\
\bottomrule
\end{tabular}
\label{Tab:MIp_23}
\end{table}

\Cref{Tab:MIp_23} presents the power performance of four variations (depending on the dimension and specification of $V$) of the $\chi^2$-test (based on the Gaussian kernel) in addition to bootstrap-based ICM tests of mean independence Gauss, MDD, and Esc6. MDD and Esc6 are based on the negative Euclidean kernel of \citet{shao2014martingale} and the kernel proposed in \citet{escanciano2006consistent}, respectively. First, one observes good size control and non-trivial power increasing in $\gamma$ for all tests. Second, increasing the dimension of $V$ appears to decrease power. A comparison of the test with $V_2$ (where the alternative is targeted) and $\{V_1,V_{1A},V_{1B},V_{1C}\}$ shows a power advantage of targeting the alternative using a parsimonious 2-dimensional $V$.

\subsection{Selection Criteria $c_n$ }\label{subsec:robustness_cn}

In all preceding implementations of the proposed $\chi^2$-test, the tuning parameter in the regularized $\widehat{\Omega}_{h,n}$ is set to $c_n=\widetilde{\lambda}_1n^{-1/3}$. This subsection conducts a robustness exercise to examine the sensitivity of the empirical size and power performance to the tuning rule $c_n$. Two scenarios are considered. 

\subsubsection{Scenario 1}

The first scenario concerns sensitivity to the constant $\iota $ in the rule $c_n=\widetilde{\lambda}_1n^{-\iota }$. For the implementation, the set $\iota\in \{ 2/5,1/3,1/4,1/6\} $ is considered with $p_v=2$,  and $V=V_1$ using the following DGP:
\begin{enumerate}
    \item[MI 7:] $U = 1 + \gamma Z^2 + \mathcal{E} $, $Z \sim\mathcal{N}(0,1)$, and $\mathcal{E} \sim \mathcal{U}[-\sqrt{3},\sqrt{3}] $.
\end{enumerate}

\begin{table}[!htbp]
\centering
\caption{DGP MI 7 - Sensitivity to $\iota $}
\begin{tabular}{@{}clcccc@{}}
\toprule
                    $\gamma$    & Sig-Lev & \multicolumn{4}{c}{$ c_n = \widetilde{\lambda}_1n^{-\iota } $} \\ \cmidrule(lr){1-1} \cmidrule(lr){2-2} \cmidrule(lr){3-6} 
                &  & $ \iota  = \frac{2}{5} $ & $ \iota  = \frac{1}{3}$ & $ \iota =\frac{1}{4} $ & $ \iota  = \frac{1}{6} $  \\ \midrule
      &10\%   &0.098 &0.098 &0.098 &0.098 \\
0.0   &5\%    &0.049 &0.049 &0.049 &0.049 \\
      &1\%    &0.010 &0.010 &0.010 &0.010 \\
\midrule
      &10\%   &0.485 &0.485 &0.485 &0.485 \\
0.2   &5\%    &0.355 &0.355 &0.355 &0.355 \\
      &1\%    &0.150 &0.150 &0.150 &0.150 \\
\midrule
      &10\%   &0.919 &0.919 &0.919 &0.919 \\
0.4   &5\%    &0.871 &0.871 &0.871 &0.871 \\
      &1\%    &0.655 &0.655 &0.655 &0.655 \\
\midrule
      &10\%   &0.991 &0.991 &0.991 &0.991 \\
0.6   &5\%    &0.982 &0.982 &0.982 &0.982 \\
      &1\%    &0.927 &0.927 &0.927 &0.927 \\
\midrule
      &10\%   &0.999 &0.999 &0.999 &0.999 \\
0.8   &5\%    &0.998 &0.998 &0.998 &0.998 \\
      &1\%    &0.987 &0.987 &0.987 &0.987 \\
\midrule
      &10\%   &1.000 &1.000 &1.000 &1.000 \\
1.0   &5\%    &1.000 &1.000 &1.000 &1.000 \\
      &1\%    &0.996 &0.996 &0.996 &0.996 \\
\bottomrule
\end{tabular}
\label{Tab:MI_cn_robust}
\end{table}

\noindent \Cref{Tab:MI_cn_robust} presents results that compare the performance of the $\chi^2$-test by different choices of $c_n$ in the first scenario. A clear conclusion is that the results are robust to the choice of $\iota $ in the rule $ c_n = \widetilde{\lambda}_1n^{-\iota } $ as there are negligible numerical differences in the empirical size and power across different valid choices of $\iota  \in (0,1/2) $.

\subsubsection{Scenario 2}

The second scenario compares the $\chi^2$-test with, in addition to $c_n$ in Scenario 1 above, suitable selection criteria typically used for truncated singular value decomposition -- see \citet{falini-2022-review} for a review. The setting adopted in this scenario is DGP MI 6, $p_v=10$, and $V=V_{1C}$ from Section \ref{subsec:pv_more}. Let $p(c_n)$ denote the number of non-zero eigenvalues in the regularized $\widehat{\Omega}_{h,n}$. Define 

    \begin{equation*}
        E_l:= - \frac{1}{\log(p_v)} \sum_{l'=1}^{l} \widetilde{f}_{l'} \log(\widetilde{f}_{l'}) \text{ where } \widetilde{f}_l:= \widetilde{\lambda}_l^2/\sum_{l'=1}^{p_v} \widetilde{\lambda}_{l'}^2.
    \end{equation*}

\noindent The following suitable SVD selection criteria are defined in terms of $p(c_n)$.

\begin{enumerate}[(1)]
    \item R-B -- ratio-based selection; $ \displaystyle p(c_n) = \argmin_{1\leq l \leq p_v} \frac{\widetilde{\lambda}_{l+1}}{\widetilde{\lambda}_l}  $; references: \citet{lam2011estimation}, \citet{lam2012factor}, and \citet[eqn. 6]{lee2018martingale}.
    
    \item E$\iota $ -- entropy-based selection; $ \displaystyle p(c_n) = \min\Big\{1 \leq l \leq p_v: \ E_l \geq \iota E_{p_v} \Big\} $; references: \citet{alter2000singular} and \citet[Sect. 2.6]{falini-2022-review}.

    \item TV$\iota $ -- Total Variance based selection; $ \displaystyle p(c_n) = \sum_{l=1}^{p_v} \indicator{ \widetilde{f}_l \geq \iota  }  $; references: \citet{suhr2005principal} and \citet[Sect. 2.6]{falini-2022-review}.

    \item CTV$\iota $ -- Cumulative percentage of Total Variance based selection; $ \displaystyle p(c_n) = \min\Big\{1 \leq l \leq p_v: \ \sum_{l'=1}^{l} \widetilde{f}_{l'} \geq \iota  \Big\} $; references: \citet[Chapter 6]{jolliffe2002principal} and \citet[Sect. 2.6]{falini-2022-review}.
\end{enumerate}

\noindent Recall $\widetilde{\Omega}_{h,n}$ is positive semi-definite hence the $\widetilde{f}_l, \ l=1,\ldots,p_v$ are in descending order given a descending ordering of the eigenvalues $ \widetilde{\lambda}_l, \ l=1,\ldots,p_v$. This ensures that $p(c_n)$ per any of the above selection criteria corresponds to the largest $p(c_n)$ eigenvalues and the corresponding $c_n$ is implicitly defined.

\begin{table}[!htbp]
\centering
\caption{DGP MI 6 - SVD selection criteria}
\begin{tabular}{@{}clccccccccccc@{}}
\toprule
                        &           & \multicolumn{4}{c}{ ($c_n=\widetilde{\lambda}_1n^{-\iota }$) } & \multicolumn{7}{c}{Selection Criteria} \\ \cmidrule(lr){3-6} \cmidrule(lr){7-13}
$\gamma$                & Sig-Lev & $ \iota  = \frac{2}{5} $ & $ \iota  = \frac{1}{3}$ & $ \iota =\frac{1}{4} $ & $ \iota  = \frac{1}{6} $ & R-B & E$.7$ & E$.9$ & TV$.05$ & TV$.10$ & CTV$.7$ & CTV$.9$ \\ \midrule
      &10\%   &0.100 &0.100 &0.100 &0.118 &1.000 &0.100 &0.100 &0.100 &0.111 &0.122 &0.111 \\
0.0   &5\%    &0.050 &0.050 &0.050 &0.061 &1.000 &0.050 &0.050 &0.050 &0.055 &0.062 &0.055 \\
      &1\%    &0.007 &0.007 &0.007 &0.009 &1.000 &0.007 &0.007 &0.007 &0.008 &0.007 &0.008 \\
\midrule
      &10\%   &0.973 &0.973 &0.973 &0.981 &1.000 &0.973 &0.973 &0.973 &0.979 &0.988 &0.979 \\
0.2   &5\%    &0.945 &0.945 &0.945 &0.963 &1.000 &0.945 &0.945 &0.945 &0.958 &0.971 &0.958 \\
      &1\%    &0.822 &0.822 &0.822 &0.853 &1.000 &0.822 &0.822 &0.823 &0.841 &0.901 &0.841 \\
\midrule
      &10\%   &1.000 &1.000 &1.000 &1.000 &1.000 &1.000 &1.000 &1.000 &1.000 &1.000 &1.000 \\
0.4   &5\%    &1.000 &1.000 &1.000 &1.000 &1.000 &1.000 &1.000 &1.000 &1.000 &1.000 &1.000 \\
      &1\%    &1.000 &1.000 &1.000 &1.000 &1.000 &1.000 &1.000 &1.000 &1.000 &1.000 &1.000 \\
\midrule
      &10\%   &1.000 &1.000 &1.000 &1.000 &1.000 &1.000 &1.000 &1.000 &1.000 &1.000 &1.000 \\
0.6   &5\%    &1.000 &1.000 &1.000 &1.000 &1.000 &1.000 &1.000 &1.000 &1.000 &1.000 &1.000 \\
      &1\%    &1.000 &1.000 &1.000 &1.000 &1.000 &1.000 &1.000 &1.000 &1.000 &1.000 &1.000 \\
\midrule
      &10\%   &1.000 &1.000 &1.000 &1.000 &1.000 &1.000 &1.000 &1.000 &1.000 &1.000 &1.000 \\
0.8   &5\%    &1.000 &1.000 &1.000 &1.000 &1.000 &1.000 &1.000 &1.000 &1.000 &1.000 &1.000 \\
      &1\%    &1.000 &1.000 &1.000 &1.000 &1.000 &1.000 &1.000 &1.000 &1.000 &1.000 &1.000 \\
\midrule
      &10\%   &1.000 &1.000 &1.000 &1.000 &1.000 &1.000 &1.000 &1.000 &1.000 &1.000 &1.000 \\
1.0   &5\%    &1.000 &1.000 &1.000 &1.000 &1.000 &1.000 &1.000 &1.000 &1.000 &1.000 &1.000 \\
      &1\%    &1.000 &1.000 &1.000 &1.000 &1.000 &1.000 &1.000 &1.000 &1.000 &1.000 &1.000 \\
\bottomrule
\end{tabular}
\label{Tab:MI_criteria_robust}
{\footnotesize

\textit{Notes}: The first four columns use the regularization technique used in this paper with $c_n=\widetilde{\lambda}_1n^{-\iota }$, $\iota \in \{2/5,1/3,1/4,1/6\} $, respectively. R-B is the ratio-based selection criterion, E$\iota $, $\iota \in \{.7,.9\}$ is the $\alpha$ fraction of total entropy selection criterion, TV$\iota $, $\iota \in \{.05,.10\}$ is the $\alpha$ of total variance selection criterion, CTV$\iota $, $\iota \in \{.7,.9\}$ is the cumulative percentage of the total variance selection criterion.
}
\end{table}

\Cref{Tab:MI_criteria_robust} compares the size control and power performance of the $\chi^2$-test with different choices of the regularization parameter $c_n$. Besides the ratio-based estimator, which fails to deliver a $\chi^2$-test that controls size, the other choices lead to meaningful size control. One observes non-trivial power under $\mathbb{H}_a$. This exercise and that of Scenario 1 confirm the reliability and robustness of the selection rule $c_n = \widetilde{\lambda}_1 n^{-1/3} $ used in this paper.

\subsection{Test of Nullity }\label{subsec:sim_nullity}\label{subsec:nullity}
$\mathbb{H}_o^*: \ \E[U \mid Z] = \E[U] = 0 \ a.s. $ is violated if either $\E[U] \neq 0$ or $\mathbb{P}\big(\E[U \mid Z] = \E[U] \big) < 1 $. To compare the performance of the $\chi^2$-test of the hypothesis of nullity $\mathbb{H}_o^* $, we take the following modified versions of DGP MI 6.

\begin{enumerate}
    \item[MI 6$'$:] $U = 2\gamma\sqrt{3/5} + \mathcal{E} $, $Z \sim\mathcal{N}(0,1)$ and $\mathcal{E} \sim \mathcal{U}[-\sqrt{3},\sqrt{3}] $.
    \item[MI 6$''$:] $U = 2\gamma\exp(-Z^2/3) + \mathcal{E} $, $Z \sim\mathcal{N}(0,1)$ and $\mathcal{E} \sim \mathcal{U}[-\sqrt{3},\sqrt{3}] $.
\end{enumerate}
\noindent $\gamma \neq 0$ corresponds to $\E[U] \neq 0$ in MI 6$'$ and to  $ \mathbb{P}\big\{\E[U \mid Z] = \E[U] \big\} < 1 $ in DGP MI 6$''$.

\begin{table}[!htbp]
\centering
\caption{DGPs MI 6$'$ and MI 6$''$ }
\begin{tabular}{@{}clcccc@{}}
\toprule
                        &           & \multicolumn{4}{c}{DGP MI 6$'$, $ c_n = \widetilde{\lambda}_1n^{-\iota } $} \\ \cmidrule(lr){3-6} 
$\gamma$                & Sig-Lev & $ \iota  = \frac{2}{5} $ & $ \iota  = \frac{1}{3}$ & $ \iota =\frac{1}{4} $ & $ \iota  = \frac{1}{6} $ \\ \midrule
      &10\%   &0.094 &0.094 &0.094 &0.094 \\
0.0   &5\%    &0.049 &0.049 &0.049 &0.049 \\
      &1\%    &0.008 &0.008 &0.008 &0.008 \\
\midrule
      &10\%   &1.000 &1.000 &1.000 &1.000 \\
0.2   &5\%    &1.000 &1.000 &1.000 &1.000 \\
      &1\%    &0.998 &0.998 &0.998 &0.998 \\
\midrule
      &10\%   &1.000 &1.000 &1.000 &1.000 \\
0.4   &5\%    &1.000 &1.000 &1.000 &1.000 \\
      &1\%    &1.000 &1.000 &1.000 &1.000 \\
\midrule
      &10\%   &1.000 &1.000 &1.000 &1.000 \\
0.6   &5\%    &1.000 &1.000 &1.000 &1.000 \\
      &1\%    &1.000 &1.000 &1.000 &1.000 \\
\midrule
      &10\%   &1.000 &1.000 &1.000 &1.000 \\
0.8   &5\%    &1.000 &1.000 &1.000 &1.000 \\
      &1\%    &1.000 &1.000 &1.000 &1.000 \\
\midrule
      &10\%   &1.000 &1.000 &1.000 &1.000 \\
1.0   &5\%    &1.000 &1.000 &1.000 &1.000 \\
      &1\%    &1.000 &1.000 &1.000 &1.000 \\
\bottomrule
\end{tabular}
\quad
\begin{tabular}{@{}clcccc@{}}
\toprule
                        &           & \multicolumn{4}{c}{DGP MI 6$''$, $ c_n = \widetilde{\lambda}_1n^{-\iota } $} \\ \cmidrule(lr){3-6} 
$\gamma$                & Sig-Lev & $ \iota  = \frac{2}{5} $ & $ \iota  = \frac{1}{3}$ & $ \iota =\frac{1}{4} $ & $ \iota  = \frac{1}{6} $ \\ \midrule
            
      &10\%   &0.094 &0.094 &0.094 &0.094 \\
0.0   &5\%    &0.049 &0.049 &0.049 &0.049 \\
      &1\%    &0.008 &0.008 &0.008 &0.008 \\
\midrule
      &10\%   &1.000 &1.000 &1.000 &1.000 \\
0.2   &5\%    &1.000 &1.000 &1.000 &1.000 \\
      &1\%    &0.998 &0.998 &0.998 &0.998 \\
\midrule
      &10\%   &1.000 &1.000 &1.000 &1.000 \\
0.4   &5\%    &1.000 &1.000 &1.000 &1.000 \\
      &1\%    &1.000 &1.000 &1.000 &1.000 \\
\midrule
      &10\%   &1.000 &1.000 &1.000 &1.000 \\
0.6   &5\%    &1.000 &1.000 &1.000 &1.000 \\
      &1\%    &1.000 &1.000 &1.000 &1.000 \\
\midrule
      &10\%   &1.000 &1.000 &1.000 &1.000 \\
0.8   &5\%    &1.000 &1.000 &1.000 &1.000 \\
      &1\%    &1.000 &1.000 &1.000 &1.000 \\
\midrule
      &10\%   &1.000 &1.000 &1.000 &1.000 \\
1.0   &5\%    &1.000 &1.000 &1.000 &1.000 \\
      &1\%    &1.000 &1.000 &1.000 &1.000 \\
\bottomrule
\end{tabular}
\label{Tab:MI_nullity}
\end{table}

\noindent \Cref{Tab:MI_nullity} presents results on DGPs MI $6'$ and MI $6''$ using the framework of \Cref{Tab:MI_cn_robust} but with a focus on the power performance under either violation of $\mathbb{H}_o^*$. From both sets of columns corresponding to DGPs MI $6'$ and MI $6''$, one observes from \Cref{Tab:MI_nullity} that the $\chi^2$-test of nullity has good size control and non-trivial power under violations of $\E[U] = 0$ or $\E[U \mid Z] = \E[U] \ a.s. $

{\section{Monte Carlo Experiments - Specification Test - Non-linear Models}\label{Sect_Suppl:MC_NLM}
In this section, we extend the simulation results in the main text to non-linear models. Specifically, we consider the problem of specification testing of propensity scores as studied in \citet{sant2019specification}. The treatment variable is generated as 
$$
\text{DGP NLM:  }  Y= \mathbbm{1}\Big\{ \theta_c + \sum_{l=1}^5 \big( X_l\theta_l + \gamma (X_l^2 -1 )\big) \geq U \Big\},
$$
 where $ X = Z $. $Z$ and the parameters $ [ \theta_c, \theta_1,\ldots,\theta_5 ]^\top $ are specified as in \Cref{Sect:MC_Sim_Spec} { of} the main text. $U$ follows the standard logistic  distribution. {The logit model is estimated via maximum likelihood. To mitigate the impact of scale differences---often substantial in non-linear models---\(U\) and \(h(Z)\) are normalized to have standard deviations of one and two, respectively, in constructing \(V_h\).\footnote{This roughly preserves the scale ratio used in the linear models.} This normalization ensures the procedure is scale-invariant.
} In the following simulation results, the proposed $\chi^2$-test is compared to the Multiplier Bootstrap test of \citet{escanciano-2024-gaussian} and the propensity score specification test of \citet{sant2019specification} (SS). 

\begin{figure}[H]
\centering 
\caption{DGP NLM -- Gaussian Kernel -- $n=400$.}
\begin{subfigure}{0.32\textwidth}
\centering
\includegraphics[width=1\textwidth]{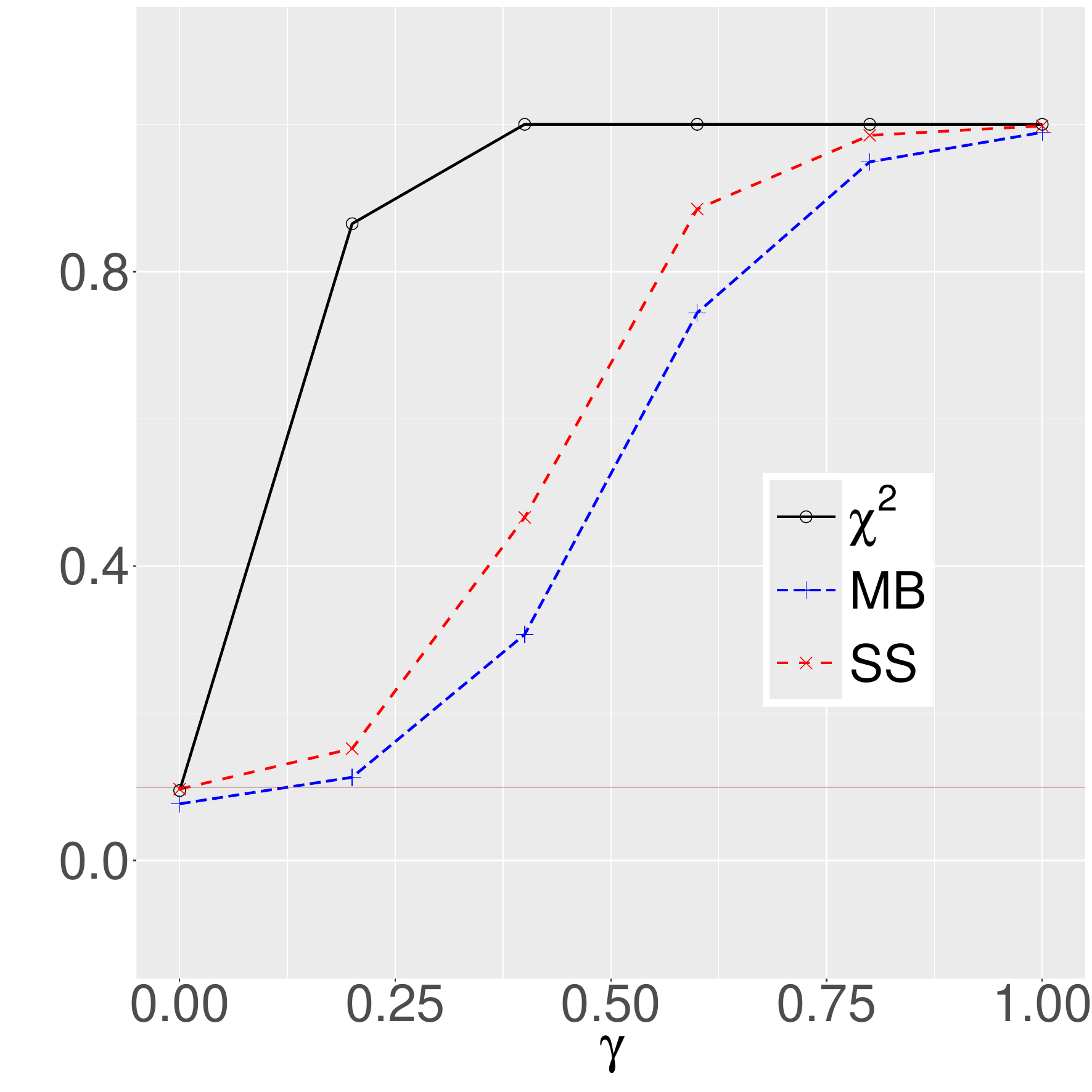}
\caption{10\%}
\end{subfigure}
\begin{subfigure}{0.32\textwidth}
\centering
\includegraphics[width=1\textwidth]{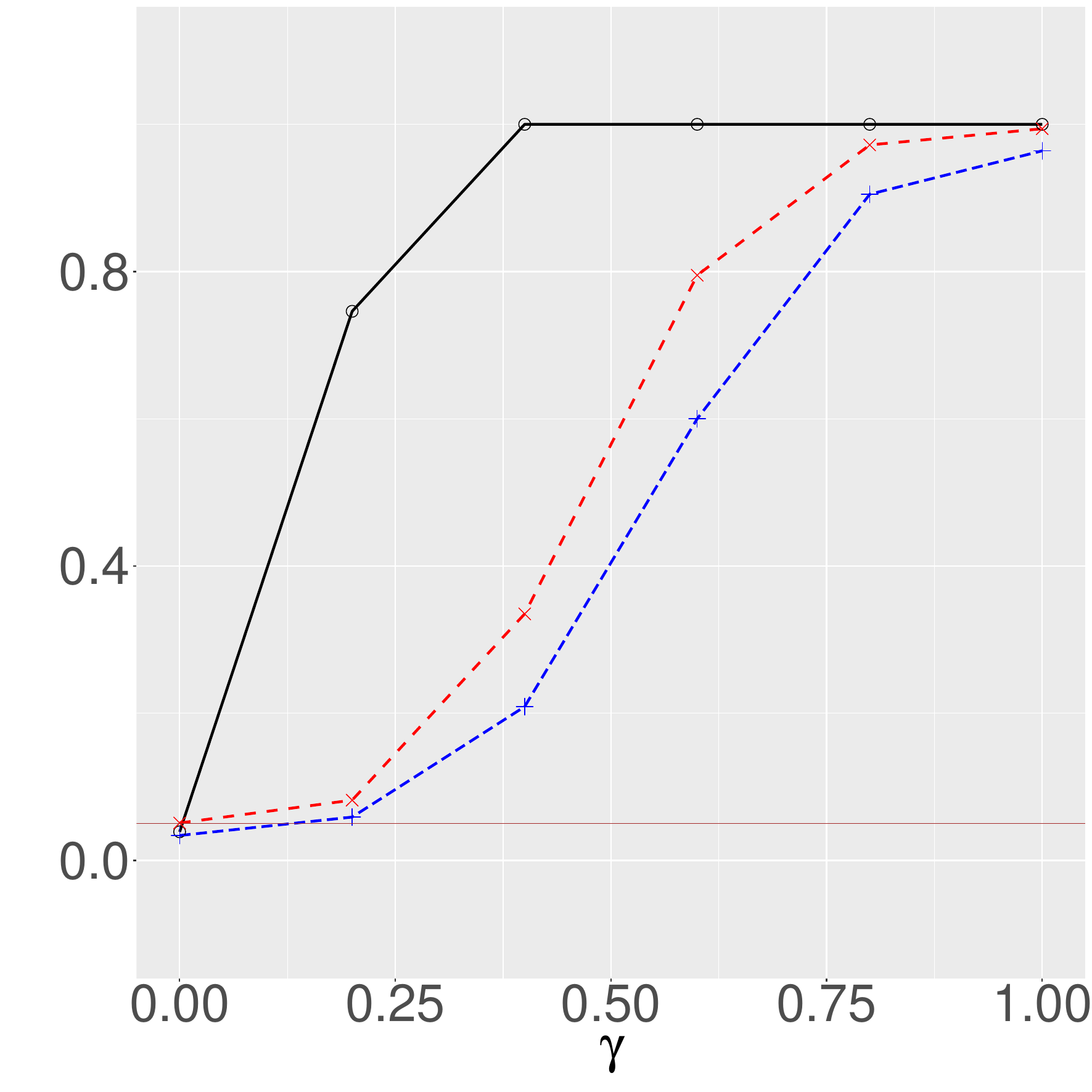}
\caption{5\%}
\end{subfigure}
\begin{subfigure}{0.32\textwidth}
\centering
\includegraphics[width=1\textwidth]{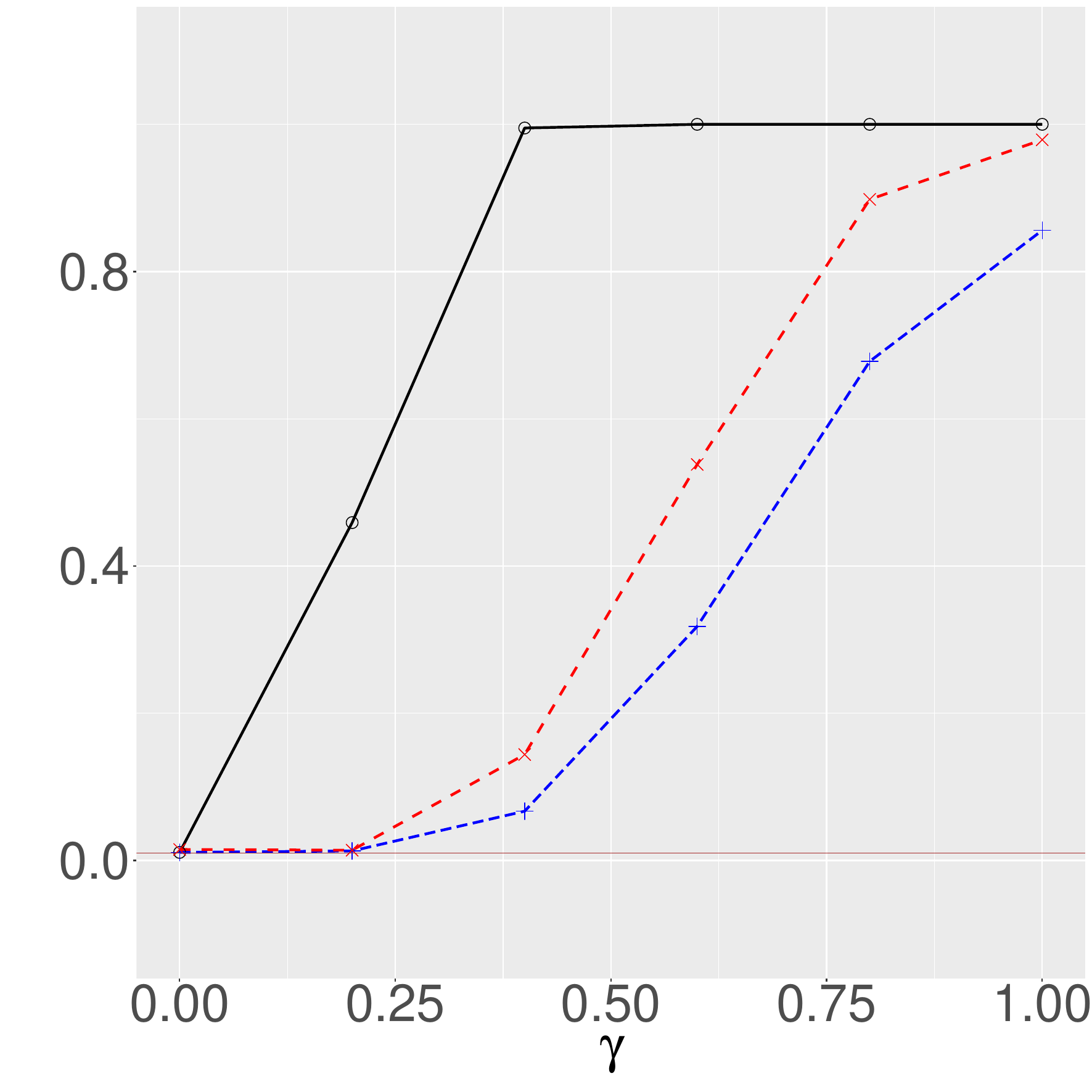}
\caption{1\%}
\end{subfigure}
\label{Fig:NLM_Gauss}
\end{figure}

From the results above in \Cref{Fig:NLM_Gauss}, all tests demonstrate reasonable size control and non-trivial power under the alternative. Quite importantly, the proposed $\chi^2$-test continues to perform in this non-linear setting with a binary limited dependent outcome. Observe that the wild bootstrap-based ICM specification test of, e.g., \citet{su2017martingale} is not directly applicable in this case since the wild bootstrap therein cannot replicate outcomes that retain the two-point support of the {binary} outcome.}

\section{Relation to CM tests}\label{Sect_Appendix:CM_Tests}
The proposed test is rooted in ICM tests, but it also shares the advantages of CM tests  \citep{newey1985maximum,tauchen1985diagnostic}, which are powerful if prior information on $\mathbb{H}_a$ is available. For example, if $\E[U \mid Z]$ can only take certain types of alternatives $f_1(Z),\cdots, f_{p_f}(Z),\ p_f\geq 1$, then setting weight functions in CM tests along the span of these alternatives may yield optimal power \citep{newey1985maximum}. Such power enhancement is also allowed in the proposed test by augmenting $V$ with a vector-valued function of $Z$. In the case of the bivariate $V_h = [h(Z), U-h(Z)]^\top$ in \Cref{lem_key}, power enhancement is also achievable by using $h(Z)$ to target alternatives. CM tests, which are closely related to the proposed test, are based on estimates of the form 
\[
\mathcal{T}_n^{CM}=\sum_{i=1}^n \widetilde{m}(Z_i)\widetilde{U}_i,
\]

\noindent where $\widetilde{m}(\cdot)$ is a vector of non-degenerate weight functions. Although one may argue that $m_{\widetilde{V}}(Z):=\E\big[K(Z,Z^\dagger)(V_h^\dagger - \E V_h) \mid Z\big] $ in our case plays a role similar to $ \widetilde{m}(Z_i) $ in CM tests, there are fundamental differences. 

First, CM tests are not omnibus for any  $\widetilde{m}(Z)$ of fixed dimension. There always exist certain forms $f_1(Z),\cdots, f_{p_g}(Z)$ in $\E[U\mid Z]$ under $\mathbb{H}_a$ such that $ \mathcal{T}_n^{CM}$ has no power; this occurs when $U$ is orthogonal to $\widetilde{m}(Z)$ under $\mathbb{H}_a$. This drawback drew much criticism from the literature and may have triggered the rapid development of ICM tests, see, e.g.  \citet{bierens1982consistent,bierens1990consistent,bierens1997asymptotic,delgado2006consistent}. Although the omnibus property for CM tests can be approximately attained by increasing the dimension of $\widetilde{m}(Z_i)$ via non-parametric techniques such as kernel smoothing (which is effectively what non-parametric tests do, e.g., \textcite{wooldridge1992test,yatchew1992nonparametric,zheng1996consistent}), our proposed specification test remains omnibus with the dimension of $V$ fixed; the proposed $\chi^2$-test can therefore be viewed as a \emph{consistent} CM test. 

Second, our specification test allows $V$ to be linearly dependent on $U$ but CM tests do not. This also distinguishes our test from CM tests as $\mathrm{ICM}(U \mid Z)$ is key to justifying the omnibus property of our test, see the proof of \Cref{lem_key}. Two independent copies $(U,V,Z)$ and $(U^{\dagger},V^{\dagger},Z^{\dagger})$ are jointly included in $\delta_h$ while $\E [\mathcal{T}_n^{CM}] = \E[ \widetilde{m}(Z)U]$ involves only a single copy. If $U$ is linearly included in the construction of $\widetilde{m}(\cdot)$, then most likely $\mathcal{T}_n^{CM}$ is non-null even under $\mathbb{H}_o$. A common feature shared by the proposed test and CM tests is the pivotal limiting distribution of the test statistic. This is achieved thanks to the non-degeneracy of $ \widetilde{m}(Z) $ and $ m_{\widetilde{V}}(Z) $.

\printbibliography
\end{refsection}
\end{document}